\documentclass[review]{elsarticle}
\journal{Information Systems}

\newtheorem{theorem}{Theorem}
\newtheorem{coro}{Corollary}

\newcommand{\D}{\mathcal{D}}
\newcommand{\svs}{\textit{svs}}
\newcommand{\bys}{\textit{bys}}
\newcommand{\merge}{\textit{merge}}

\newcommand{\lookup}{\textit{lookup}}
\newcommand{\repair}{Re-Pair}
\newcommand{\lzend}{LZ-End}
\newcommand{\lzma}{LZMA}

\newcommand{\vbyte}{\texttt{Vbyte}}
\newcommand{\vbyteB}{\texttt{VbyteB}}
\newcommand{\rice}{\texttt{Rice}}
\newcommand{\riceB}{\texttt{RiceB}}
\newcommand{\simplen}{\texttt{Simple9}}

\newcommand{\pfordelta}{\texttt{PforDelta}}
\newcommand{\qmx}{\texttt{QMX}}
\newcommand{\riceRuns}{\texttt{Rice-Runs}}

\newcommand{\vbyteCM}{\texttt{Vbyte-CM}}
\newcommand{\vbyteCMB}{\texttt{Vbyte-CMB}}
\newcommand{\vbyteST}{\texttt{Vbyte-ST}}
\newcommand{\vbyteSTB}{\texttt{Vbyte-STB}}
\newcommand{\repairNo}{\texttt{RePair}}
\newcommand{\repairSkip}{\texttt{RePair-Skip}}
\newcommand{\repairSkipCM}{\texttt{RePair-Skip-CM}}
\newcommand{\repairSkipST}{\texttt{RePair-Skip-ST}}
\newcommand{\vbyteLZMA}{\texttt{Vbyte-LZMA}}
\newcommand{\vbyteLzend}{\texttt{Vbyte-Lzend}}
\newcommand{\rlcsa}{\texttt{RLCSA}}
\newcommand{\wcsa}{\texttt{WCSA}}
\newcommand{\slp}{\texttt{SLP}}
\newcommand{\wslp}{\texttt{WSLP}}
\newcommand{\lzindex}{\texttt{LZ77-index}}
\newcommand{\lzendindex}{\texttt{LZend-index}}

\newcommand{\interpolative}{\texttt{Interpolative}}
\newcommand{\efopt}{\texttt{EF-opt}}
\newcommand{\optpfd}{\texttt{OPT-PFD}}
\newcommand{\varint}{\texttt{varintG8IU}}

\usepackage{fullpage}
\usepackage{comment}
\usepackage{url}
\usepackage{colortbl}
\usepackage[usenames]{xcolor}

\newcommand{\highlight}[1]{\textcolor{red}{#1}}

\begin{document} 
\begin{frontmatter}	
	\title{Universal Indexes for Highly Repetitive Document Collections\tnoteref{t1}}
	\tnotetext[t1]{Funded in part by European Union's
Horizon 2020 research and innovation programme under the Marie
Sk{\l}odowska-Curie grant agreement No 690941, by Fondecyt Grant 1-140796
(Conicyt, Chile), by  MINECO (PGE and FEDER)  grants TIN2013-47090-C3-3-P,
TIN2015-69951-R, and TIN2013-46238-C4-3-R; by MINECO and CDTI grants IDI-20141259, ITC-20151305;
by ICT COST Action IC1302; and  by Xunta de Galicia (co-founded
with FEDER)  grant GRC2013/053 (Spain).
An early partial version of this article appeared in {\em Proc. CIKM 2011.}}

	\author[udp]{Francisco Claude}
	\ead{fclaude@recoded.cl}

	\author[udc]{Antonio Fari\~na\corref{cor1}}
	\ead{fari@udc.es}

	\author[uva]{Miguel A. Mart\'inez-Prieto}
	\ead{migumar2@infor.uva.es}

	\author[uchile]{Gonzalo Navarro}
	\ead{gnavarro@dcc.uchile.cl}

	\cortext[cor1]{Corresponding author}	

	\address[udp]{Escuela de Inform\'atica y Telecomunicaciones,
Universidad Diego Portales, Chile.\\}
	\address[udc]{Database Laboratory, University of A Coru\~na, Spain.\\}
	\address[uva]{DataWeb Research, Department of Computer Science, 
	  University of Valladolid, Spain.}		
	\address[uchile]{Department of Computer Science, University of Chile, Chile.}

	\begin{abstract}
Indexing highly repetitive collections has become a relevant problem
with the emergence of large repositories of versioned documents, among
other applications. These collections may reach huge sizes, but are formed
mostly of documents that are near-copies of others. Traditional techniques 
for indexing these collections fail to properly exploit their regularities in
order to reduce space.

We introduce new techniques for compressing inverted indexes that 
exploit this near-copy regularity. They are based on run-length,
Lempel-Ziv, or grammar compression of the differential inverted lists, 
instead of the usual practice of gap-encoding them. We show that, in this 
highly repetitive setting, our compression methods significantly reduce the
space obtained with classical techniques, at the price of moderate slowdowns.
Moreover, our best methods are universal, that is, they do not need to know
the versioning structure of the collection, nor that a clear versioning 
structure even exists.

We also introduce compressed self-indexes in the comparison. These are 
designed for general strings (not only natural language texts)
and represent the text collection plus the index 
structure (not an inverted index) in integrated form. We show that these
techniques can compress much further, using a small fraction of the space 
required by our new inverted indexes. Yet, they are orders of magnitude 
slower.
\bigskip
\noindent

{\bf Keywords:} {\em repetitive collections, inverted index, self-index.}
	\end{abstract}

\end{frontmatter}

%%%%%%%%%%%%%%%%%%%%%%%%%%%%%%%%%%%%%%%%%%%%%%%%%%%%%%%%%%%%%%%%%%%%%%%%%%%%%%%%%%%%%%%%% 
%%%%%%%%%%%%%%%%%%%%%%%%%%%%%%%%%%%%%%%%%%%%%%%%%%%%%%%%%%%%%%%%%%%%%%%%%%%%%%%%%%%%%%%%% 
\section{Introduction}
Large versioned document collections, such as {\em Wikipedia}
({\tt www.wikipedia.org}) and the {\em Wayback Machine} from the
{\em Internet Archive} ({\tt www.archive.org/web/web.php}),
are examples of the emergence of highly repetitive document collections, where 
most documents are near-duplicates of others. Apart from versioned document 
collections, other applications where this situation arises are software
repositories (where a tree of versions is maintained), biological databases
(where many DNA or protein sequences from organisms of the same or related 
species are maintained), periodic technical publications (where the same data,
with small updates, are published over and over), and so on.

These collections may be very large, but at the same time are highly 
compressible. While Lempel-Ziv compressors \cite{ZL77} are successful in 
capturing their repetitiveness, such compression is suitable only for archival 
purposes. Version control systems offer, in addition, efficient direct access to
individual versions. This was done, since their early beginnings, by storing
edits with respect to some previous version \cite{Roc75}.
The applications we have enumerated require even more: fast searching 
capabilities on all the versions. Thus, we need to compress not only the data, 
but also the indexes built on them to speed up searches.

There is a burst of recent activity in exploiting repetitiveness at the
indexing structures, in order to provide fast searches in the collection 
within little space. Both inverted indexes for word and phrase queries over
natural language texts \cite{AF92,BEFHLMQS06,HYS09,YDS09,HZS10,He:Sigir12}, 
and other indexes for general string collections 
\cite{MNSV09,CFMPN10,CN11,CN12,KNtcs12,GGKNP12,GGKNP14,DJSS14,BCGPR15}, 
have been pursued.

The focus of this paper is on natural language text collections, which can be
decomposed into words, and queried for words or phrases. The classical data
structure to index such collections is the inverted index, where a list of the
occurrences of each distinct word is maintained. The variant where
the lists are sorted by increasing document identifier has gained relevance, 
since such ordering is most useful for list intersections. 
Intersections of inverted lists arise as a fundamental task under the 
Google-like policy of treating bag-of-word queries as ranked 
AND-queries. Therefore, intersections form the heaviest part of the search 
process, and relevance ranking is done as a postprocessing step 
\cite{YDS09sigir,YDS09,DAS10} or via on-the-fly filtration 
\cite{DS11,DNS13,KNCLO13}. Intersections are also fundamental for solving 
phrase queries.

In this context, there are two different types of indexes.
{\em Non-positional} indexes find, given a word or bag-of-words query, the 
documents containing all the words. They store, for each word, the increasing 
list of documents containing it. {\em Positional} indexes retrieve the precise 
positions in each document where a word or phrase query appears. They store, 
in addition to the document identifiers, the word offsets of the occurrences 
within each document.

Traditional techniques to compress inverted indexes 
\cite{WMB99,BYMN02,ZM06,BCC10,BYRN99}
represent the differences between consecutive document or position values.
Many of those differences tend to be small, and thus they
are encoded in a way that favors small numbers. While very effective for
traditional collections, this compression technique generally fails to 
exploit the repetitiveness that arises in versioned collections.

In this paper we introduce new techniques for compressing inverted indexes on 
highly repetitive collections. Instead of using the classical encoding of 
differences, we adapt and apply run-length, Lempel-Ziv \cite{ZL77}, and grammar
\cite{LM00} compression on the lists of differences. Run-length compression
simply exploits successive differences equal to 1 in the lists, and works if the
similar documents are grouped in the ordering. Lempel-Ziv compression captures
more general repetitive patterns that appear within the list of the same word,
and needs to decompress the whole list to access any position of it. Grammar
compression can be applied to the whole set of lists because it can decompress 
portions of any list independently. Therefore, it can also exploit repetitions
across lists. These repetitions are very interesting
because they capture the case of pairs of words that appear in almost the same
documents. We enrich grammar compression with additional information on the
nonterminal symbols, which allows us skip them without decompressing when
a particular document is sought during list intersections.

Our techniques clearly outperform classical inverted index 
compression in this scenario. For example, our Lempel-Ziv-based index is 15
times smaller than a Rice-encoded index (the best choice for classical text
collections), at the modest price of being at most 70\% slower on word and 
conjunctive queries. Our grammar-based compressed index is up to 30 times 
smaller and at most 3 times slower. 
Our experiments also show that some encodings designed for typical collections,
such as Partitioned Elias-Fano \cite{OV14} and PforDelta 
\cite{Hem05,ZHNB06,OV14}, improve substantially on repetitive collections, yet 
they still require 2.5--7 times more space than ours. These modern encodings
are also 2--9 times faster than Rice codes on conjunctive queries.
Other recent codes built on the last generation of SIMD-based techniques 
\cite{trotman2014} are an order of magnitude faster, but they are shown to 
use too much space on repetitive collections.

Except for our run-length compression, which is not the best technique, our
compressed representations are {\em universal}, that is, they do not need to 
identify which document is a version of which. Indeed, there is no need that 
the documents have a well-defined versioning structure at all: it is sufficient
that most documents contain long parts that appear in others. 

We also apply our compression techniques in the positional scenario, where
the lists are less compressible. This time, the method that performs best in 
space is Lempel-Ziv compression, which is around 3.5 times smaller than 
classical inverted indexes, but also 2.5--10 times slower on 
the most common queries. %1-word y frases de 2
The grammar-compressed index is twice as large as the Lempel-Ziv index, but it 
is significantly faster to solve queries, 2--5 times slower
than classical indexes.

In this scenario we also consider {\em self-indexes}, which are compressed 
indexes (not based on inverted indexes) that encode both the text and the index.
We use, adapt, or implement existing self-indexes designed for highly repetitive
collections (mostly for computational biology scenarios) 
\cite{MNSV09,CFMPN10,CN11,CN12,KNtcs12}. Self-indexes obtain sharp space reductions,
being up to 12 times smaller than our Lempel-Ziv-compressed 
inverted index. Their query times, however, are 10--200 times 
slower on the most typical queries. While the time difference is large, this
shows that there is still much redundancy to exploit in the positional scenario.

We have left our codes and experimental testbeds available at
{\tt https://github.com/migumar2/uiHRDC}.

\subsection{Related work}

The most relevant previous work targeting highly repetitive collections of
natural language text is by He et al.~\cite{HYS09,HZS10}. They presented 
alternative compression methods for non-positional indexes on versioned
collections. Their approach, called {\em two-level indexing},
merges all the versions of each document for creating the inverted lists.
A secondary index stores, for each entry of the main inverted list,
a bitmap indicating the versions
of the document that contain the term. They convert previous ``one-level'' 
techniques \cite{AF92,BEFHLMQS06} into two-level methods, and also study 
methods for reordering the versions in order to improve compression. 

They obtained excellent compression results on a non-positional inverted index 
built over subsets of Wikipedia and the Internet Archive collections. The
authors attribute the success of their technique to the effective management 
of the bitmaps in the second level of the index.

Our experiments show that
our techniques still do not match the performance of He et al.'s methods.
However, these work under a restricted model where there exists a set 
of independent documents, each of which has a number of versions, and this 
versioning information is known for indexing. Our universal techniques, 
instead, also 
work on settings where the versions form a tree structure (as in collaborative 
document creation, software repositories, or phylogenetic trees), or where the
versions form a continuous stream of incremental changes 
(as in periodic publications of 
technical data), or where it is unknown or unclear which documents are versions
of which (as in DNA sequence databases, or near-duplicate pages in Web crawls). 

He and Suel \cite{He:Sigir12} also designed a positional inverted index for the
repetitive scenario. They apply a previous technique to partition documents
into fragments \cite{ZS:www07} and then use their non-positional approach
\cite{HZS10} on the fragments. They focus on answering top-k queries,
by first obtaining the top-$k'$ ($k'>k$) documents over the non-positional 
index and then re-ranking them using the positional information in order to 
return the top-$k$ results. This is faster than using the whole positional 
information in the first stage \cite{He:Sigir12,Shieh:2003}.

Although their index serves a different kind of queries than those we study here,
a rough comparison is possible. The space they obtain with their best result
(ZS-FREQ) is about half the 
space of our Lempel-Ziv inverted indexes, whereas their time to extract each
position from the index is around the microsecond, that is, about 5
times slower than our Lempel-Ziv index. Our self-indexes are still up to 6 
times smaller than ZS-FREQ, yet also around 20 times slower. 
Once again, their index is not universal, as it makes explicit use of a known
and specific versioning structure.

%OJO: en la tabla 1 del paper  \cite{He:Sigir12} dicen que nuestro Ziv-lempel obtiene un indice de 1954MB fren al suyo que estaria por 1420Mb, y que finalmente su mejor tecnica (tabla 4) alcanza 1133Mb!). Pero a mi esos numeros no me cuadran del todo: por que dicen que nuestro vbyte-lzma consiguen dejar el indice posicional en <2000Mb, cuando a mi sus numeros me darian unos 5GB!. 
%En cualquier caso, si a su mejor opcion ZS-FREQ le sumamos el tamanho del texto comprimido siempre estarian por encima de los 2.5\% (el texto comprimido con repair-accesible ocupa un 1.21\%)... lo cual de nuevo hacer que VBYTE-lzma ocupa aprox. el doble. Por otra parte,  nuestros autoindices (LZ77-index y WSLP...) por lo menos les ganarian claramente en espacio (alcanzan tamanhos del 0.5\%)... asi que ahi si que cuadran los numeros }

%%%%%%%%%%%%%%%%%%%%%%%%%%%%%%%%%%%%%%%%%%%%%%%%%%%%%%%%%%%%%%%%%%%%%%%%%%%%%%%%%%%%%%%%% 
%%%%%%%%%%%%%%%%%%%%%%%%%%%%%%%%%%%%%%%%%%%%%%%%%%%%%%%%%%%%%%%%%%%%%%%%%%%%%%%%%%%%%%%%% 

%%%%%%%%%%%%%%%%%%%%%%%%%%%%%%%%%%%%%%%%%%%%%%%%%%%%%%%%%%%%%%%%%%%%%%%%%%%%%%%%%%%%%%%%% 
%%%%%%%%%%%%%%%%%%%%%%%%%%%%%%%%%%%%%%%%%%%%%%%%%%%%%%%%%%%%%%%%%%%%%%%%%%%%%%%%%%%%%%%%% 
\section{Basic Concepts} \label{sec:related}

In this section we briefly review the best known strategies to intersect
inverted lists, and then consider compression methods that are compatible 
with those intersection algorithms. We then present Re-Pair and Lempel-Ziv 
compression methods, which are the base of the best inverted list 
representations we propose in Sections~\ref{sec:lists} and 
\ref{sec:list-repair}. Finally, we provide a brief introduction to the
compressed self-indexes that are well-suited to handle repetitive data, which
we will adapt to our positional scenario in Appendix~\ref{sec:self}.

\subsection{Intersection algorithms for inverted lists}
\label{sec:iialg}

The intersection of two inverted
lists can be done in a merge-wise fashion (which is the best
choice if both lists are of similar length), or using a
set-versus-set (\svs) approach where the longer list is searched for
each of the elements of the shortest, to check which should
appear in the result. Either binary or exponential (also called galloping or
doubling) search are
typically used in \svs. The latter checks the list at positions
$i+2^j$ for increasing $j$, to find an element known to be after position
$i$ (but probably close). All these approaches assume that the lists
to be intersected are given in sorted order.

Algorithm \bys~\cite{BY04} is based on binary searching the longer list
$n$ for the median of the smallest list $m$. If the median is found, it is
added to the result set. Then the algorithm proceeds recursively on the left
and right parts of each list. At each new step the longest sublist is
searched for the median of the shortest sublist. Results showed that \bys\
performs about the same number of comparisons than \svs\ with binary search.
As expected, both \svs\ and \bys\ improve upon \merge\ algorithm when 
$|n|\gg |m|$ (actually from $|n| \approx 20 |m|$).

Multiple lists can be intersected using any pairwise 
approach (iteratively intersecting the two shortest lists, and then the
result against the next shortest one, and so on). Other algorithms are
based on choosing the first element of the smallest list as an {\em eliminator}
that is searched for in the other lists (usually keeping track of the
position where the search ended). If the eliminator
is found, it becomes a part of the result. In any case,
a new eliminator is chosen. Barbay et al. \cite{BLOLS09} compared four multi-set
intersection algorithms: {\em i)} a pairwise \svs-based
algorithm; {\em ii)} an eliminator-based algorithm \cite{BK02} (called
{\em sequential}) that chooses the eliminator
cyclically among all the lists and exponentially searches for
it; {\em iii)} a multi-set version of \bys; and {\em iv)} a
hybrid algorithm (called {\em small-adaptive}) based on \svs\ and on
the so-called {\em adaptive algorithm} \cite{DM00}. The adaptive algorithm
recomputes at each step the list ordering according to the elements not 
yet processed, chooses the eliminator from the shortest list, and tries the 
others in order.
The simplest pairwise \svs-based approach (with exponential search)
performed best in practice \cite{BLOLS09}.

\subsection{Data structures for inverted lists}
\label{sec:iids}

The previous algorithms require that lists can be accessed at any
given position (for example those using binary or exponential search)
and/or that, given a value, its smallest successor from a list can
be obtained. Those needs interact with the inverted list compression
techniques.

The compression of inverted lists usually represents each list
$\langle p_1,p_2,p_3,\ldots,p_\ell\rangle$ as a sequence of d-gaps $\langle
p_1,p_2-p_1, p_3-p_2,\ldots,p_\ell-p_{\ell-1}\rangle$, and uses
a variable-length encoding for these differences, for example
$\gamma$-codes, $\delta$-codes, Rice codes, etc.~\cite{WMB99}.
Those methods assign shorter codes to smaller numeric values, this way
taking advantage of the fact that, on longer lists, the d-gaps are shorter.

If the inverted lists stay on disk, minimizing I/O is the key to improve 
performance, and thus the codes achieving the least space are 
preferable. Rice codes are usually the best choice for compressing inverted 
lists in this case \cite{WMB99}.

Recently, there has been an increasing interest on inverted index structures 
designed to reside in main memory (possibly distributed across
several processors) \cite{SC07,CM10,TS10,BCC10}. While space-efficient
representations are still important to reduce communication and number of
processors, the CPU time for traversing the lists in memory
becomes relevant as well.

A proposal in this line uses byte-aligned codes \cite{CM10}, which lose
little compression and are faster at decoding. We will use in particular
\vbyte~\cite{WZ99}, which splits a number into 7-bit chunks and places 
each chunk in a byte, using the highest bit to signal the end of the codeword.

Other representations achieve space even closer to that of Rice \cite{ZLS08}. 
\simplen~\cite{AM05} packs consecutive d-gaps into a 32-bit word. The first 4 
bits signal the type of packing done, depending on how many bits the next 
d-gaps need: we can pack 28 1-bit numbers, or 14 2-bit numbers, and so on (9
combinations in total). 
%\simpled~\cite{ZLS08} slightly optimizes the unused bits of \simplen. 
\pfordelta~\cite{Hem05,ZHNB06} extends this
idea by packing many more numbers, namely 128, while allowing for 10\% of
exceptions that need more bits than the core 90\% of the numbers. The exceptions
are then coded separately afterwards using, say, a variant of \simplen.

In recent years, new word-wise integer representations that take advantage of the large registers (typically 64 or 128-bit) of modern processors and their SIMD-based instruction set have been developed \cite{Anh:2010:ICU:1712666.1712668, Schlegel:2010, Stepanov:2011, trotman2014, Lemire2015:simdInt}. Most of these works mainly target at improving the decoding speed of existing representations by using implementations that take advantage of the capabilities of modern CPUs.

%A variant of the classical \simplen\ using  64-bit machine words permits to improve (mainly) decoding speed on the longest inverted lists, yet the overall impact is reduced.
%or even 128-bit (SIMD-registers) \cite{trotman2014, Lemire2015:simdInt} 

Intersections can be carried out by traversing the lists sequentially.
When one list is much shorter than the other, it is advantageous to provide
direct access so that the longer list can be searched for the elements of the 
shorter one. As said before, it was shown experimentally \cite{BLOLS09} that in practice the 
best is to sort the lists by length, taking the shortest as the ``candidate'' 
list, and iteratively intersect the candidate list with longer and longer 
lists, shortening the candidate list at each step. 

One of the best techniques to intersect 
two lists of very different length \cite{CM10} samples regularly the compressed
list and stores separately the array of samples, which is searched with 
exponential search. 
Given a sampling 
parameter $k$, a list of length $\ell$ is sampled every $k \log_2 \ell$ 
positions. 
Very long lists (more precisely, longer than $u/8$, where $u$ is the largest
document identifier) are replaced by a bitmap \cite{CM10}, which marks which
documents are present in the list. This is both space and time efficient when
the bitmap is sufficiently dense.
Another good method \cite{TS10} regularly samples the universe of 
positions, so that the exponential search is avoided. Given a parameter $B$, 
it samples the universe of size $u$ at intervals 
$2^{\lceil \log_2 (uB/\ell) \rceil}$.

\subsection{Re-Pair compression algorithm}
\label{sec:repair}

{\em \repair} \cite{LM00} consists of repeatedly
finding the most frequent pair of symbols in a sequence of integers and
replacing it with a new symbol, until no more replacements are useful.
More precisely, \repair\ over a sequence $L$ works as follows:
(1) It identifies the most frequent pair $ab$ in $L$; 
(2) It adds the rule $s\rightarrow ab$ to a dictionary $R$, where $s$ is a
new symbol not appearing in $L$; 
(3) It replaces every occurrence of $ab$ in $L$ by $s$;
%\footnote{As far as
%possible, e.g., one cannot replace both occurrences of $aa$ in $aaa$.}
(4) It iterates until every pair in $L$ appears once.

\repair\ can be implemented in linear time \cite{LM00}. We call $C$ the sequence resulting from $L$ after compression. Every symbol in
$C$ represents a {\em phrase} (a substring of $L$), which is of length $1$
if it is an original symbol (called a {\em terminal}) or longer if it is an
introduced one (a {\em nonterminal}). Any phrase can be recursively expanded
in optimal time (i.e., proportional to its length). Note that replaced
pairs can contain terminal and nonterminal symbols.

Larsson and Moffat \cite{LM00} proposed a method to compress the set of rules
$R$. In this work we prefer another method \cite{GNFjea14}, which is not so
effective but allows accessing any rule without decompressing the whole set of
rules. It represents the DAG of rules as a set of trees. Each tree
is represented as a sequence of leaf values (collected into a sequence $R_S$)
and a bitmap that defines the tree shapes in preorder (collected into a bitmap
$R_B$). Nonterminals are represented by the starting position of their tree (or
subtree) in $R_B$. In $R_B$, internal nodes are represented by 1s and leaves
by 0s, so that the value of the leaf at position $i$ in $R_B$ is found at
$R_S[rank_0(R_B,i)]$. Operation $rank_b$ counts the number of $b$s in $R_B[1,i]$
and can be implemented in constant time, after a linear-time preprocessing that
stores only $o(|R_B|)$ bits of space
\cite{Cla96} on top of the bitmap. 
To expand a nonterminal, we traverse $R_B$ and extract the leaf
values, until we have seen more 0s than 1s. Leaf values corresponding to
nonterminals must be recursively expanded. 

\begin{figure*}[htbp]
\begin{center}
\includegraphics[width=0.60\textwidth]{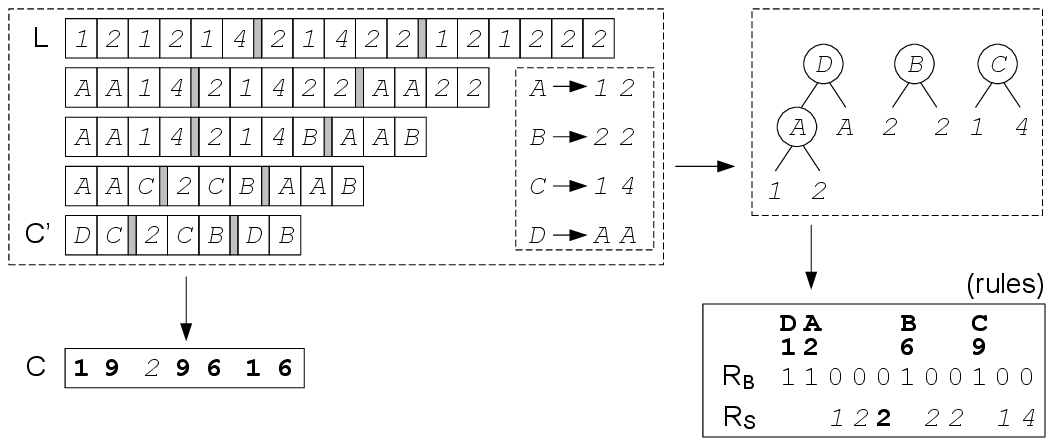}
\caption{Example of \repair\ compression.
Solid boxes ($C$ and {\em rules}) enclose the data represented. Bold numbers (nonterminals) in the final $C$ sequence
and $R_S$ refer to positions in $R_B$, whereas slanted ones (terminals) refer to original values.
To distinguish them, the maximum offset $u$ is added to the bold numbers.}
\label{fig:example1}
\end{center}
\end{figure*}

Figure~\ref{fig:example1} shows an example. 
 %Consider the second box (gaps) as the text to be compressed. 
 %
Consider $L$ as the sequence to be compressed (where the shaded boxes indicate that 
pairs cannot include symbols crossing such boxes; we will need this in Section~\ref{sec:list-repair}).
The most frequent pair in $L$ is $\langle \mathit{1},
\mathit{2}\rangle$. Hence we add rule $A \rightarrow \mathit{1\,2}$ to the
dictionary $R$ and replace all the occurrences of $\mathit{1\,2}$ by nonterminal
$A$. We go on replacing pairs; note that the fourth rule $D \rightarrow AA$
replaces nonterminals. In the final sequence $D\,C\,\mathit{2}\,C\,B\,D\,B$,
no repeated pair appears. We now represent the dictionary of four rules as
a forest of four small subtrees. Now, as nonterminal $A$ is used in the
right-hand side of another rule, we insert its tree as a subtree of one such
occurrence, replacing the leaf. Other occurrences of $A$ are kept as is (see the
rightmost box in the first row). This will save one integer in the
representation of $A$. The final representation is shown in the rightmost box
of the second row.
In $R_B$, the shape of the first subtree (rooted at $D$) is represented by
$11000$ (the first 1 corresponds to $D$ and the second to $A$); the other two
($B$ and $C$) are $100$. These nonterminals will be further identified by the
position of their 1 in $R_B$: $D=\mathbf{1}$, $A=\mathbf{2}$, $B=\mathbf{6}$,
$C=\mathbf{9}$. Each 0 (tree leaf) corresponds to an entry in $R_S$, containing
the leaf values: $\mathit{12}A = \mathit{12}\mathbf{2}$ for the first subtree,
and $\mathit{22}$ and $\mathit{14}$ for the others. Nonterminal positions (in
boldface) are in practice distinguished from terminal values (in italics) by
adding them the largest terminal value. Finally, sequence $C$ is
$\mathbf{1\,9}\,\mathit{2}\,\mathbf{9\,6\,1\,6}$.%
\footnote{By adding the 
largest terminal value ($l=4$) to the values in $C$ we would obtain $C'= 5\, 13\, 2\, 13\, 10\, 7\, 10$. Values
$C'[i] >4$ are the nonterminals $C'[i]-4$, and the others are terminals.} 
To expand, say, its sixth
position ($C[6]=\mathbf{1}$), we scan from $R_B[1,\ldots]$ until we see more 0s
than 1s, i.e., $R_B[1,5] = 11000$. Hence we have three leaves, namely the first
three 0s of $R_B$, thus they correspond to the first three positions of
$R_S$, $R_S[1,3] = \mathit{12}\mathbf{2}$. Whereas $\mathit{12}$ are already
final (i.e., terminals), we still have to recursively expand $\mathbf{2}$.
This corresponds to subtree $R_B[2,4] = 100$, that is, the first and second 0
of $R_B$, and thus to $R_S[1,2] = \mathit{12}$. Concatenating, $C[6]$ expands
to $\mathit{1212}$.

\subsection{Lempel-Ziv compression}
\label{sec:lzend}

The Lempel-Ziv 1977 ({\em LZ77}) compression algorithm \cite{ZL77} works by decomposing
 the text from left to right into phrases. Each phrase $i$ is represented as a 
triplet $\langle k,l,a \rangle$: a pointer to a previous position $k$ in the text, the length $l$ 
of the portion to be copied, and a new symbol $a$ to be added. This means 
that %if phrase $i$ points to position $k$ in the text, has length $\ell$, and new symbol $a$, 
% then phrase $i$ represents the substring of length $\ell$ at position $k$ with symbol $a$ added at the end. 
phrase $i$ represents the substring $text[k,k+l-1]$ with symbol $a$ added at the end.
The parsing maximizes the length of the phrases being added in order to obtain a small encoding. 
The main drawback of the LZ77 parsing is that it does not support random access, 
and even constructing the substring represented by a single phrase can be expensive. The 
pointers point to text positions and not phrases, thus obtaining more than just the last 
symbol of a phrase might be arbitrarily costly. Only decompression from the beginning is efficient.

An alternative parsing called {\em \lzend} \cite{KNtcs12} solves this 
problem and achieves constant time per symbol when retrieving a whole phrase 
or a suffix of it. The main idea is to limit the positions where the source of 
a phrase may end, so that sources can only end where a previous phrase ends. 
Although the parsing is stricter than that of LZ77, the compression was shown
to be competitive on highly repetitive collections \cite{KNtcs12}.

The representation of the \lzend\ parsing that supports random access to a 
sequence $L[1,n]$ is as follows. We keep the triplets corresponding to the 
phrases, but the pointer only indicates the phrase number where the
source ends, and lengths are replaced by a bitmap $B[1,n]$ that marks where 
each phrase ends. This sparse bitmap is represented with gap-encoding of the 
distances between consecutive ones, plus some sampled absolute values, so as
to support $rank$ and $select$ queries \cite{Cla96}. Operation $select_b(j)$ 
obtains the position where the $j$-th bit $b$ appears.
To extract the content of the phrase $p$, with pair $\langle k,a\rangle$, we
recursively extract the content of phrase $k$ and then output $a$ (the
recursion terminates when $k$ is void). This takes $O(1)$ time per character
extracted. In general, for extracting a snippet $L[i,j]$, we extract the longer
one $L[i,j']$, where $j'=select_1(B,p)$ is the final position of the $p$th 
phrase, and $p=rank_1(B,j)+1$ is the phrase where position $j$ falls. As we
have now a sequence of complete phrases, possibly started by a phrase suffix, 
we can extract the string in time $O(j'-i)$. This is generally efficient if 
$j-i$ is not too small compared to the average phrase length.

\subsection{Compressed self-indexes}
\label{sec:indexes}

Self-indexes are data structures that enable efficient searches over an arbitrary string 
collection (called the text), and also replace the text by supporting extraction of arbitrary 
snippets or documents. The supported searches obtain all the positions of a substring in the
collection. This enables self-indexes to compete in the positional setting.

Self-indexes have undergone much progress in the last decade \cite{NM07}. Recently they have 
been adapted to index highly repetitive sequences 
\cite{MNSV09,CFMPN10,CN11,CN12,KNtcs12,GGKNP12,GGKNP14,DJSS14,BCGPR15}. 
While general self-indexes have been successful by targeting statistical compression, they have 
been proved insufficient on highly repetitive collections \cite{MNSV09,KNtcs12}. 

The first self-index successfully capturing high repetitiveness was the \rlcsa~\cite{MNSV09} (for 
Run-Length Compressed Suffix Array). This index adapts the well-known \verb|CSA| of Sadakane \cite{Sad03} 
to better cope with the regularities that arise when indexing highly repetitive sequences. A \verb'CSA' 
variant aimed at indexing natural language, \wcsa~\cite{FBNCPR12} (for Word CSA), regards the text as a 
sequence of words and separators instead of characters.
Another index aimed at repetitive collections is the \slp~\cite{CN11,CFMPN10}, which exploits the regularities 
of highly repetitive sequences because its structure is determined by a grammar-based compressor (\repair). We 
adapt the \slp\ to words (\wslp) in this paper. 
Finally, other strong indexes for repetitive sequences are \lzindex\ and \lzendindex~\cite{KNtcs12}, which are
based on the LZ77 or \lzend\ compression algorithms, respectively. The LZ77 parsing is at least as powerful as 
any grammar representation \cite{Ryt02}, and thus, becomes also a good candidate for highly repetitive sequences.
All these self-indexes are explained in Appendix \ref{sec:self}.

%%%%%%%%%%%%%%%%%%%%%%%%%%%%%%%%%%%%%%%%%%%%%%%%%%%%%%%%%%%%%%%%%%%%%%%%%%%%%%%%%%%%%%%%% 
%%%%%%%%%%%%%%%%%%%%%%%%%%%%%%%%%%%%%%%%%%%%%%%%%%%%%%%%%%%%%%%%%%%%%%%%%%%%%%%%%%%%%%%%% 
\section{New List Representations} \label{sec:lists}

We present inverted list compression schemes for repetitive collections
capturing progressively more sophisticated regularities. First, we consider
Run-Length and Lempel-Ziv compression, and in Section~\ref{sec:list-repair} we
use grammar compression (\repair, precisely). The latter is enriched so that
the list can be processed in compressed form, without the need to fully
decompress it when carrying out intersections. 

Both non-positional and positional indexes will consist of essentially
lists of increasing numbers. In the first case this will be a list of document
identifiers, and list intersections will correspond to conjunctive queries. 

For positional indexes, the lists will be sequences of word positions. 
We consider the collection as a concatenation $\D$ 
of texts, with separators between consecutive documents to avoid 
false matches. Then our positional indexes store the positions (word offsets) 
where each word occurs in $\D$. 

To find a phrase on a positional index, we 
use a modified intersection process to take into account the (word) offsets. 
Consider a phrase $w_1 w_2 \ldots w_m$; we need to find the positions 
$p_1,\ldots,p_k$ in the list of $w_1$ such that $p_i+j$ is in the list of 
$w_{j+1}$ for $1\le j< m$. For simplicity, in our descriptions 
we will consider plain intersections, yet adapting them to phrase queries
is straightforward.

The absolute positions resulting from a query on a positional index are
translated back to a document number and a word offset within the document by
means of an array where the document starting positions are stored in plain
form. Once the index returns the increasing list of positions where
a word or phrase appears, the list is traversed and each element is found in
the array of document starting positions using exponential search, starting
from the position of the previous translated occurrence. This way, $o$
occurrences are translated in time $O(o(1+\log\frac{d}{o}))$, where
$d$ is the number of documents.

Our basic idea, for both non-positional and positional indexes, 
is to differentially encode the 
inverted lists, transforming a sequence $\langle p_1,p_2,p_3,\ldots,p_\ell
\rangle$ into the d-gap sequence $\langle p_1,p_2-p_1,p_3-p_2,\ldots,
p_\ell-p_{\ell-1}\rangle$, and then apply a general compression algorithm to 
the sequence formed by the concatenation of all the lists. Each
vocabulary word will store a pointer to the beginning of its list in the 
compressed data that permits us to fetch any inverted list individually.

\label{sec:simple}

\subsection{Using run-length compression}
\label{sec:rle}

A typical regularity in highly repetitive collections arises when the 
versions of a document happen to receive consecutive identifiers. As most
of the words in such documents will appear in all versions, a consecutive 
sequence of numbers will appear in each inverted list of non-positional
indexes. Consider the word $w_t$ appearing in documents $d_i,\ldots,d_{i+k}$: 
the d-gapped list for $w_t$ will contain $k-1$ consecutive ones. 

In this representation we use any variable-length encoding for the 
differences. However, when this difference is 1, the next encoded number
is the number of 1s in that run ($k$, in the previous paragraph). This
number is encoded in the same way as the d-gaps.

This encoding allows us to skip whole runs in a single operation when
processing intersections.
It also allows combining with sampling techniques that support intersection
methods other than the sequential one \cite{CM10,TS10}.
In this paper we combine run-length compression with Rice coding, giving
rise to method \riceRuns.

Note that run-length compression works well only under the assumption that the 
documents can be linearized so that close documents receive consecutive
numbers. While such kinds of assumptions are used in previous work
\cite{HYS09,HZS10}, we aim to handle more general cases in
this paper. Moreover, this technique can be efficient only for non-positional
indexes.

\subsection{Using LZMA} \label{sec:lzma}

This representation (already used for compressing q-gram indexes on DNA
\cite{CFMPN10}) encodes each d-gap list with \vbyte\ and then compresses it
with the {\em \lzma} variant of LZ77 ({\tt www.7-zip.org}). 
\lzma\ is applied only on the lists where it reduces space. Otherwise, the 
plain \vbyte\ encoded sequence is stored. A bitmap indicates which inverted 
lists were stored compressed with \vbyte\ plus \lzma, and which only with 
\vbyte.

This representation, called \vbyteLZMA, only supports extracting a list from 
the beginning, that is, we cannot jump to a random position on the list, and 
thus the only intersection algorithm supported is the sequential one. Moreover,
unlike run-length compression, it cannot skip a compressed subsequence without 
fully processing it.

\lzma\ handles more complex regularities than run-length compression. In 
particular, it also works well on positional indexes: Consider 
a long substring $S$ that occurs in $r$ similar documents across the 
collection. For each word $w_t$ in $S$, with occurrences at relative positions 
$i_1,i_2,\ldots,i_k$ in $S$, the sublist $i_2-i_1,\ldots,i_k-i_{k-1}$ appears 
$r$ times in the list of word $w_t$. Hence, \lzma\ will capture this repetition
and represent $r-1$ of those sublists with just one reference. Note that this
will occur independently of whether the versions are consecutive, and even
without any need to know which documents are close versions of which.

\subsection{Using LZ-End}

This method works similarly to the previous one, but instead of compressing 
the lists with \lzma, it uses \lzend. Since \lzend\ allows random access, the 
sequence of all the concatenated lists is compressed as a whole, not list-wise. 
Therefore, we first concatenate the \vbyte\ representation of all the posting
lists (keeping track of where each posting list started in the \vbyte\ sequence), 
and then use \lzend\ to represent it. This makes up our  \vbyteLzend\ representation.

Note that, since the phrases of the parsing are not limited by the ends of lists, this
technique does not use an array of pointers from words to compressed data
to mark the beginning of the inverted lists, but pointers to the positions
in the original (\vbyte) sequence. Then, the \lzend\ capability to extract arbitrary
substrings is used.

Although \lzend\ is weaker than \lzma, it has the potential of spotting 
inter-list regularities, whereas \lzma\ is limited to intra-list regularities. 
In the same example of Section~\ref{sec:lzma},
consider a long substring $S$ occurring $r$ times in the collection, and that 
we have a phrase $w_{t_1} w_{t_2} \ldots w_{t_s}$ occurring $k$ times in $S$,
at relative offsets $i_1,i_2,\ldots,i_k$. Then the sequence 
$i_2-i_1,\ldots,i_k-i_{k-1}$ will appear 
$r$ times inside each of the $s$ lists, and \lzend\ compression will be able 
to replace $rs-1$ of the occurrences by a single reference. \lzma\ would
have replaced only $rs-s$.

\lzend\ also captures these regularities in the non-positional case.
Let $d_{j_1},\ldots,d_{j_r}$ be the $r$ documents where the phrase 
$w_{t_1} w_{t_2} \ldots w_{t_s}$ appears. Then, we will have the sequence
$d_{j_2}-d_{j_1},\ldots,d_{j_r}-d_{j_{r-1}}$ repeated in the lists of the
words that do not appear in other documents in between. Again, by compressing 
globally, \lzend\ is exploiting inter-list similarities that \lzma\ could not 
detect.

%%%%%%%%%%%%%%%%%%%%%%%%%%%%%%%%%%%%%%%%%%%%%%%%%%%%%%%%%%%%%%%%%%%%%%%%%%%%%%%%%%%%%%%%% 
%%%%%%%%%%%%%%%%%%%%%%%%%%%%%%%%%%%%%%%%%%%%%%%%%%%%%%%%%%%%%%%%%%%%%%%%%%%%%%%%%%%%%%%%% 

\section{Re-Pair Compressed Lists}
\label{sec:list-repair}
As in the representation based on \lzend, we aim at globally compressing the
list of d-gaps. However, we will use a grammar compressor (\repair) and will
operate over the integer values rather than over their \vbyte\ encodings %
%\footnote{Operating over \vbyte\ increased the space by 10\% in preliminary
%experiments.}
%\lzma\ is limited to spotting intra-list regularities. A grammar-based compressor
%aimed at globally compressing the lists of d-gaps could spot inter-list 
%similarities as well, as in the case of \lzend. In the same example of Section~\ref{sec:lzma}, consider 
%that we have a phrase $w_{t_1} w_{t_2} \ldots w_{t_s}$ occurring $k$ times in 
%$S$. Then the sequence $i_2-i_1,\ldots,i_k-i_{k-1}$ will appear $r$ times 
%inside each of the $s$ lists, and a grammar compressor could be able to 
%replace $rs-1$ of the occurrences by a single reference. \lzma\ would have 
%replaced only $rs-s$.
%
%We will use \repair\ as our grammar compressor. We will
%operate over the integer values and not over their \vbyte\ encodings.%
(using the latter increased the space by 10\% in preliminary
experiments).
%
%477947236 uint32 (comprimido con IRePair)
%=========================================
%2121832 uint32.C
%1665764 uint32.R
%
%120039246 vbyte (comprimido con RePair)
%=======================================
%2960372 2gbnopositional.posts.vbyte.C
%1232172 2gbnopositional.posts.vbyte.R

\begin{figure*}[tbp]
\begin{center}
\includegraphics[width=0.7\textwidth]{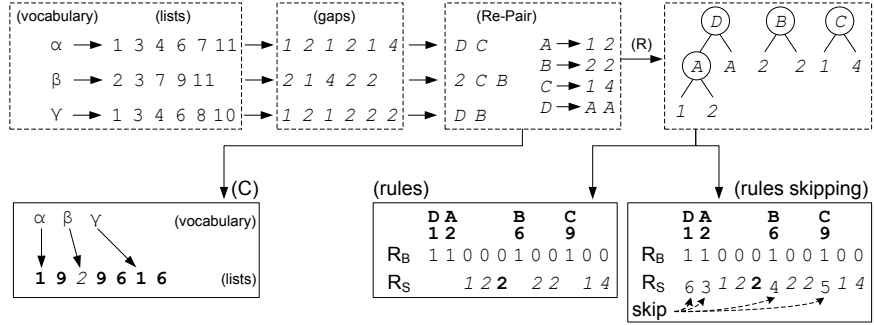}
\caption{Example of inverted lists compressed with our \repair-based method.
Solid boxes enclose the data we finally keep. We include both variants of compressed rules with data
aligned to the 1s in $R_B$ (skipping) or not. Again, bold numbers (nonterminals) in the lists
and $R_S$ refer to positions in $R_B$, whereas slanted ones (terminals)
refer to gap values.}
%\caption{Example of inverted lists compressed with our \repair-based method.
%Solid boxes enclose the data we represent. We include both variants of data
%aligned to the 1s in $R_B$. Bold numbers (nonterminals) in the lists
%and $R_S$ refer to positions in $R_B$, whereas slanted ones (terminals)
%refer to gap values. To distinguish them, the maximum offset $u$ is actually
%added to the bold numbers.}

\label{fig:example}
\end{center}
\end{figure*}

We prevent phrases from spanning multiple lists, which can be easily enforced
at compression time. We simply have to add a separator to mark the beginning of each posting
list ($-1$ for the first posting list, $-2$ for the second, $-3$ for the third, and so on) prior to
performing the Re-Pair process. Since those separators occur only once, \repair\ will not
use them in any phrase. Therefore, we can remove them directly from the compressed sequence $C$ after 
Re-Pair completes (see the shaded boxes in Figure~\ref{fig:example1}).
We can store pointers from the vocabulary to the 
points in the compressed sequence 
$C$ where the lists begin, so that any list 
can be expanded in optimal time (i.e., proportional to its uncompressed 
size), by expanding all the symbols of $C$ from its vocabulary 
pointer to the next. We store the \repair\ dictionary in the compact 
format described in Section~\ref{sec:repair}. 

The terminal symbols are directly the corresponding 
differential values, for example, value $C[3]=\mathit{2}$ is represented by the 
terminal integer value 2. This saves table accesses at decompression time.
Figure~\ref{fig:example} shows the complete inverted lists representation. Note that the \repair\ process 
is identical to that in Figure~\ref{fig:example1}.

\subsection{Using skipping data}
\label{sec:skipping}

An attractive feature of grammar compression is that we can add 
extra information to nonterminals that enables fast skipping over the 
compressed list data without decompressing. This yields much faster sequential
list intersections.

The key idea is that {\em
nonterminals also represent differential values}, namely the sum of the
differences they expand to. We call this the {\em phrase sum} of the
nonterminal. In our example of Figure~\ref{fig:example}, as 
$D = \mathbf{1}$ expands into $\mathit{1212}$, its
phrase sum is $1+2+1+2 = 6$. If we store this sum associated to $D$, we can
skip it in the lists without expanding it, by knowing that its symbols add up 
to 6.

Phrase sums will be stored in sequence $R_S$, aligned with the 1s of
sequence $R_B$. Thus $rank$ is not anymore necessary to move from one sequence
to the other. The 0s in $R_B$ are aligned in $R_S$ to the leaf data, and the 1s
to the phrase sums of the corresponding nonterminals.

In order to find whether a given document $d$ is in the compressed list, we
first scan the entries in $C$, adding up in a sum $s$ the value $C[i]$ if it
is a terminal, or $R_S[C[i]]$ if it is a nonterminal. If at some point we
reach $s=d$, then $d$ is in the list. If instead we reach $s>d$, we
consider whether the last $C[i]$ processed is a terminal or not. If it is a
terminal, then $d$ is not in the list. If it is a nonterminal, we restart the
process from $s-C[i]$ and process the $R_S$ values corresponding to the 0s
in $R_B[C[i],\ldots]$, recursing as necessary until we get $s=d$ or $s>d$
after reading a terminal.

In our example of Figure~\ref{fig:example}, assume we want to know whether
document 9 is in the list of word $\beta$. We scan its list $\mathit{2}\,C\,B =
\mathit{2}\,\mathbf{9\,6}$, from sum $s=0$. We process $\mathit{2}$, and since
it is a terminal we set $s = s+2 = 2$. Now we process $\mathbf{9}$, and since
it is a nonterminal, we set $s = s + R_S[9] = s+5 = 7$ (note the $5$ is
correct because $\mathbf{9} = C$ expands to $\mathit{1\,4}$). Now we process
$\mathbf{6}$, setting $s = s + R_S[6] = s+4 = 11$. We have exceeded $d=9$,
thus we restart from $s=7$ and now process the zeros in $R_B[6,\ldots] =
100\ldots$. The first 0 is at $R_B[7]$, and since $R_S[7] = \mathit{2}$ is a
terminal, we add $s = s +R_S[7] = 9$, concluding that $d=9$ is in the list.
The same process would have shown that $d=8$ was not in the list.

\subsection{Using sampling}
\label{sec:sampling}

%\no{
  %We return to sampling now.
  Apart from the skipping capabilities above, we can add sampling to speed up
  the access to sequence $C$.  
 Depending on whether we want to use strategies of type \svs\ or \lookup\ for
  the search, we can add the corresponding sampling of absolute values to the
  \repair\ compressed lists. For \svs\ we will sample $C$ at regular positions,
  %(i.e., (a)-sampling),
  and will store the absolute values preceding each sample. The pointers to $C$
  are not necessary, as both the sampling and the length of the entries of
  $C$ are regular. This is a plus compared to classical gap encoding
  methods. %\highlight{OJO: en la pr\'actica si que hay ptrs para poder acceder
  %a los samples de cada lista individualmente, pues en svs-CM el periodo de muestreo
  %depende de la longitud de cada lista. En nuestro caso, del número de entradas en $C$
  %que corresponden a cada lista}
  %
  Strategy \lookup\ will insert a new sample each time the absolute
  value surpasses a new multiple of a sampling step.
  % (i.e., (b)-sampling). 
  Now we need to store
  pointers to $C$ (as in the original method) and also the absolute values
  preceding each sample (unlike the original method). The reason is that the
  value to sample may be inside a nonterminal of $C$, and we will be able to point
  only to the (beginning of the) whole nonterminal in $C$. Indeed, several
  consecutive sampled entries may point to the same position in $C$.

  In Figure~\ref{fig:example}, imagine we wish to do the second kind of sampling on list $\gamma$,
  for $2^k=4$ (including the first element too). Then the 
  samples should point at positions 1, 3, and 5, of the
  original list $\gamma = \langle  \underline{1}~ 3~  \underline{4}~ 6~  \underline{8}~ 10  \rangle$. 
  But this list is compressed into $D\,B$, so the first two
  pointers point to $D$, and the latter to $B$. That is, the sampling array stores the pairs  
  $[(0,1),\,(0,1),\,(6,2)]$. Its third entry $(6,2)$, for example, means that the first 
  element $\ge 8$
  is at the 2nd entry ($B$) in its compressed list, and that we should start from value
  6 when processing the differences. For example, if we wish to access the first
  list value exceeding 4, we should start from $(0,1)$, that is, accumulate
  differences from $D$ (the first entry), starting from value 0, until exceeding 4.
%}

\subsection{Intersection algorithm}

To intersect several lists, we sort them in increasing order of their
{\em uncompressed} length (which we store separately). Then we
proceed iteratively, searching in step $i$ the list $i+1$ for the elements of
the candidate list (recall Section~\ref{sec:iids}). 

At each intersection, the candidate list is sequentially traversed.
Let $x$ be its current element. We skip phrases of list $i+1$, accumulating
gaps until exceeding $x$, and then consider the previous and current cumulative
gaps, $x_1 \le x < x_2$. Thus, the last phrase represents the range $[x_1,x_2)$.
If $x_1=x$ we report $x$ and shift to the next element in the candidate list.
In either case, we advance in the candidate list until finding the largest 
$x' < x_2$. Then, we process the whole interval $[x,x']$ within the phrase 
representing $[x_1,x_2)$ in $R_S$ in a recursive fashion, until one of the
two lists gets empty.

This procedure can be speeded up by adding samples. The most promising choice
is the one \cite{CM10} that
simply stores the absolute value 
of every $s$-th entry in $C$. Then, instead of sequentially traversing the 
phrases in list $i+1$ looking for $x$, we can exponentially search for $x$ in 
the samples and then sequentially traverse only one chunk of length at most 
$s$. Samples that are regular on the universe \cite{TS10} are not so efficient
because \repair\ does not give direct access to arbitrary list elements.
Moreover, note that once we switch to the recursive search inside $R_S$ we 
do not have any sampling to speed up the scanning, so the impact of 
sampling is limited.

\begin{comment}
Note that these samplings also enable implementing intersection algorithms that 
require direct accesses to the inverted lists. These methods would, however,
experiment a slowdown because, as explained, we can only have direct access 
to the \repair\ phrase beginnings, that is, to integers in the compressed 
sequence $C$. 
\end{comment}

\begin{comment}
\highlight{OJO, revisar}
In the experimental section, we will mainly focus on a variant 
that stores skipping information on nonterminals (\repairSkip) and one 
that does not (\repairNo). Yet, we will also include some experiments with
variants \repairSkipCM\ and \repairSkipST\ to show that, as expected 
in a repetitive scenario (where a single entry in $C$ could expand into 
a large number of terminals), adding sampling does not actually improve the
results of simply using skipping information.
\end{comment}

In the experimental section, we will present four \repair\ variants. The first
one (\repairSkip)
stores skipping information on nonterminals and no sampling. The second
one (\repairSkipCM) uses skipping data and sampling at regular positions of $C$ to permit exponential
search within the list $i+1$ as shown above. The third alternative (\repairSkipST) uses sampling at
domain positions and adapts the {\em lookup} search strategy. Finally, we also provide
a simpler variant with neither skipping nor sampling data  (\repairNo) where the 
intersection is done by first decompressing the whole lists and then applying a merge-type algorithm.

 Note that, even though the use of sampling enables 
direct accesses to the inverted lists, these methods would
experiment a slowdown compared to their uncompressed counterpart. This is because, as explained, we can only have direct access 
to the \repair\ phrase beginnings, that is, to integers in the compressed 
sequence $C$. This worsens even more in a repetitive scenario,
where a single entry in $C$ may expand to a large number of terminals, and 
may make the advantage of sampling vanish in practice.

\subsection{Analysis}

One can achieve worst-case time $O(m (1+\log \frac{n}{m}))$ to intersect two
lists of length $m < n$, for example
by binary searching the longer list for the median of the shortest and dividing
the problem into two \cite{BY04}, or by exponentially searching the longer
list for the consecutive elements of the shortest \cite{CM10}. This is a lower
bound in the comparison model, as one can encode any of the ${n \choose m}$
possible subsets of size $m$ of a universe of size $n$ via the results of the
comparisons of an intersection algorithm, so these must be
$\ge \log_2 {n \choose m} \ge m\log_2\frac{n}{m}$ in the worst case, and the
output can be of size $m$. Better results are possible for particular classes
\cite{BLOLS09}.

We now analyze our skipping method in an idealized scenario where we assume
that (1) we represent the dictionary as a binary tree, not using $R_S$ and $R_B$
(this can be done at the expense of worsening compression); (2) the derivation 
trees of our rules have logarithmic depth (this can be achieved \cite{Sak05},
in particular the RePair implementation we use usually satisfies this property);
and (3) we use an appropriate
sampling (which, again, impacts on the space). We aim to show that
our \repair\ compression is able to achieve optimum performance in theory,
even if in practice we opt for a more space-efficient representation.

We expand the 
shortest list if it is compressed, at $O(m)$ cost, and use skipping to find its
consecutive elements on the longer list, of length $n$ but compressed to
$n' \le n$ symbols by \repair. Thus, we pay $O(n')$ time for skipping over all
the symbols of $C$. Now, consider that we have to expand $C[j]$, to its
length $n_j$, to find $m_j$ symbols of the shortest list, 
$\sum_{j=1}^{n'} n_j = n$, $\sum_{j=1}^{n'} m_j = m$. 
Assume $m_j>0$ (the others are absorbed in the $O(n')$ cost). In the worst 
case we will traverse all the $O(m_j)$ nodes of the derivation tree up to level 
$\log_2 m_j$, and then carry out $m_j$
individual traversals from that level to the leaves, at depth $O(\log n_j)$.
%In the first part, we pay $O(2^i\log\frac{m_j}{2^i})$ for the $2^i$ binary
%searches within the corresponding subinterval $[x,x']$ of $m_j$ at that level
%(an even partition of the $m_j$ elements into $m_j/2^i$ is the worst case),
%for $0\le i \le \log_2 m_j$. This adds up to $O(m_j)$. 
For the second part, we have $m_j$ individual searches for one element $x$,
which costs $O(m_j(\log n_j - \log m_j))$. All adds up to
$O(m_j (1+\log\frac{n_j}{m_j}))$. Added over all $j$, this is
$O(m(1+\log\frac{n}{m}))$, as the worst case is $n_j = \frac{n}{n'}$, $m_j =
\frac{m}{n'}$.

\begin{theorem}\label{thm:inter}
The intersection between two lists $L_1$ and $L_2$ of length $n$ and $m$ 
respectively, with $n>m$, can be computed in time 
$O\left(n' + m(1+\log\frac{n}{m})\right)$, where \repair\ compresses $L_1$ to $n'$ 
symbols using rules of depth $O(\log n)$.	
\end{theorem}

To achieve the optimal worst-case complexity we need to add sampling.
One absolute sample out of $\log_2\frac{n}{n'}$ positions in $C$
multiplies the space by $1+\frac{1}{\log_2\frac{n}{n'}}$,
and reduces the $O(n')$ term to $O(m\log\frac{n}{n'})$. This is absorbed by 
the optimal complexity as this matters only when $m \le n'$. 

\begin{coro}
With an $1+\frac{1}{\log_2\frac{n}{n'}}$ extra space factor, the algorithm of Theorem \ref{thm:inter}
takes $O\left(m(1+\log\frac{n}{m})\right)$ time. 
\end{coro}

%%%%%%%%%%%%%%%%%%%%%%%%%%%%%%%%%%%%%%%%%%%%%%%%%%%%%%%%%%%%%%%%%%%%%%%%%%%%%%%%%%%%%%%%% 
%%%%%%%%%%%%%%%%%%%%%%%%%%%%%%%%%%%%%%%%%%%%%%%%%%%%%%%%%%%%%%%%%%%%%%%%%%%%%%%%%%%%%%%%% 
%%%%%%%%%%%%%%%%%%%%%%%%%%%%%%%%%%%%%%%%%%%%%%%%%%%%%%%%%%%%%%%%%%%%%%%%%%%%%%%%%%%%%%%%% 
%%%%%%%%%%%%%%%%%%%%%%%%%%%%%%%%%%%%%%%%%%%%%%%%%%%%%%%%%%%%%%%%%%%%%%%%%%%%%%%%%%%%%%%%% 
%%%%%%%%%%%%%%%%%%%%%%%%%%%%%%%%%%%%%%%%%%%%%%%%%%%%%%%%%%%%%%%%%%%%%%%%%%%%%%%%%%%%%%%%% 
%%%%%%%%%%%%%%%%%%%%%%%%%%%%%%%%%%%%%%%%%%%%%%%%%%%%%%%%%%%%%%%%%%%%%%%%%%%%%%%%%%%%%%%%% 
%%%%%%%%%%%%%%%%%%%%%%%%%%%%%%%%%%%%%%%%%%%%%%%%%%%%%%%%%%%%%%%%%%%%%%%%%%%%%%%%%%%%%%%%% 
%%%%%%%%%%%%%%%%%%%%%%%%%%%%%%%%%%%%%%%%%%%%%%%%%%%%%%%%%%%%%%%%%%%%%%%%%%%%%%%%%%%%%%%%% 

%%%%%%%%%%%%%%%%%%%%%%%%%%%%%%%%%%%%%%%%%%%%%%%%%%%%%%%%%%%%%%%%%%%%%%%%%%%%%%%%%%%%%%%%% 
%%%%%%%%%%%%%%%%%%%%%%%%%%%%%%%%%%%%%%%%%%%%%%%%%%%%%%%%%%%%%%%%%%%%%%%%%%%%%%%%%%%%%%%%% 
\section{Experimental results} \label{experiments}

We experimentally study the space/time tradeoffs obtained with the described 
inverted list representations, in both the non-positional and positional 
scenarios. In the positional scenario we also add a comparison with the 
self-indexes proposed in Appendix~\ref{sec:self}.

The machine used  has an Intel(R) Xeon(R) E5520 CPU (2.27GHz, 
8 MB cache, 4 cores) and 72 GB of DDR3@800MHz memory. 
It runs Ubuntu GNU/Linux version 9.10 (kernel 2.6.31-19-server-64 bits) and 
\verb|g++| compiler version 4.4.1 (unless stated otherwise). Our code was compiled with the \verb|-O9| directive\footnote{Some recent
state-of-the-art techniques \cite{OV14} used libraries and software we were unable to install in our base server. In those cases, we set a temporary Ubuntu GNU/Linux version 14.04 (kernel 3.13.4.49-generic-64 bits) on the same machine to run those experiments.}. We measure CPU user-times.

We used the 108.5 GB 
Wikipedia collection described by He et al.~\cite{HZS10}, which contained 10\% of 
the complete English Wikipedia from 2001 to mid 2008. This collection is formed by 240,179 articles, 
each of which has a number of versions. Table~\ref{tab:coll} shows its statistics.

\begin{table*}[t]
  {\small
  \begin{center}
  \begin{tabular}{l|r|r|r|r|r}
  Subset & Size~~ & Articles & Number of & Versions / & Avg bytes / \\
	 & (GB)~  &          &  Versions & Article~~~ & Version  \\
  \hline
  Full   & 108.50 & 240,179 & 8,467,927 & 35.26 & 13,757 \\
  Non-pos  & 24.77 &   2,203 &   881,802 &400.27 & 23,782 \\
  %20GB   & 24.79 &  70,190 & 2,501,233 & 35.64 &  8,386 \\
  Pos    & 1.94 &   4,327 &   149,761 & 34.61 & 13,941 

  \end{tabular}
  \end{center}
  }
  \caption{Some characteristics of the Wikipedia subcollections used.}
  \label{tab:coll}
\end{table*}

We also chose two %random 
subsets of the 
articles, and collected all the versions of the chosen articles. Each version 
is considered as a document in our collection (we do not mark which versions belong to which article). For the non-positional setting 
our subset contains a prefix of 24.77 GB of the full collection, whereas for positional indexes we
chose 1.94 GB of random articles. 
Table~\ref{tab:coll} also provides the statistics of our two subsets. As it can be 
seen, the 24.77 GB prefix is more repetitive than the full collection (yet, in 
Section~\ref{comparisonSUEL} we will also
present some results over the full collection for the non-positional scenario in order to
compare with the approach of He et al.~\cite{HZS10}), whereas
the 1.94 GB subcollection is similar to the global collection.

To compare the space/time tradeoffs of the different indexing alternatives we used four 
different query sets, each of them containing 1,000 queries.
The first two consist of one-word patterns that were chosen at random from the 
vocabulary of the indexed subcollection. They differ in the number of occurrences of
the patterns. The first one (low-frequency scenario) includes infrequent words, which 
occur less than 1,000 times in the subcollection. The second query set (high-frequency scenario) 
includes only very frequent words, which occur more than 1,000 times in the subcollection.

The last two query sets consist of 1,000 phrases of 2 and 5 
words that were chosen at random from the text of the subcollection. When dealing 
with non-positional indexes, such phrases are taken as conjunctive (AND) queries. In the positional scenario, we will take them as phrase queries, hence returning the documents where the words of each query occur at consecutive positions.

Finally, for the self-indexes used in the positional scenario we also include experiments 
to check the speed to recover the original documents. We choose two sets
of random snippets of length 80 (around one line) and 13,000 (around one document, in our
collection) characters. 

%The first two query sets consist of one-word 
%We consider four query sets, with 1000 queries per set. Two sets are one-word 
%searches, chosen at random from the vocabulary of the indexed subcollection. In 
%the first one we only take infrequent words, with up to 1000 occurrences in the 
%subcollection, whereas in the second we take frequent words, with more than 
%1000 occurrences. The other two query sets correspond to phrases of 2 and 5 
%words chosen at random from the text of the subcollection. For non-positional 
%indexes this is taken as an AND query, whereas for the positional ones it is 
%taken as a phrase query.

\subsection{Non-positional indexes}

Our experiments compare the space/time tradeoffs of several variants of 
non-positional inverted indexes over the highly repetitive 24.77 GB subcollection described above. We include in this comparison some of the best classical encodings to represent d-gaps covered in Section~\ref{sec:related}, such as \rice, \simplen, 
\pfordelta, and \vbyte\ with no sampling to speed up intersections (thus only merge-wise intersections are feasible). 
We also include two alternatives using \vbyte\ coupled with list sampling \cite{CM10} (\vbyteCM) with $k=\{4,32\}$, or domain sampling \cite{TS10} (\vbyteST) with $B=\{16,128\}$. In addition, we include the hybrid variant of \vbyteCM\ that uses bitmaps to represent the largest inverted lists (\vbyteCMB) \cite{CM10}. For completeness we used the same approach on \vbyteST, to build \vbyteSTB, and also included variants \vbyteB\ and \riceB\ with no sampling.
We also tested the novel \qmx\footnote{\url{http://www.cs.otago.ac.nz/homepages/andrew/papers/QMX.zip}. } technique \cite{trotman2014} that uses SIMD-instructions to boost decoding of large lists, and coupled it with an intersection algorithm \cite{Lemire2015:simdInt} that also benefits from SIMD-instructions.\footnote{ \url{https://github.com/lemire/SIMDCompressionAndIntersection}. In this case we had to use \texttt{g++} version 4.7 with the \texttt{-O3 -msse4} flags, according to authors' sources.}
 %
 %/usr/bin/g++-4.7 -O3 -msse4 -mavx -std=c++11  -Weffc++  -ggdb -DDEBUG=1 -D_GLIBCXX_DEBUG 
 % gcc version 4.7.3

We also tested the recent Partitioned Elias-Fano indexes \cite{OV14}, and used the best/optimized variant from that paper (\efopt). The source code is available at authors' website.\footnote{\url{https://github.com/ot/partitioned_elias_fano.}}
From  the same authors \cite{OV14}, we also included in our experiments the variants \optpfd\ (optimized PForDelta variant \cite{YDS09}), \interpolative\ (Binary Interpolative Coding \cite{MS00}), and \varint\ (SIMD-optimized Variable Byte code \cite{Stepanov:2011}). To match their software requirements, we set our system to Ubuntu 14.04 and used \texttt{CMake} 2.8.12.2 (Release mode), \texttt{g++} version 4.8.2 with options (\texttt{-msse4.2 -std=c++11}), and {\tt libboost} library version 1.54.0 (available with apt). All the experiments run on our Ubuntu 14.04 system will be marked with an `\texttt{*}' in the figures. 

 %
 % CMAKE -DCMAKE_BUILD_TYPE=Release 
 % Options -msse4.2 -std=c++11
 % /usr/bin/g++-4.8 
 % gcc version 4.8.2

We compare all those techniques in Section~\ref{exp:nopos:others}, to determine which of the traditional techniques perform best in the repetitive scenario. 
Then, in Section~\ref{exp:nopos:ours} we compare them with the new variants designed to deal with repetitive data we proposed in Sections~\ref{sec:lists} and \ref{sec:list-repair}. In particular, we include \riceRuns,
\vbyteLZMA,  \vbyteLzend, \repairNo, \repairSkip. We add no sampling to them. Therefore, only \repairSkip\ can benefit of additional data to boost the intersections (which are performed sequentially). In addition, we show the \repair\ variants using sampling, \repairSkipCM\ (with $k=\{1,64\}$) and \repairSkipST\ (with $B=\{16, 256, 1024\}$). For \vbyteLzend\ we will obtain different space/time tradeoffs by tuning its {\em delta-codes-sampling} parameter $ds$ (see \cite{KNtcs12} for details) to $ds=\{4,16,64,256\}$.

Later, in Section~\ref{comparisonSUEL}, we will compare the approach of He et al.~\cite{HZS10} with our new representations, over the same data used in their article. To make a fairer comparison, during the parsing stage of the construction of our inverted indexes we use exactly their same parsing
of words. Therefore, we apply case folding, we do not consider stemming, and remove the 20 most common stopwords. 
%
 %In those inverted indexes we used exactly the same parsing of words of He et 
%al.~\cite{HZS10}, which involves applying case folding, removing 20 very common 
%stopwords, and no stemming.
%
After this parsing, the original 108.5 GB are reduced to
around 85.55 GB, and the 24.77 GB of the indexed subcollection becomes around 19.54 GB. 
Anyway, from here on, the space results reported for the indexes are shown as a percentage 
of the index size with respect to the size of the original [sub]collection in plain text
($index\_size / original\_size \times 100$). Note that we are not considering in
this experiment the compressed representation of the text. 
Times are shown in microseconds per occurrence.
%Yet we report the space results with respect to the original text size.

\subsubsection{Using traditional techniques in a repetitive scenario} \label{exp:nopos:others}
%The space is given as a percentage of that used by the text in plain form (we are not considering in this experiment the compressed representation of the text itself). The time is shown in microseconds per occurrence.

\begin{figure}[t]
\begin{center}
\includegraphics[angle=-90,width=0.49\textwidth]{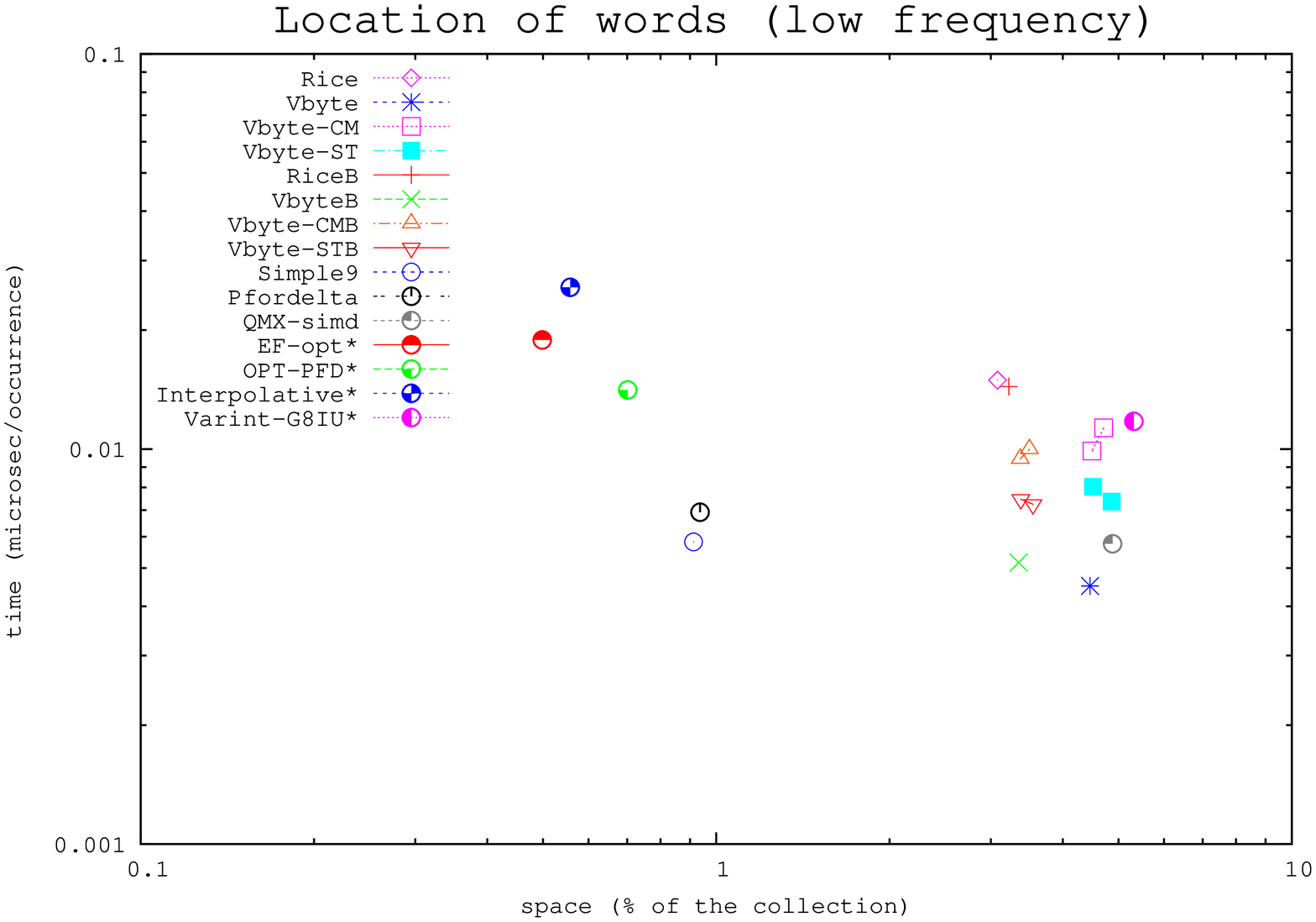}
\includegraphics[angle=-90,width=0.49\textwidth]{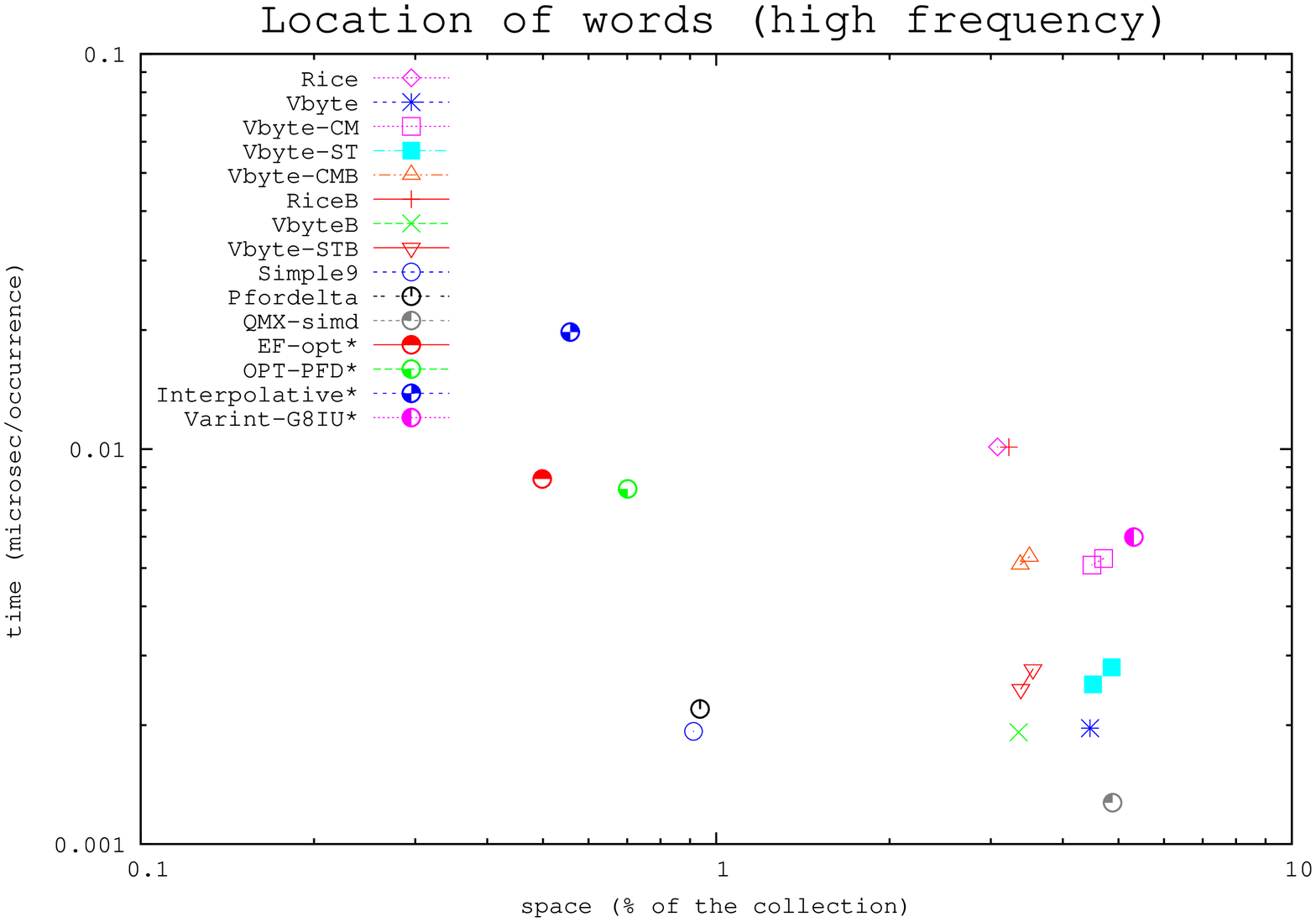}
\includegraphics[angle=-90,width=0.49\textwidth]{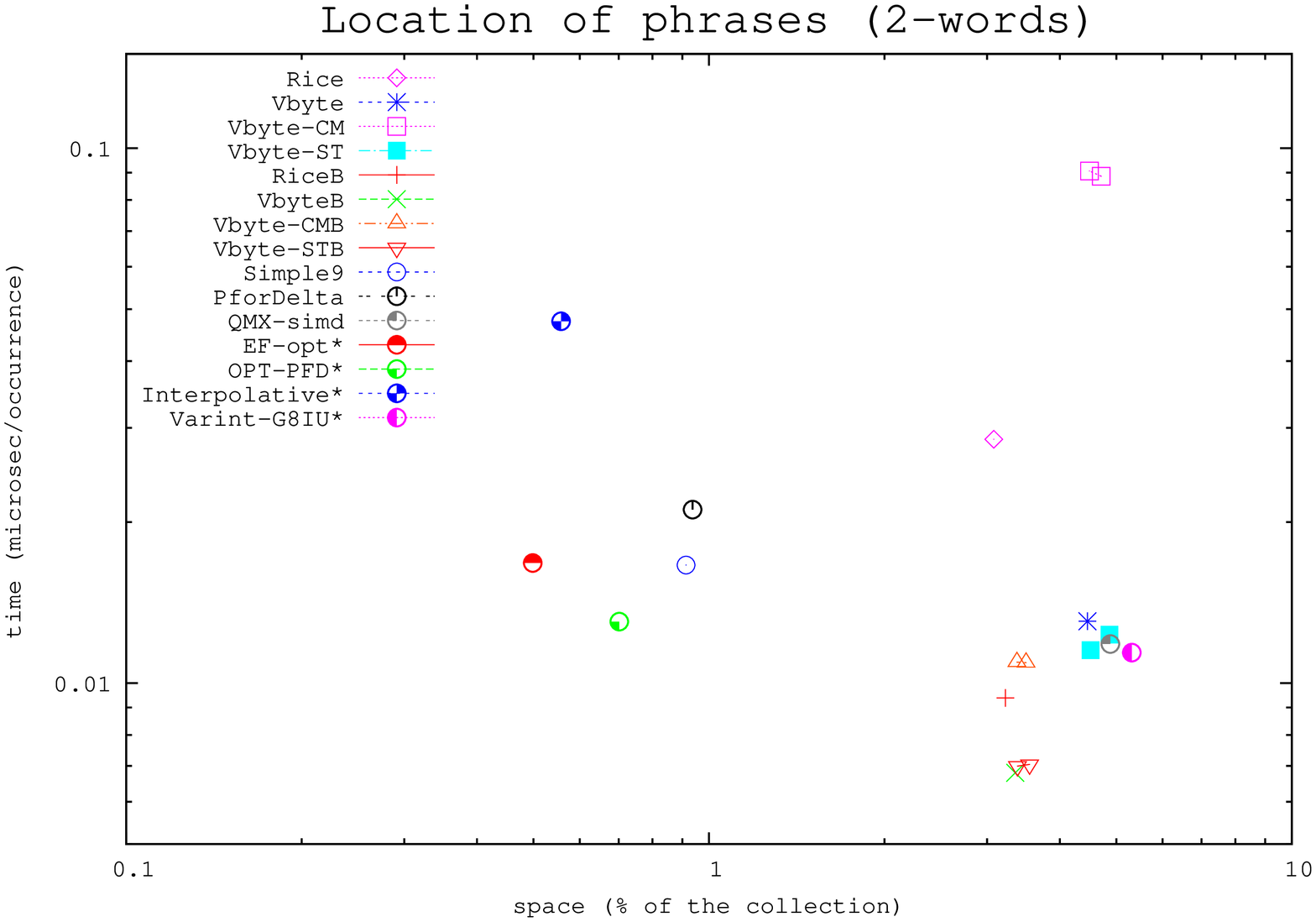}
\includegraphics[angle=-90,width=0.49\textwidth]{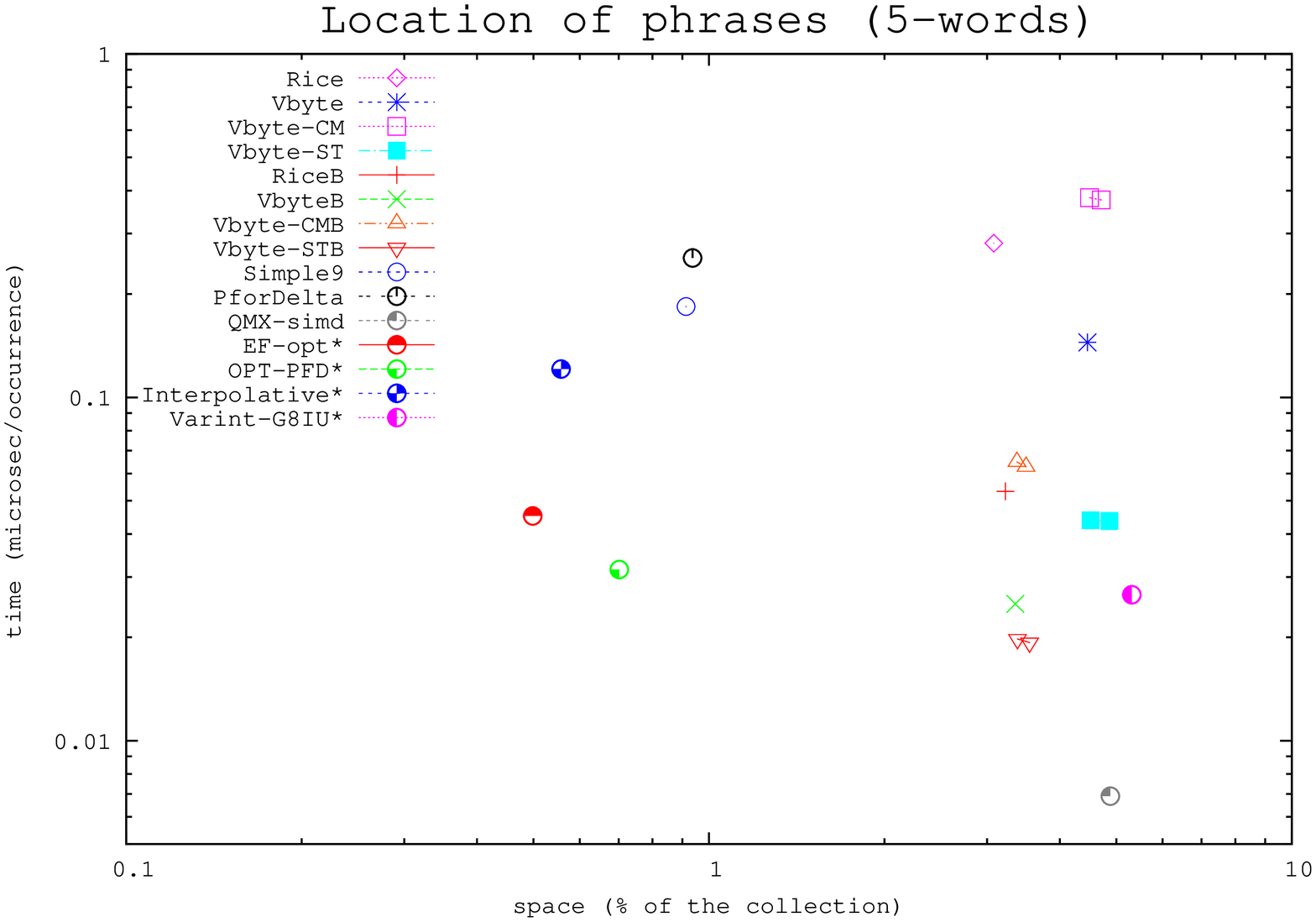}
\caption{Space/time tradeoffs for non-positional indexes using traditional
techniques. Logscale.}
\label{fig:nonpos}
\end{center}
\end{figure}

As indicated above we include a comparison of well-known techniques that were initially developed
for non-repetitive scenarios, now operating on repetitive collections.
Figure \ref{fig:nonpos} shows the space/time tradeoffs for 
non-positional indexes using those techniques to represent posting lists. 

We can see that, among classical compression methods, both 
\simplen\ and \pfordelta\ are much better in space than the older techniques like 
\rice\ (one third of the space) and \vbyte\ (one fifth of the space). In typical 
collections \rice\ achieves the best space, but these newer methods take 
advantage of the many runs of 1s. They are also several times faster than
\rice, and roughly as fast as \vbyte\ on word queries. On conjunctive queries, 
however, \vbyte\ is faster. In those queries,
adding samples to support \lookup-type intersections is advantageous: \vbyteST\ 
is significantly faster than \vbyte\ (more than 3 times faster on 5-word
queries). Surprisingly, in our experiments \vbyteCM\ (sampling to support \svs\ with exponential
search on the longest list) turned out to be a bad choice, obtaining worse results than the simple
\vbyte\ with no sampling. Note that the increase of time needed to recover an inverted list is due to 
the fact that values stored in the samples are removed from the differential sequence, what adds an
additional branch at decompression time.
Using hybrid techniques (representing the longest inverted lists with bitmaps and \vbyte\ for the others) turned out to be a good choice, as both space and time are improved with respect to  the non-hybrid \vbyte\ variants. In the case of \rice, the \riceB\ hybrid counterpart did not improve space (as expected, since \rice\ is a bit-wise technique), yet the intersection times benefit significantly from the fast direct access to the longest lists.

The novel \qmx\ shows to be an extremely fast technique when we deal with long posting lists (otherwise it cannot benefit from SIMD-optimized decompression) and the SIMD-based intersection algorithm \cite{Lemire2015:simdInt} outperforms the others for long conjunctive queries. Unfortunately, its space is high, even worse than \vbyte.

The comparison with the Elias-Fano indexes \cite{OV14} shows that the recent \efopt\ performs very well in the repetitive scenario. It obtains the best overall space (half the space of the above \pfordelta\ variant). It also outperforms \interpolative\ in time, and is close to the times of \optpfd\ (which requires around 20\% more space). Technique \varint\ is the fastest of this group \cite{OV14}, yet as expected its space usage is far from the best ones (it is slightly worse than \vbyte). The conjunctive queries using these techniques are very fast. The use of a two-level structure enables an intersection algorithm (similar to {\em sequential}) where the candidate/eliminator from the shortest list is searched for in the others using {\em{next\_geq()}} built-in operator. Their implementation is however somehow slow on word queries, mainly because list elements are recovered one at a time by using an operator {\em next}. Note that \varint\ (the SIMD-optimized \vbyte) is slightly slower than our implementation of \vbyte\ even when fetching the posting lists of frequent words.

 %elige la más corta como candidato y las busca en las demás usando el operador next_geq
 
For the comparison with our new techniques, considering the scope of our
article, we have retained three of those techniques, whose space usage is under 1\%: {\em i)} \simplen, which offers good performance both for word and 2-word conjunctive queries; {\em ii)} \efopt, which obtains the best space values and still good performance; and {\em iii)} \optpfd, which obtains space close to that of \efopt\ and slightly better performance both for word and conjunctive queries.

\subsubsection{Comparison with our proposals} \label{exp:nopos:ours}

\begin{figure}[t]
\begin{center}
\includegraphics[angle=-90,width=0.49\textwidth]{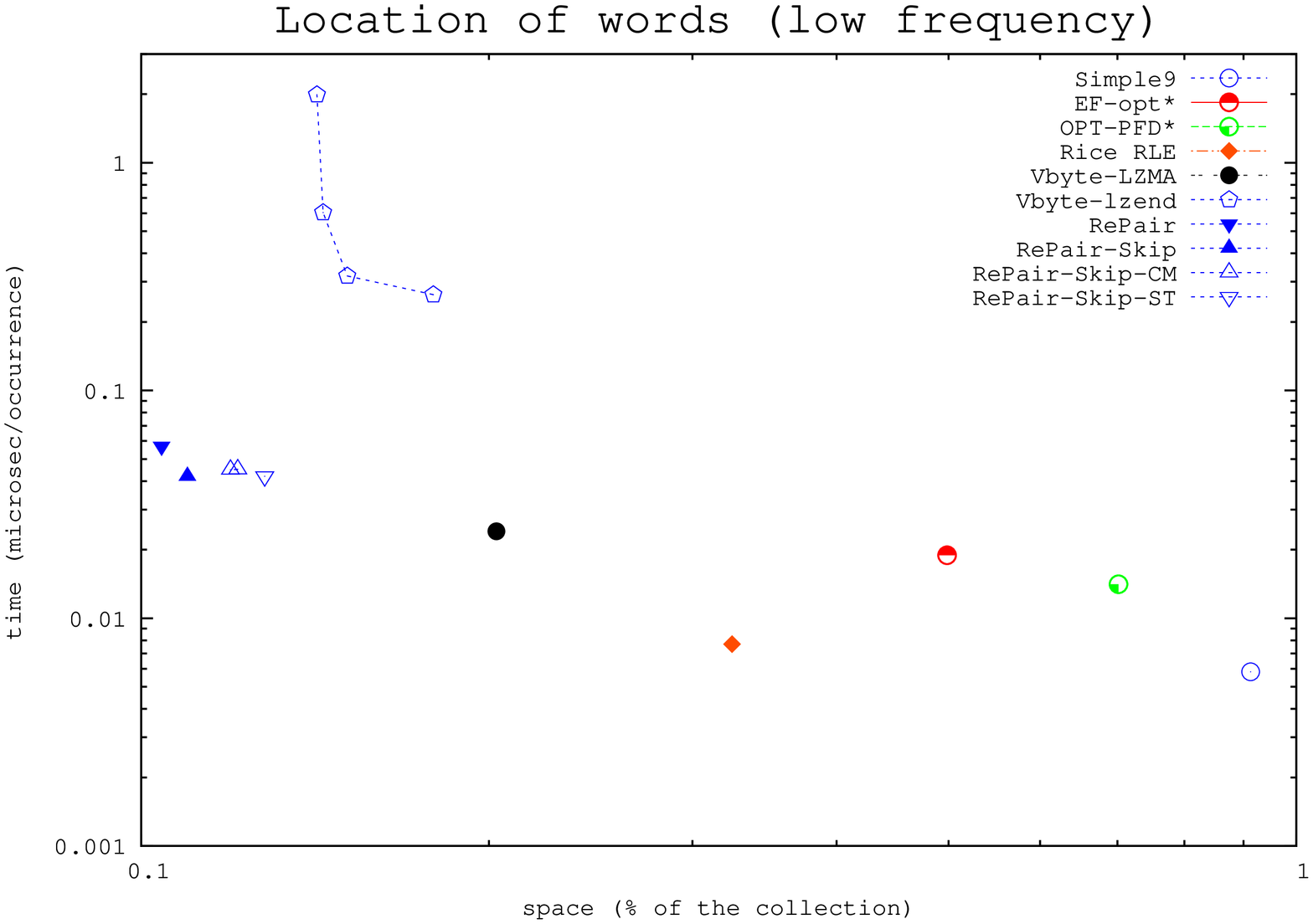}
\includegraphics[angle=-90,width=0.49\textwidth]{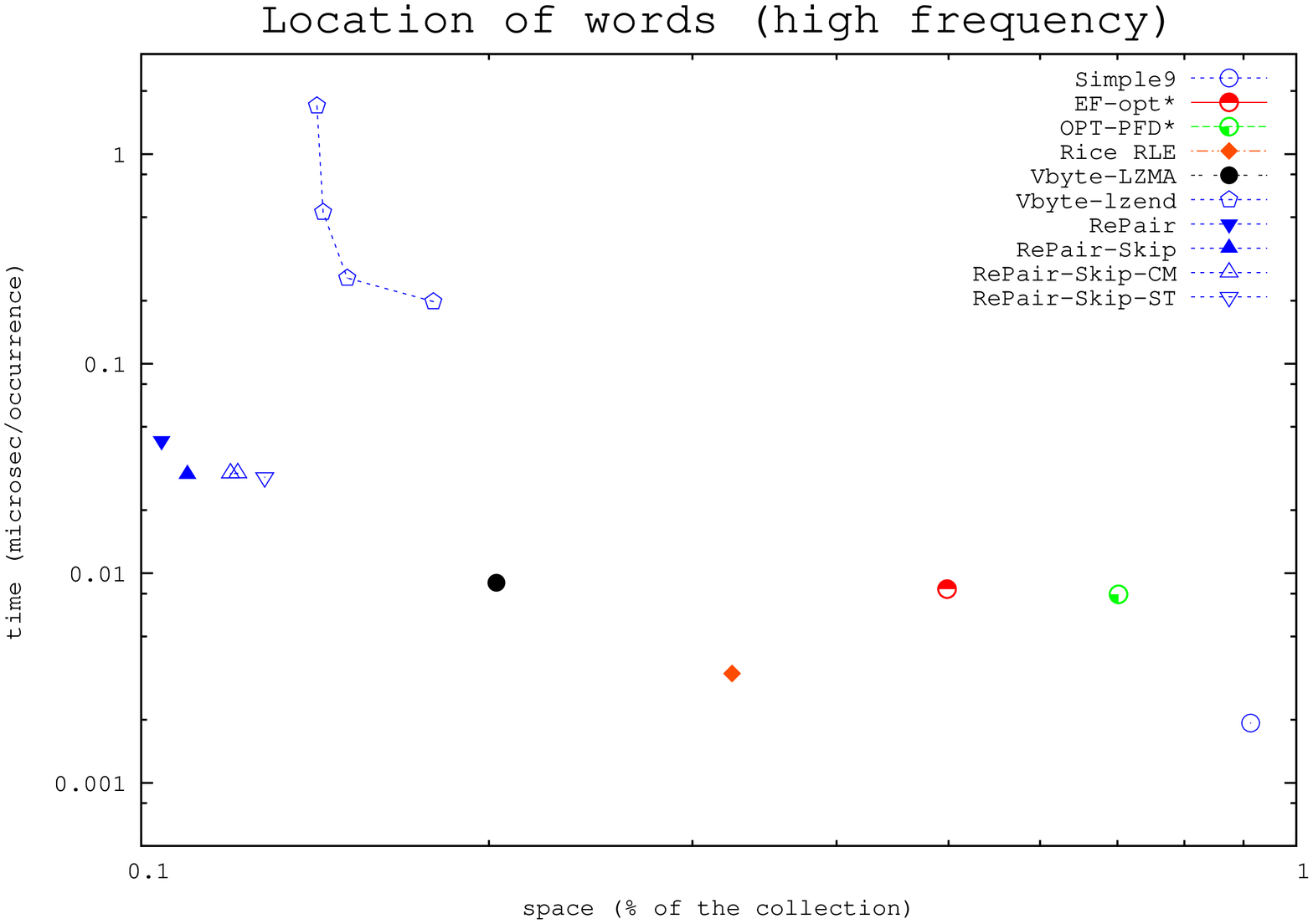}
\includegraphics[angle=-90,width=0.49\textwidth]{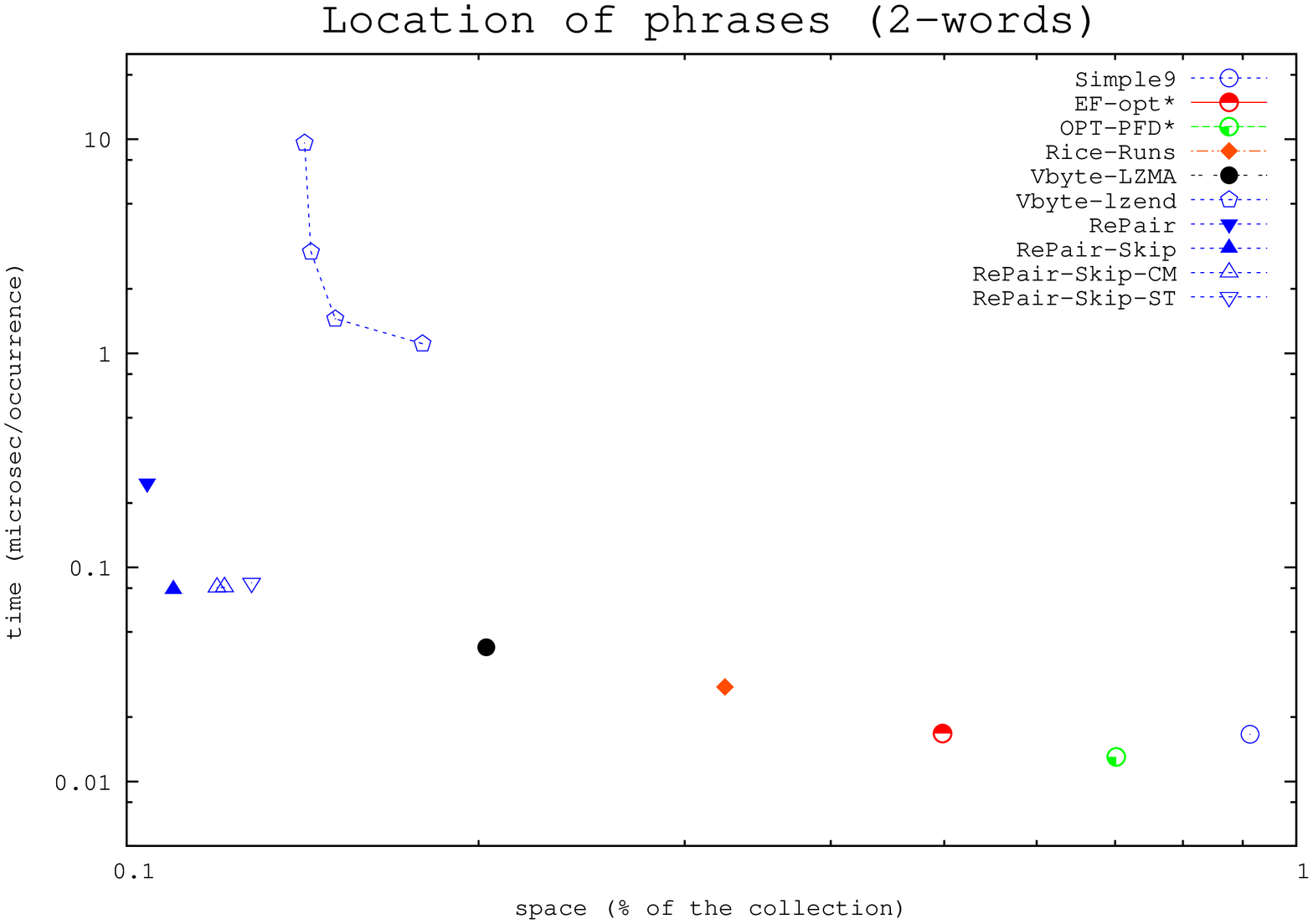}
\includegraphics[angle=-90,width=0.49\textwidth]{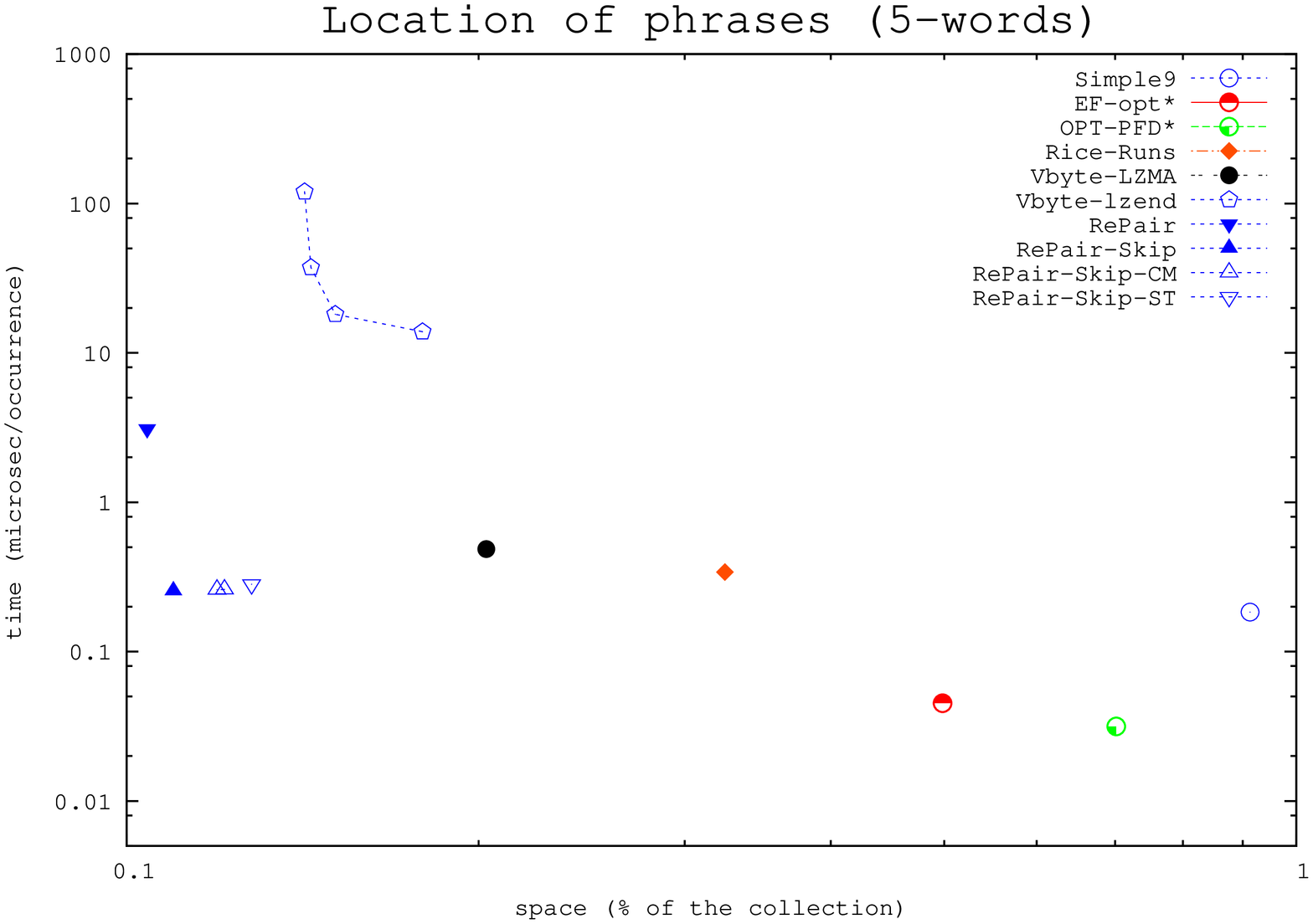}
\caption{Space/time tradeoffs for non-positional indexes, comparing the best
classical techniques with our new ones. Logscale.}
\label{fig:nonpos2}
\end{center}
\end{figure}

Our next experiments compare our proposals with the best counterparts from the previous section. Figure \ref{fig:nonpos2} shows the space/time tradeoffs for all the resulting non-positional indexes. 
 %
%All the other methods (except \vbyteLZMA) can be combined with such samplings
%to speed up long queries.
%Yet, 
 The most important conclusion with regard to classical
encoding methods is that they are unsuitable for highly repetitive collections.
Our new techniques take one order of magnitude less space than \simplen\, and up to 5 times less space than \efopt. Yet, they are also significantly slower than the fastest classical variants. 

Our first simple method to take advantage of repetitiveness, \riceRuns, makes
an interesting leap in space, from 1\% taken by \simplen\ or 0.5\% taken by \efopt, to
around 0.3\%. If we compare it in Figure~\ref{fig:nonpos} we can see that it takes one tenth the space of plain \rice\ and is at the same time faster, as it needs much less decompression work. It does 
not, however, get close to the space of stronger methods like \vbyteLZMA, which
achieves around 0.2\% space by exploiting other types of redundancy. However,
\riceRuns\ is significantly faster than \vbyteLZMA\ (up to 3 times).

\vbyteLZMA\ is close to the smallest space that we can achieve. It is
significantly faster than \repairNo\ (up to 10 times) and than \repairSkip\ 
(up to 3 times) at word queries, as it decompresses faster the inverted list. 
However, on conjunctive queries, where many of the decoded values have to be 
discarded, the ability of \repairSkip\ to skip nonterminals without 
decompressing them finally makes it almost twice as fast as \vbyteLZMA,
and even faster than \riceRuns.

As expected, \vbyteLzend\ succeeds at exploiting inter-list regularities 
and almost halves the space of \vbyteLZMA, yet it is by far the
slowest technique in our comparison, being more than an order of magnitude slower than \vbyteLZMA.

Note that \repairNo\ obtains lower space (85\%) than \vbyteLZMA, despite the 
fact that LZ77 compression is more powerful than \repair. This is a
consequence of \lzma\ exploiting only intra-list regularities, and, as in \vbyteLZMA, shows that 
significant further repetitions are captured when considering the inter-list 
redundancies. The skipping information added to \repairNo\ adds very
little space (6\%), but significantly improves its time performance (almost 20
times faster on long phrases). This improvement occurs even on one-word queries
(up to 2.6 times faster), since \repairSkip\ does not need to carry out $rank$ 
operations on $R_B$ (recall Section~\ref{sec:repair}). Results also show that it is not
worth adding
sampling to \repairSkip. Sampling increases the space requirements, but no
improvements at intersections upon \repairSkip\ are reported by \repairSkipCM\ nor  \repairSkipST.
Note that for \repairSkipST\ we are showing only the plot corresponding to sampling 
parameter $B=1024$, which obtained the least space, since we obtain no time 
improvements and space usage grew from 25\% to 700\% using smaller values.

\subsubsection{Comparison with previous work for repetitive collections} \label{comparisonSUEL}

As described in Section~\ref{sec:iids}, the best previous work for repetitive
collections is by He et al.~\cite{HZS10}. We tried hard to compile their index
in our machine in order to carry out a direct comparison, with no success. On
the other hand, limitations in one of our codes ({\vbyteLzend}) prevented us from indexing 
their full 108.5 GB collection.

We opted for the following compromise to compare the approaches as fairly as
possible. The machine where they ran their experiments is very similar to
ours in speed, RAM, and processors (except they have 8 cores and we have 4).
%\footnote{More precisely, their experiments ran on an Intel(R) Xeon(R) E5520 CPU running at 2.27 GHz, 8 MB cache, 72 GB of RAM memory, 8 cores. The 
%operating system installed is an Ubuntu GNU/Linux version 9.10 running kernel 
%2.6.32-22-generic.}
 %
Thus times are comparable.  
We ran the same set of 9,508 queries they used in their experiments with our new techniques.
For {\vbyteLzend} (which was unable to index the whole collection) we split the 108.5 GB collection into four
subcollections of approximately equal size, indexed them separately, and added
up the space of the four parts. We ran the same set of 9,508 queries on each of the four subcollections, and added up the
times. We believe that this gives us a very tight upper bound on the space and time \vbyteLzend\ would need for the whole collection, because we repeated the same process for the other techniques and obtained negligible differences between the estimated values and the real ones.

%For {\vbyteLzend} (which was unable to index the whole collection) we \highlight{split the 108.5 GB collection into four
%subcollections} of approximately equal size, indexed them separately, and added
%up the space of the four. We ran the same set of \highlight {9,508 queries} they used
%in their experiments, on each of the four subcollections, and added up the
%times. This gives a (probably tight) upper bound on the space and time our
%techniques would need on the whole collection. For the other techniques we indexed the whole
%collection.

\begin{figure}[t]
\begin{center}
\includegraphics[angle=-90,width=0.49\textwidth]{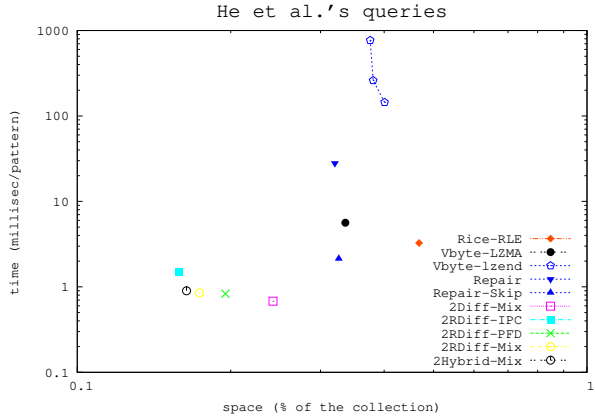}
\caption{Space/time tradeoffs for our best non-positional indexes and those
of He et al. Logscale.}
\label{fig:suelReal}
\end{center}
\end{figure}

Figure~\ref{fig:suelReal}  shows the space and time, the latter in milliseconds per
query. We consider \riceRuns, {\repairNo}, \repairSkip,  {\vbyteLZMA}, and {\vbyteLzend}, and the techniques that
performed best in their experiments \cite[Table 6]{HZS10}.

As it can be seen, techniques of He et al.\ obtain roughly half the space and 
time of \repairNo, \repairSkip, and \vbyteLZMA. The advantage of the latter is that they are universal, 
that is, they work for more general scenarios where their techniques could not be applied.

Finally, note that the gap in space between \repairNo\ and \vbyteLZMA\ is much smaller than in Section~\ref{exp:nopos:ours}. 
This is because the whole collection has lower repetitiveness than the 24.77 GB subcollection and less
inter-list regularities can be found. This also affected \vbyteLzend, whose results are now worse than those of \vbyteLZMA.

\begin{comment}
Apart from the fact that \repairSkip\ is universal and their techniques are
not, a finer analysis of the collection reveals that it has some articles with
many versions (hundreds), and many articles with few versions (even just one).
These non-versioned files pose a serious space overhead to our \repair\ 
technique. Just as He et al.\ optimize the ordering of the versions and combine
various methods, we show that our techniques have potential by carrying out a
simple combination. We separate one subcollection with the articles with at
least OJO versions, and another with the rest. The first collection is split
into OJO subcollections and these are indexed with \repairSkip, whereas the
second is split into OJO subcollections and these are indexed with 
\simplen. The result, called {\tt \repairSkip/\simplen}, is also shown in
Figure~\ref{fig:suel}. OJO comentar
\end{comment}

\subsection{Positional indexes}

\begin{comment}
For testing the positional indexes we used the smaller subcollection, 
because several self-index implementations are unable to handle 
texts larger than 2 GB. 
q

For this scenario we used an Intel(R) Xeon(R) E5335 CPU running at 2.00 GHz, 
4 MB cache, 16 GB of RAM memory, 8 cores. 
The operating system installed is an Ubuntu GNU/Linux version 8.04.4 LTS 
running kernel 2.6.24-29-server (64 bits) and \verb|g++| compiler version 
4.2.4. Our code was compiled with the \verb|-O9| directive. We measure user 
times.
\end{comment}

For testing the positional indexes we used the 1.94 GB subcollection 
because several self-index implementations are unable to handle 
texts larger than $2^{31}$ bytes. 

Since self-indexes must reproduce the precise text, we cannot apply case
folding nor any kind of filtering in this scenario. We index the original text
as is. As explained, word-based self-indexes will regard (and index) the text 
as a sequence of words and separators. For fairness, the positional inverted 
indexes will index separators as valid words as well, and phrase queries will 
choose sequences of tokens, be they words or separators. Still, we note that
character-based self-indexes will return more occurrences than word-based
self-indexes (or than inverted indexes), as they also report the
non-word-aligned occurrences. Times per occurrence still seem comparable, yet
they slightly favor character-based self-indexes since
the time per occurrence drops as more occurrences are reported (there is a
fixed time cost per query). 

We consider most of the techniques of the non-positional setting, now
operating on position lists. Yet, for \rice\ and \vbyte\ we do not include the hybrid variants using bitmaps as they obtain no space/time improvements in the positional scenario. For the \vbyte\ counterparts using sampling, we set the same sampling parameters as in the previous section: \vbyteCM\ with $k=\{4,32\}$ and \vbyteST\ with $B=\{16,128\}$.
%For \vbyte\ we only include results for the hybrid variants \vbyteB, \vbyteCMB, and \vbyteSTB\ which performed the best in the non-positional scenario.
We had to adapt \simplen\ because it is unable
to represent gaps longer than $2^{28}$. While such gaps do not arise on
document lists, they do occur in position lists. We use the gap $2^{28}-1$
as an escape code and then the next 32 bits represent the real gap. We
exclude \pfordelta\ because it has the same limitation, fixing it is more
cumbersome, and its performance is not very different from that of \simplen.
We also exclude \riceRuns, as runs do not
arise in the positional setting.
%
%We used sampling parameters $k=\{4,\underline{32}\}$ for \vbyteCMB\ and $B=\{16,\underline{64}\}$ for \vbyteSTB. Yet, for clarity, we show only the underlined values in the plots.

We did not include \vbyteLzend, as it was clearly overcome by \vbyteLZMA\ and our \repair\ variants.
For the variants of \repair\ using sampling, we set the sampling parameters to $k=\{1,64\}$ for \repairSkipCM\ and $B=\{4,256\}$ for \repairSkipST\ (yet we will again show only results for $B=256$, as using $B=4$ doubled the space and brought no time improvements).

To compete in similar conditions with self-indexes, positional inverted indexes
must be enhanced with an efficient decoding mechanism that allows any portion
of the text to be efficiently reproduced. We choose \repair\ for this purpose because 
it is well-suited for highly repetitive collections and supports fast direct access 
to the text. Because the text in this way represents a very small fraction 
of the total space, we will represent the rules as pairs of integers to speed up text 
extraction, instead of the slower $R_B$ and $R_S$ based implementation. This adds up to 
$1.21\%$ of the original text size. To further improve extraction performance, we add a regular 
sampling of the array $C$, which increases space up to $1.3\%$ for the densest sampling. As 
a comparison, {\tt p7zip} (from {\tt www.7-zip.org}), the best compressor for this type 
of repetitive texts, achieved 0.52\% space on this subcollection (albeit not providing 
direct access).

We compare the self-indexes described in Appendix~\ref{sec:self}, but first we tune \slp\ and the \verb|LZ|-based indexes to optimize their performance. Regarding 
\verb|CSA|-based self-indexes, we consider sampling rates of the form $2^i$ for $i= [5\dots 11]$ for
\rlcsa, and seven different configurations of sampling parameters for \wcsa\ 
(for the structures $\langle \psi, A_S, A^{-1}_S \rangle$, ranging from 
$\langle 8,8,8 \rangle$ to $\langle 2048,2048,2048 \rangle$). We add to both 
the inverted indexes and self-indexes the time and space required for 
converting absolute positions to (document,offset) pairs, as explianed in 
Section~\ref{sec:lists}. The extra space added by the corresponding mapping
structure is just 0.03\%.

In Section~\ref{exp:pos:others} we compare positional inverted indexes using 
state-of-the-art representations for posting lists. Then, in 
Section~\ref{exp:pos:selfindexes}, we fine-tune the self-indexes we use, to find
their best configurations. In Section~\ref{exp:pos:ours} we compare the best
state-of-the art representations and our new representations of positional
inverted lists, plus the tuned self-indexing alternatives.
Our final experiments, in Section~\ref{exp:pos:extract}, study the speed to
extract snippets and recover the original documents.

\subsubsection{Traditional inverted list representations} \label{exp:pos:others}

\begin{figure}[t]
\begin{center}
\includegraphics[angle=-90,width=0.49\textwidth]{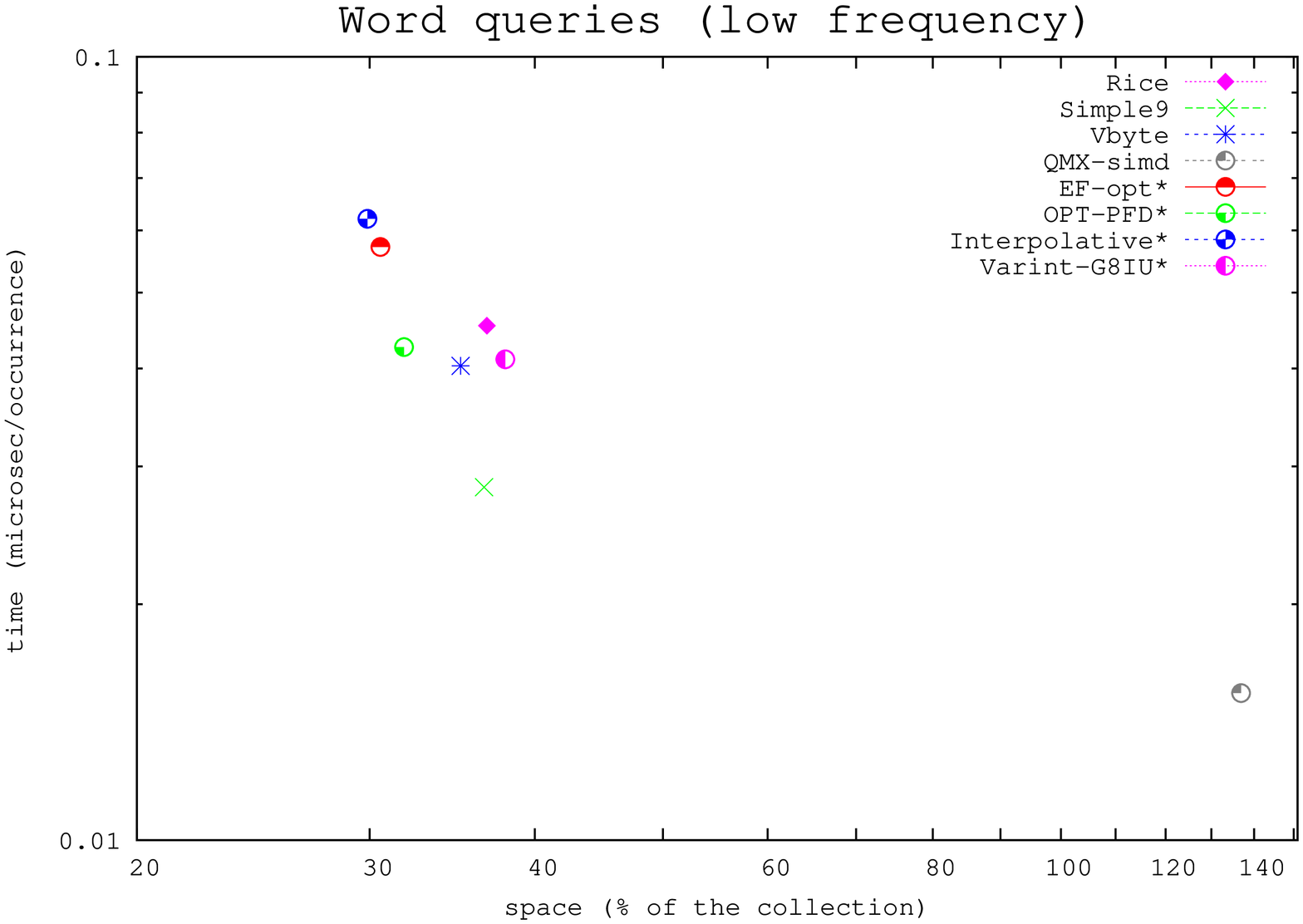}
\includegraphics[angle=-90,width=0.49\textwidth]{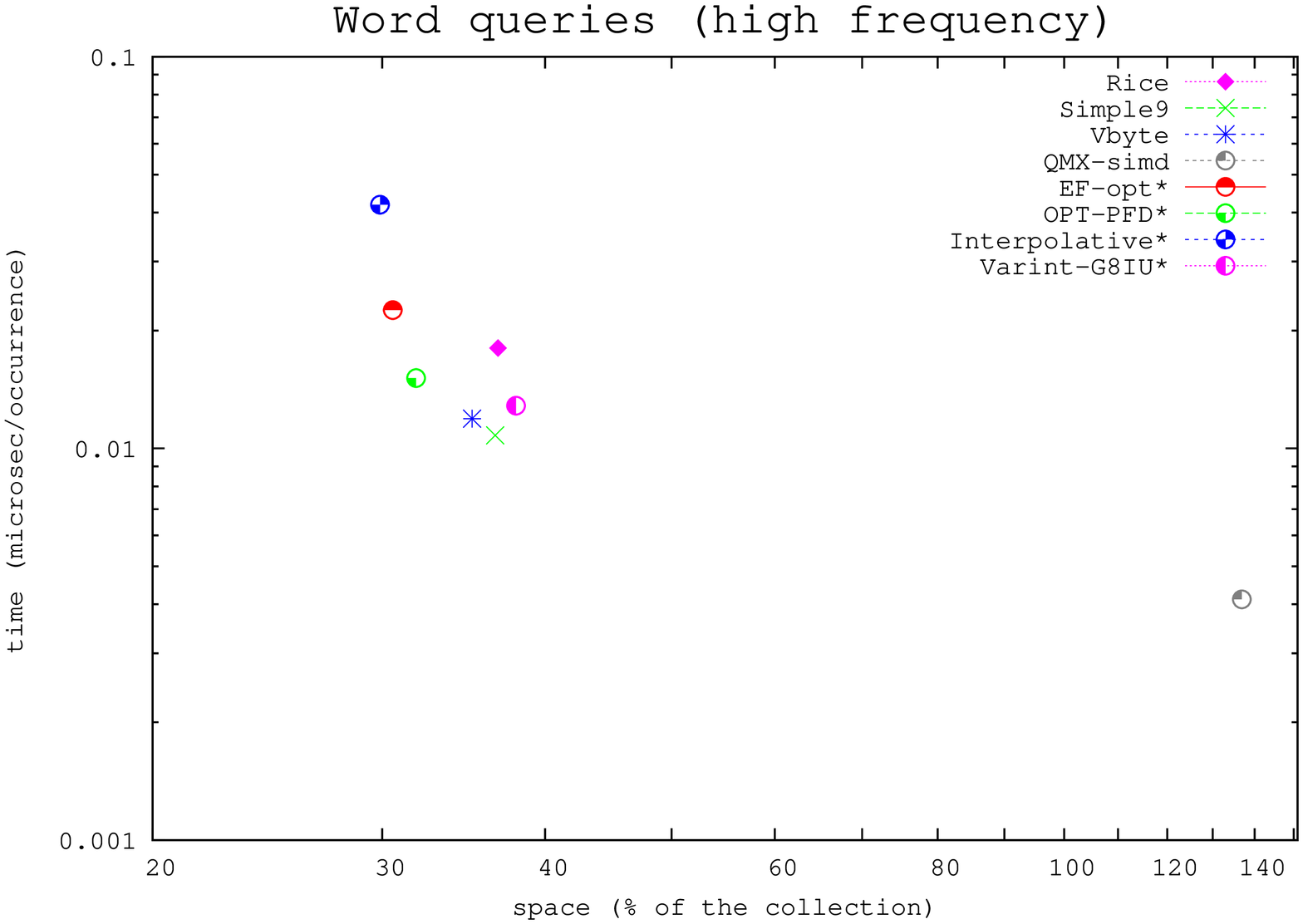}
\includegraphics[angle=-90,width=0.49\textwidth]{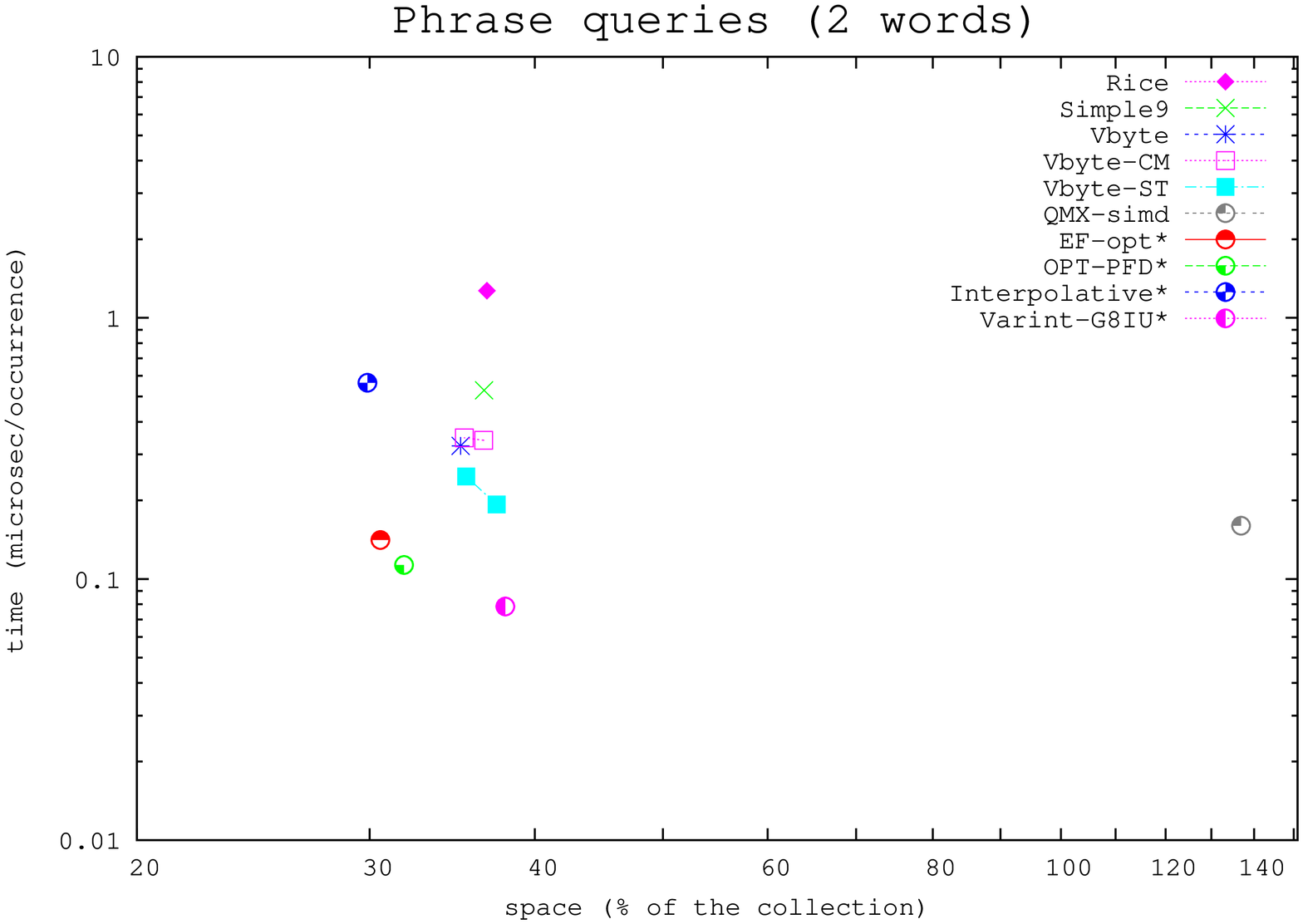}
\includegraphics[angle=-90,width=0.49\textwidth]{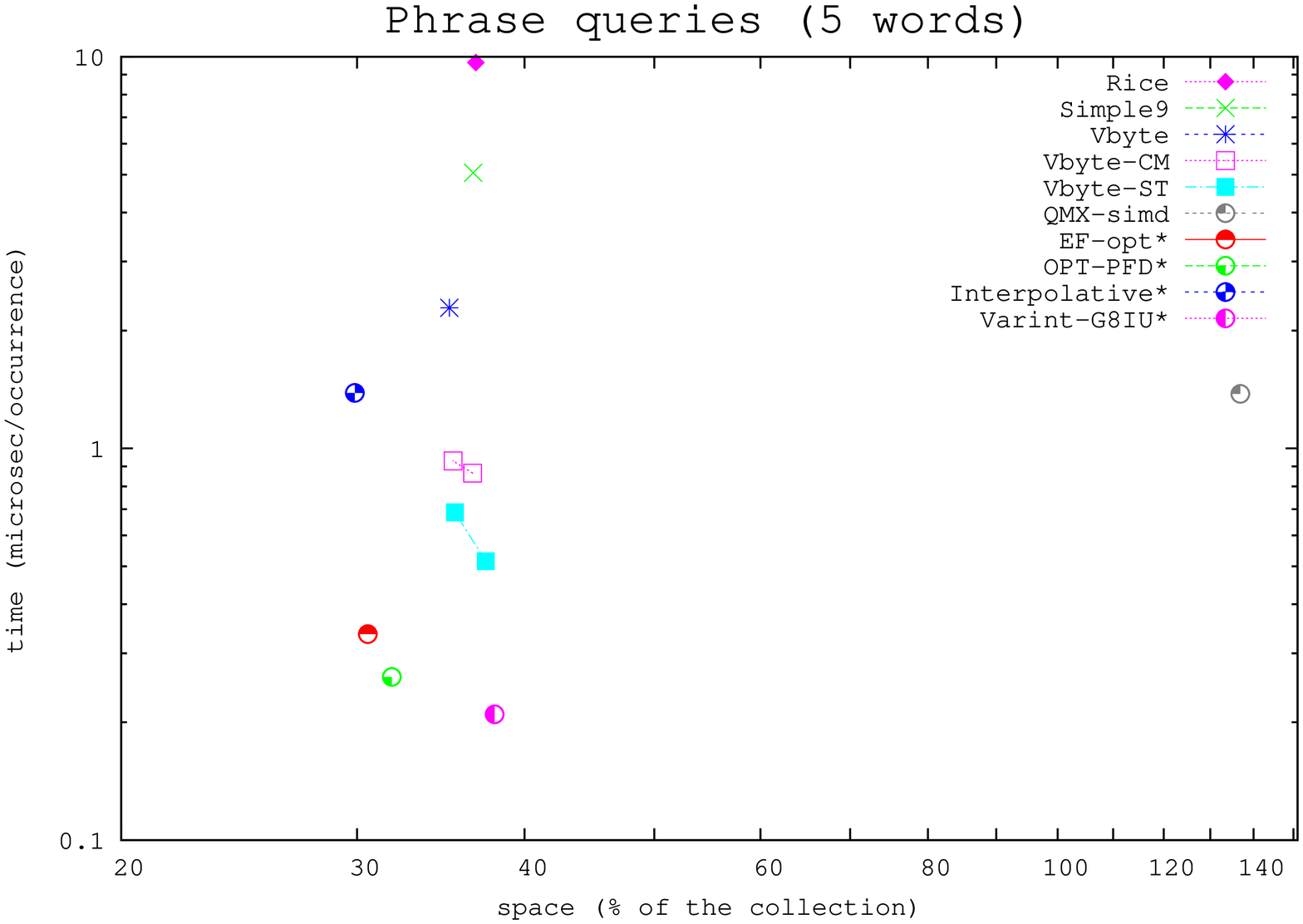}
\caption{Space/time tradeoffs for traditional representations of positional indexes. Logscale.}
\label{fig:pos}
\end{center}
\end{figure}

Figure \ref{fig:pos} shows the space/time tradeoffs achieved, for the four
types of queries, with traditional inverted index representations.
All classical inverted indexes achieve similar space, ranging from 30\% to 40\% of the text size. The more recent representations, instead, reach almost 10\%
of space. From those, \interpolative\ obtains the best compression, with a slight gap over \efopt\ and \optpfd. 

\simplen\ is slightly faster than \vbyte\ for decompressing (i.e., 
for one-word queries), yet \vbyte\ becomes faster on phrases.
Adding sampling, particularly \vbyteST, improves phrase query times
significantly while almost not affecting the space. This is much more noticeable as the 
length of the pattern increases. Note that as more terms are involved in the query, it is
more expectable that the ratio between the length of the shortest 
and longest involved lists increases. Therefore, a merge-wise intersection algorithm  becomes 
less suitable than those that look up the longer lists.
\rice\ is not competitive in this scenario. \qmx\ (which occupies more space than using uncompressed posting values) is again the fastest technique at decompression (word queries). At phrase queries it is still twice as fast as \vbyte, but is clearly overcome by the techniques using sampling. 

In particular, excluding \qmx\ due to its poor compression, the most successful techniques considering phrase queries are \efopt, \optpfd,  and \varint\ (yet it requires around 25\% more space than \efopt). At word-queries \simplen\ is the fastest representation. These four techniques will be used as the baselines to compare with our inverted list representations in Section~\ref{exp:pos:ours}.

\begin{figure*}[t]
\begin{center}
\includegraphics[angle=-00,width=0.45\textwidth]{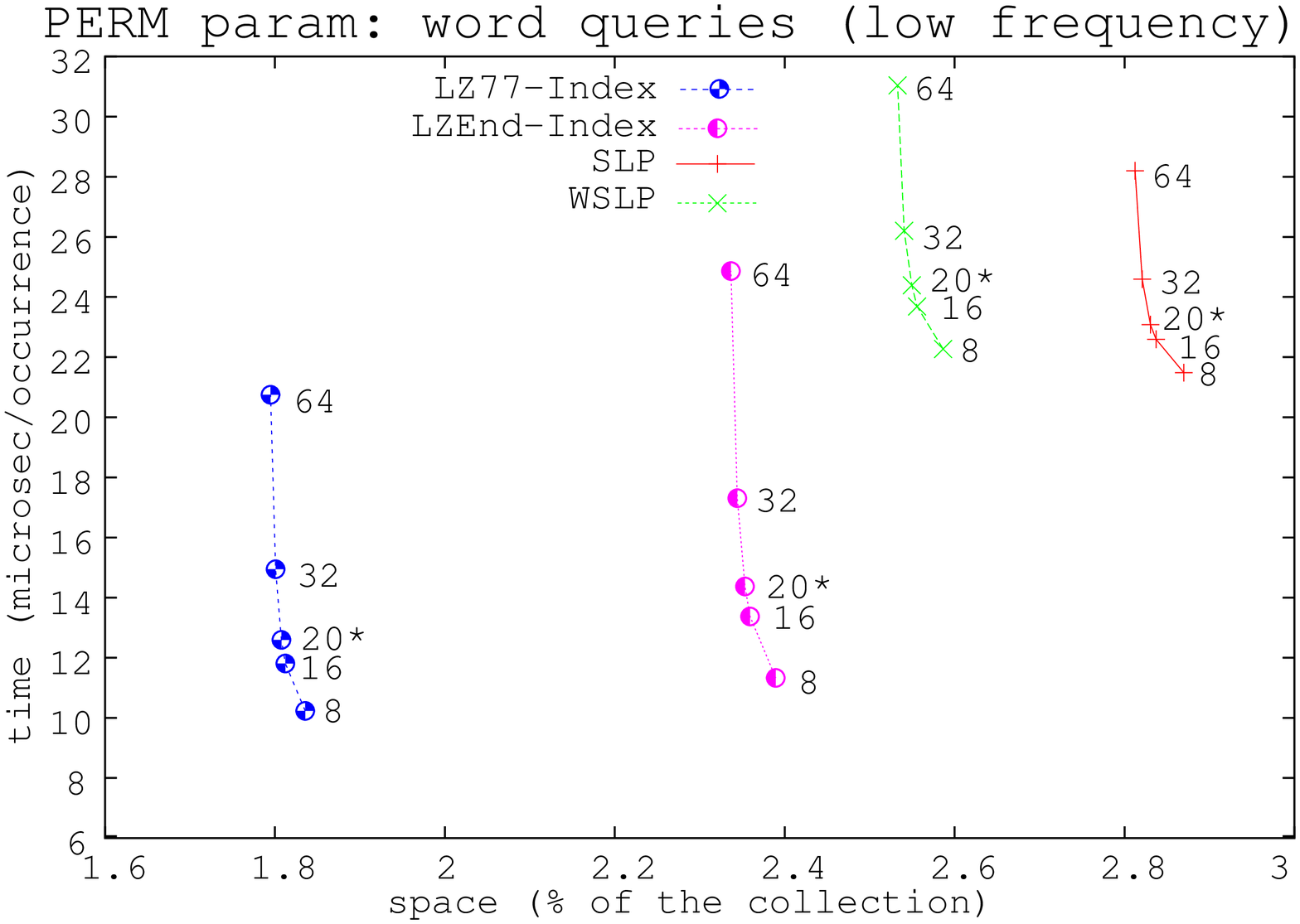}
\includegraphics[angle=-00,width=0.45\textwidth]{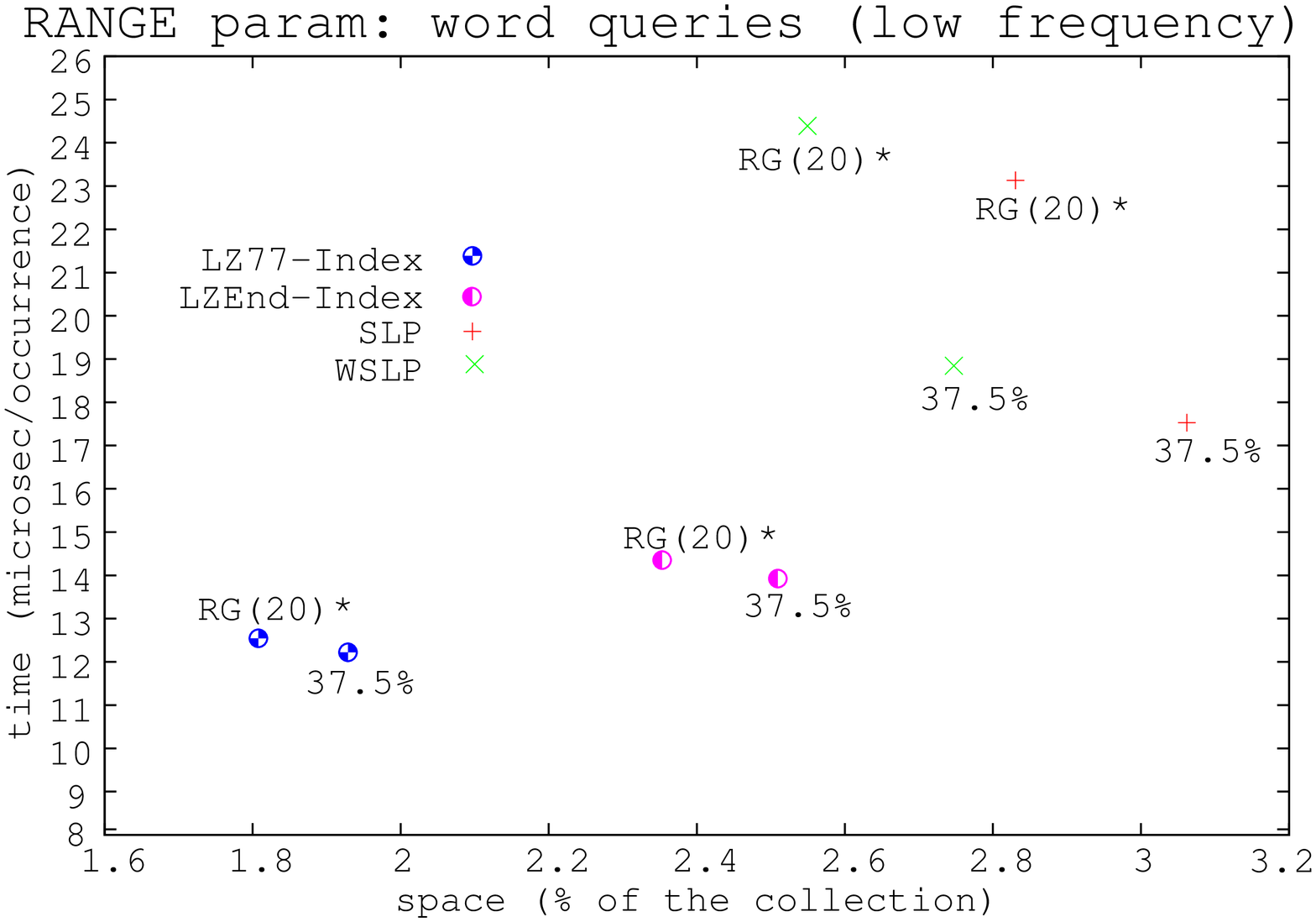}
\caption{Space/time tradeoffs for fine tuning on {\tt perm} (left plot) and {\tt range} (right plot) parameters.}
\label{fig:tuning}
\end{center}
\end{figure*}

%%%% SELF-INDEXES CONFIGURATIONS
%%%% 
%%%% LZ77-LZend:
%%%%	- Configuration 5 -> [3] (binarySst,binaryRev)
%%%%	- Delta=16 // Perm=8 // Range=RG(20)
%%%%
%%%% SLP-WSLP:
%%%%	- Bitmaps Xa/Xb = RGK(32)
%%%%	- Delta=16 // Perm=8 // Range=375

\subsubsection{Tuning self-indexes} 
\label{exp:pos:selfindexes}

Before comparing them with inverted indexes, we tune \slp\ and \verb|LZ|-based
self-indexes to improve their space/time tradeoffs.

\paragraph{SLP-based self-indexes}
We analyze four different parameters for tuning \slp\ and \wslp:

\begin{itemize}
	\item {\tt delta} establishes the sampling value used by the bitmap $B$, which records the starting positions of the symbols in $C$. The space/time tradeoffs are 
	  not greatly improved by this parameter, so we retain its original value {\tt delta=16}.
	\item {\tt perm} determines the largest allowed cycle length \cite{MRRR03} in the permutation $\pi$ that maps
	  reverse to direct lexicographic ordering. We try values from 
	  {\tt perm=64} to {\tt perm=8}. As seen in Figure \ref{fig:tuning} (left), 
	  smaller values yield better query times at the price of slight additional
	  space. On the other hand, this parameter does not have any influence on snippet 
	  extraction performance. Even so, we set {\tt perm=8} to improve all types of word/phrase 
	  queries, because it only adds $\approx 800$ KB to both self-index sizes.
	\item {\tt range} parameterizes the bitmap configuration used for implementing the two binary
	  relations in the self-index. We evaluate several configurations of well-known techniques 
	  from the state of the art \cite{GGMN05}. As shown in Figure \ref{fig:tuning} (right), 
	  the technique called $37.5\%$ (which uses $37.5\%$ extra space on top of the bitmap) 
	  outperforms by far the original {\tt range} configuration (which only adds $5\%$ extra space) 
	  for all types of queries. A similar situation arises for snippet extraction. In this case, we 
	  prefer the fastest configuration at the price of increasing the self-index sizes ($\approx 5$MB 
	  for \slp\ and $\approx 4$MB for \wslp).	  
	\item Finally, parameter {\tt dict} affects the compressed string dictionaries 
	  used for indexing $q$-gram prefixes and suffixes from $LBR$s. As explained in Appendix 
	  \ref{sec:self:slp}, these structures speed up binary searches on rows and columns of 
	  $LBR$s. We consider $q$-grams of different length (from $q=1$ to $q=12$ characters), but the 
	  best tradeoffs are reported for $q=4$. Thus, we index all prefix and suffix combinations of 
	  $4$-chars using a Plain Front-Coding (PFC) dictionary \cite{MPBCCN:16}.
\end{itemize}

%% SLP		Wl	Wh	P2	P5		s80	s13k
%% 59095601	23,08	22,00	21,11	20,73		5,07	3,10
%% 67590250	16,04	15,18	14,44	13,90		2,88	1,40
%% WSLP
%% 53232576	24,38	22,83	22,17	21,89		2,90	1,41
%% 60695504	16,43	15,42	14,87	14,30		1,77	0,61

All these decisions converge into new \slp\ and \wslp\ configurations that use $\approx14\%$ more space
than their original counterparts, but are clearly faster in all types of queries and snippet extraction operations.
More precisely, word/phrase queries are $30\%$--$35\%$ faster, while the extraction speed is doubled on average.

\paragraph{LZ-based self-indexes}
We consider five different variants of these indexes \cite{KNtcs12}. These are,
ordered by decreasing space, as follows:

\begin{itemize}
	\item {\tt Conf.\#1} uses suffix and reverse Patricia trees for representing
	  phrase boundaries.
	\item {\tt Conf.\#2} performs binary search on the {\em id} array and 
	%\item {\tt Conf.\#2} performs binary search on the explicit {\em id} array and 
	  holds the Patricia tree for the reversed phrases.
	\item {\tt Conf.\#3} holds the Patricia tree for the suffixes and performs
	  binary search on the explicit {\em rid} array.
	  %binary search on the {\em rid} array.
	\item {\tt Conf.\#4} performs binary searches on {\em id} and the explicit
	%\item {\tt Conf.\#4} performs binary searches on the explicit {\em id} and 
	  {\em rid} arrays.
	\item {\tt Conf.\#5} performs binary searches on {\em id} and the implicit {\em rid} arrays.
	%\item {\tt Conf.\#5} performs binary searches on the implicit {\em id} and
	  %{\em rid} arrays.
\end{itemize}

\begin{figure*}[t]
\begin{center}
\includegraphics[angle=-90,width=0.32\textwidth]{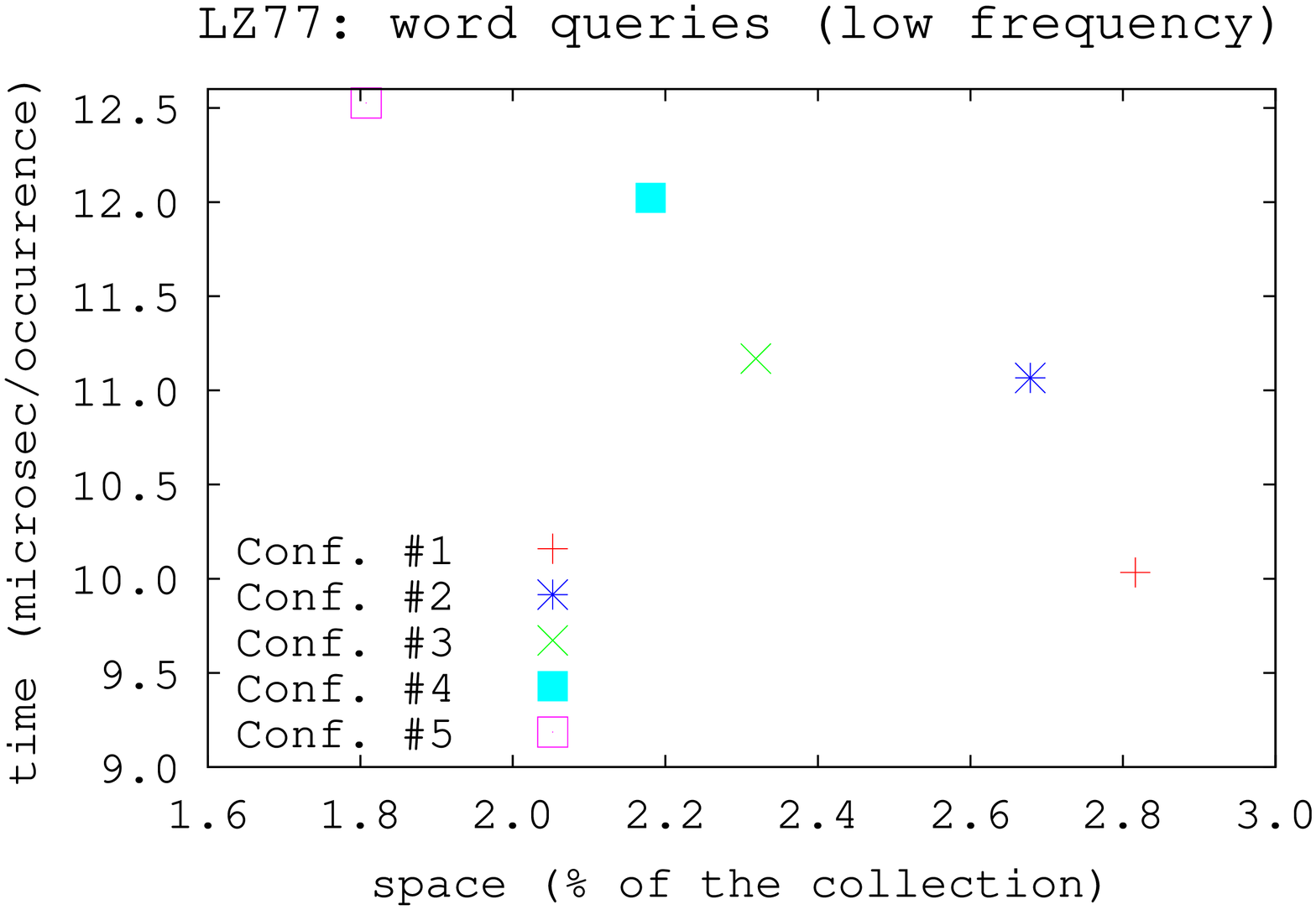}
\includegraphics[angle=-90,width=0.32\textwidth]{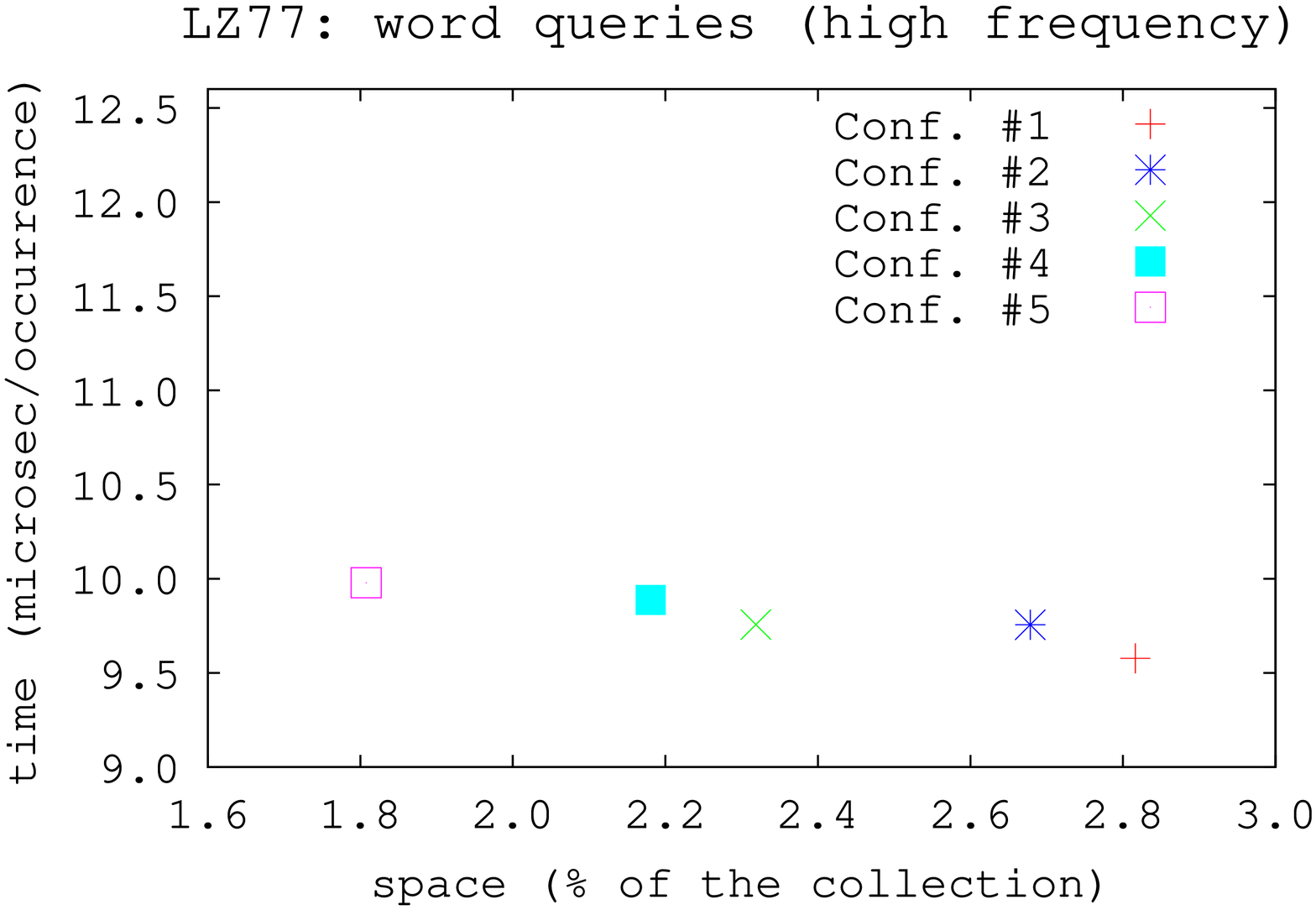}
\includegraphics[angle=-90,width=0.32\textwidth]{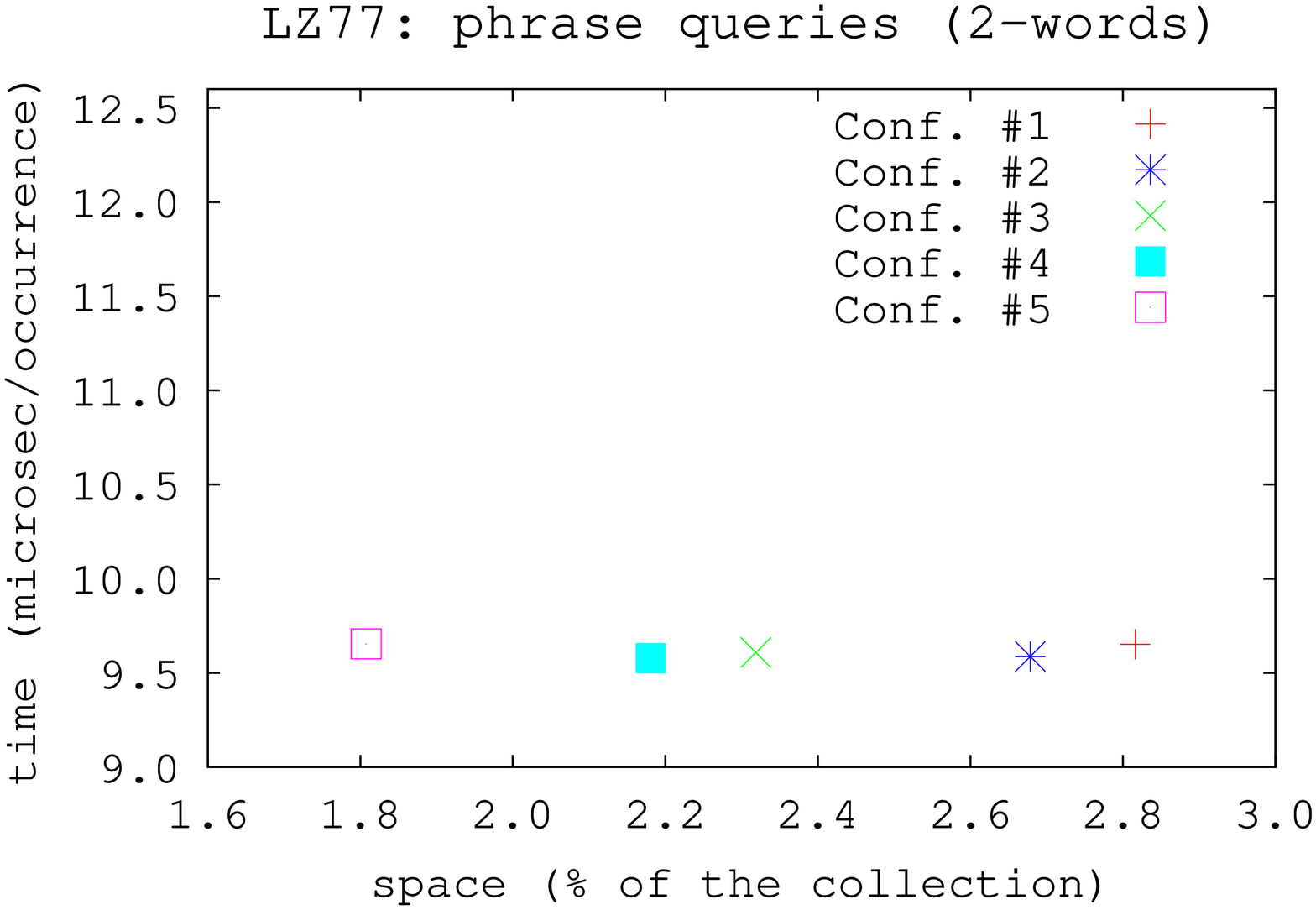}
\includegraphics[angle=-90,width=0.32\textwidth]{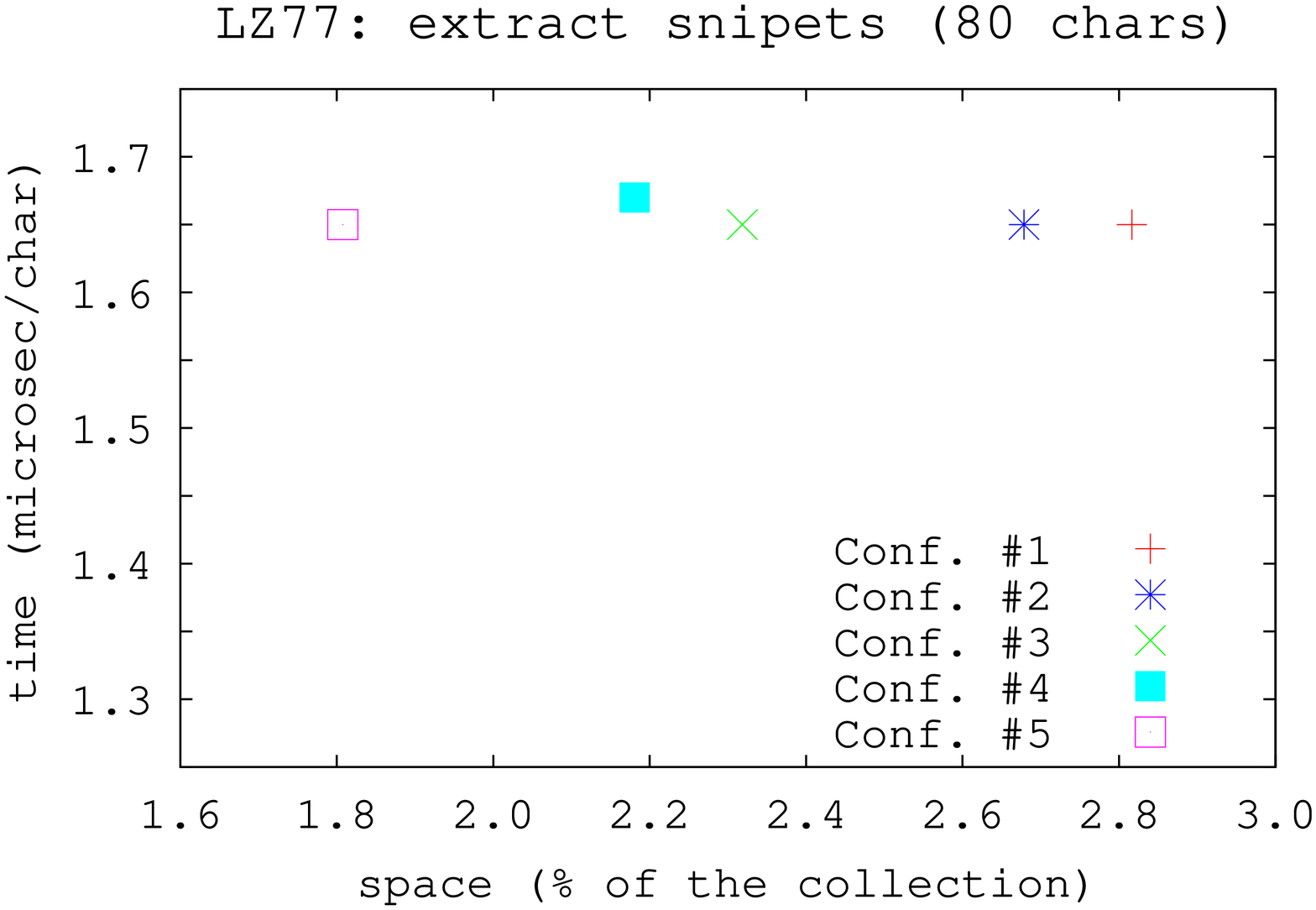}
\includegraphics[angle=-90,width=0.32\textwidth]{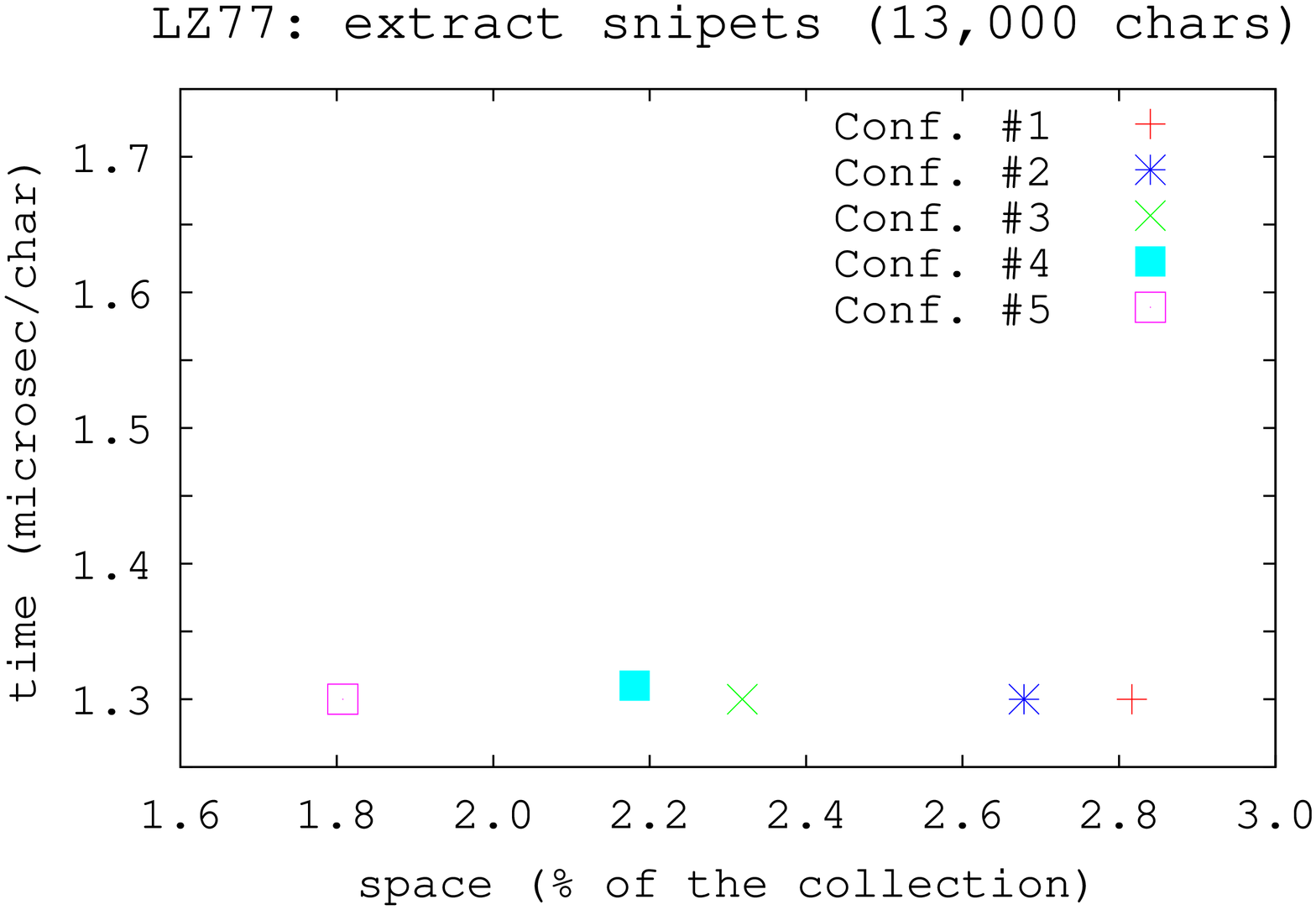}
\caption{Space/time tradeoffs for different LZ77-index configurations.}
\label{fig:pos2}
\end{center}
\end{figure*}

Figure \ref{fig:pos2} shows the space/time tradeoffs on \lzindex\ (similar conclusions are obtained on \lzendindex). In this case, the space used ranges from $1.8\%$ to $2.8\%$ of the original 
collection size. This means, in practice, that {\tt Conf.\#1} requires about $56$ MB and {\tt Conf.\#5} about
$36$MB. Regarding performance, the most noticeable difference is seen in low-frequency 
word queries (results range from $10$ to $12.5\mu s$ per pattern occurrence). In the remaining cases,
the difference between the fastest and the slowest configurations is roughly $1\mu s$ per pattern 
occurrence. Moreover, snippet extraction performance does not depend on the chosen configuration, as
can be seen in the bottom plots. Thus, we decide to use {\tt Conf.\#5}, as it clearly
reports the best compression numbers while providing very competitive performance both for query
and extraction operations.

Once the configuration is defined, we tune the same {\tt delta}, {\tt perm}, and {\tt range} 
parameters described for SLP-based indexes. In this case, {\tt delta} parameterizes the bitmaps $S$ and $B$, which encode the phrase structure.
As in the previous case, it has no relevant effect in the query tradeoffs,
but it affects extraction performance. Nevertheless, we discard reducing {\tt delta}
because it introduces a non-negligible space overhead (more than $20\%$ of the space). Thus,
the original value {\tt delta=16} is maintained. Regarding {\tt perm}, its influence is similar to that for 
\slp-indexes; see Figure \ref{fig:tuning} (left). Thus we also also choose {\tt perm=8} to 
favor query times. Finally, the {\tt range} value is not as decisive as in the previous 
case. As can be seen in Figure \ref{fig:tuning} (right), query performance is barely improved, but 
space increases as for \slp-indexes. Thus, we retain the RG(20) bitmaps for building binary relations
in both LZ-based self-indexes.

%% LZ77		Wl	Wh	P2	P5
%% 37740334	12,59	9,99	9,66	10,21
%% 38324314	10,22	7,86	7,53	 8,28
%% LZend
%% 49131858	14,37	12,06	11,67	11,67
%% 49884478	11,33	 9,31	 9,02	 9,29

Summarizing, we only change the {\tt perm} value on the {\tt Conf.\#5} variant.
This means only 
$1.5\%$ extra space, but speeds up word/phrase queries by around $20\%$ in all cases.

\subsubsection{Comparing positional inverted indexes with self-indexes} \label{exp:pos:ours}

\begin{figure}[t]
\begin{center}
\includegraphics[angle=-90,width=0.49\textwidth]{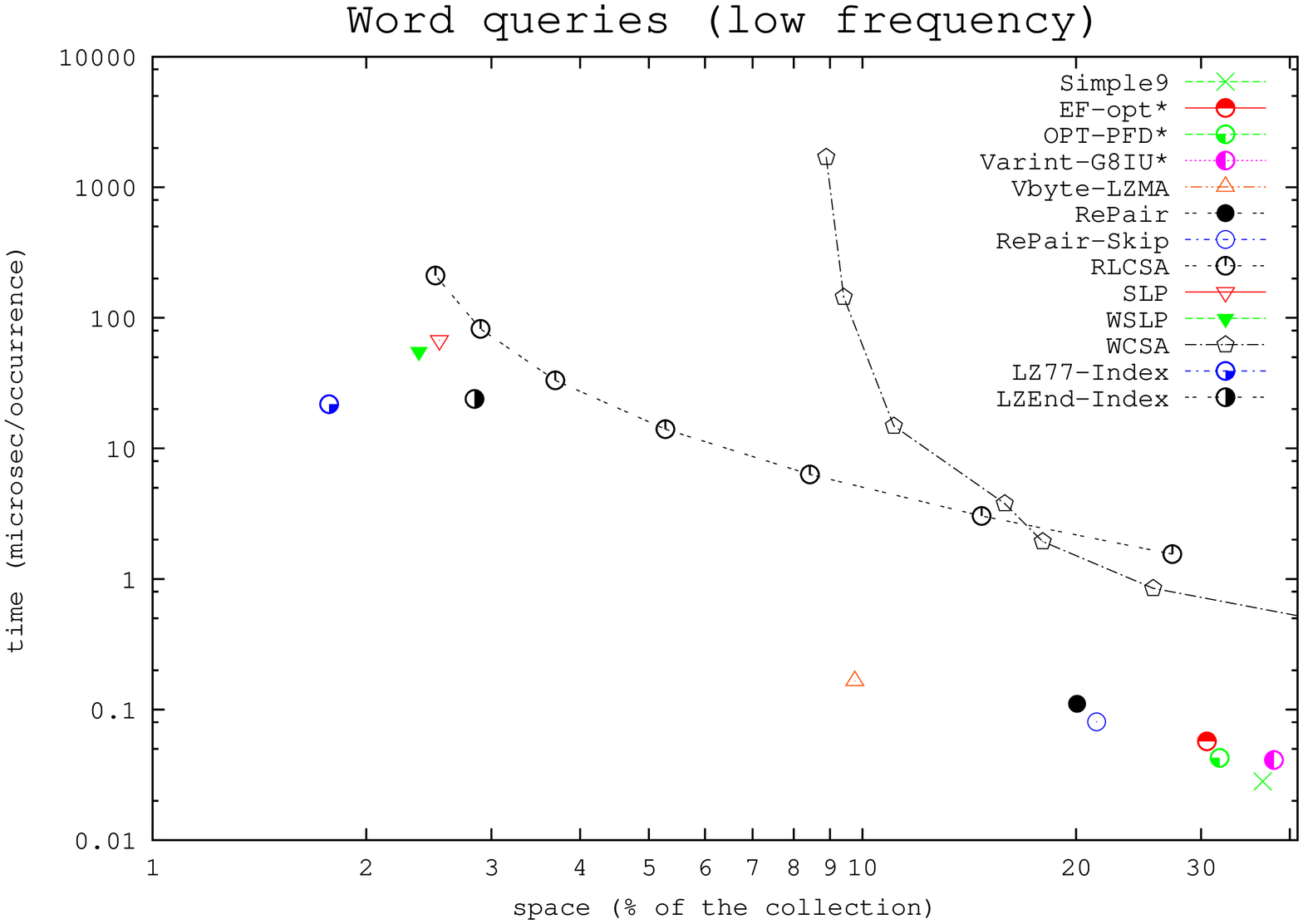}
\includegraphics[angle=-90,width=0.49\textwidth]{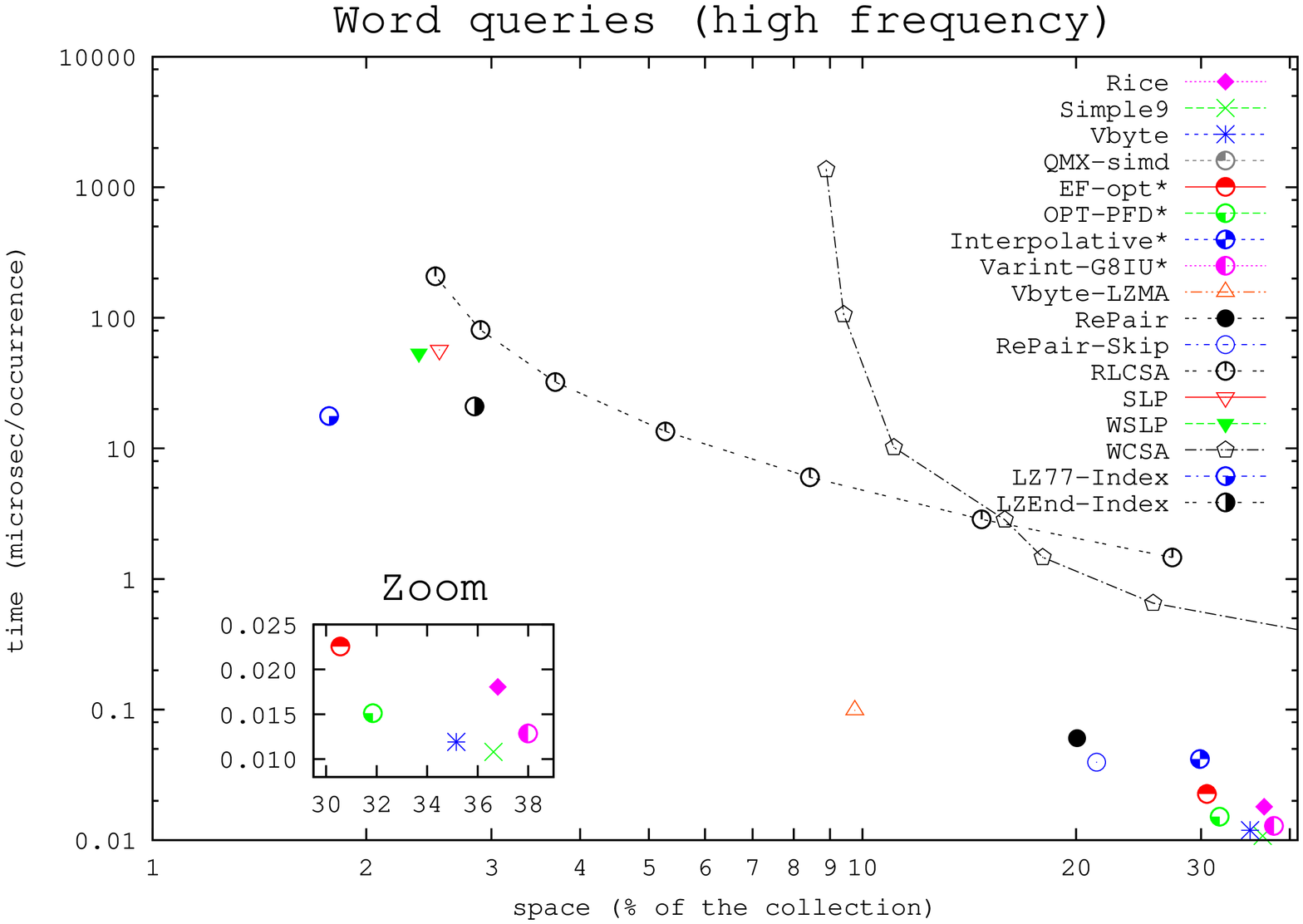}
\includegraphics[angle=-90,width=0.49\textwidth]{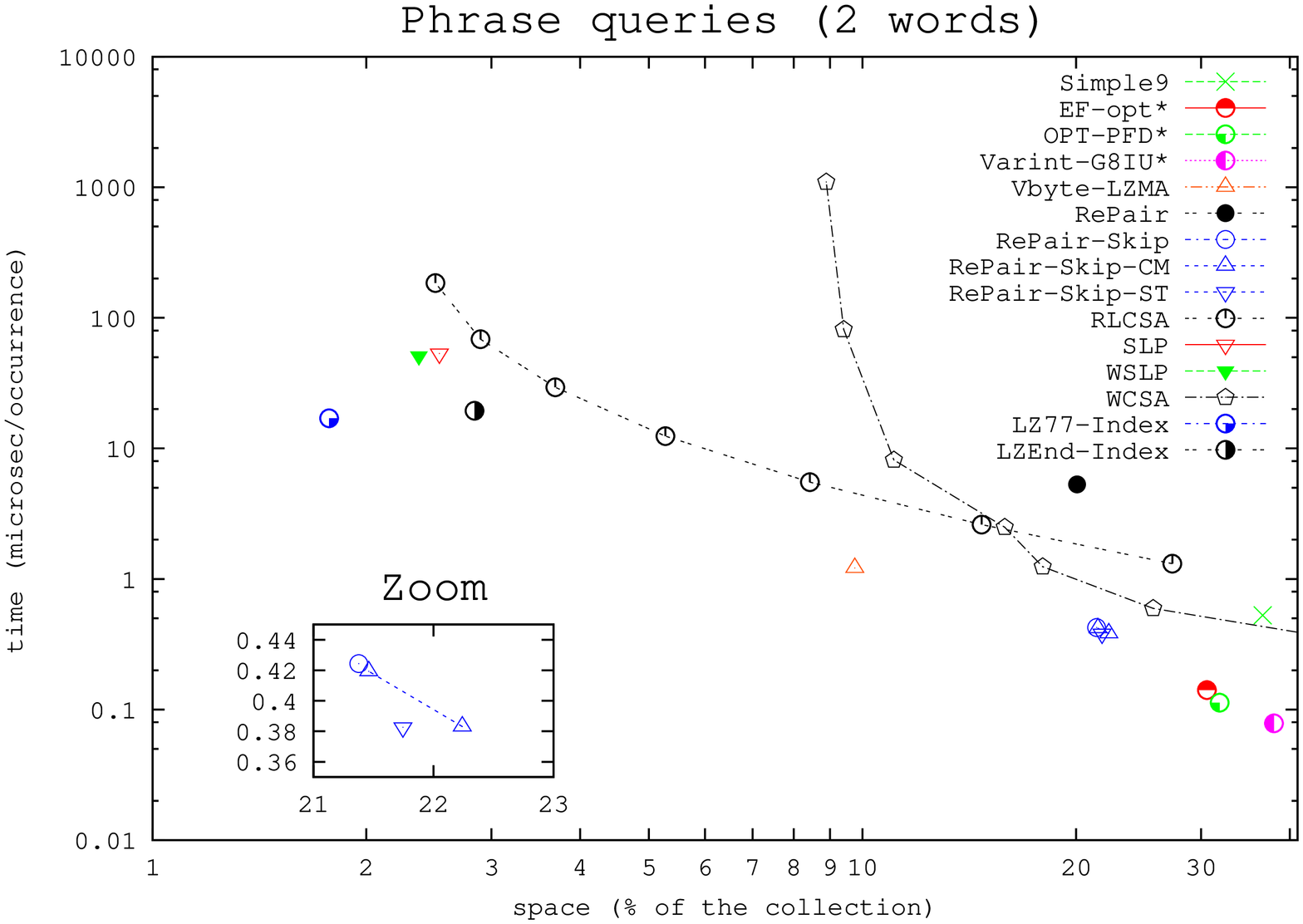}
\includegraphics[angle=-90,width=0.49\textwidth]{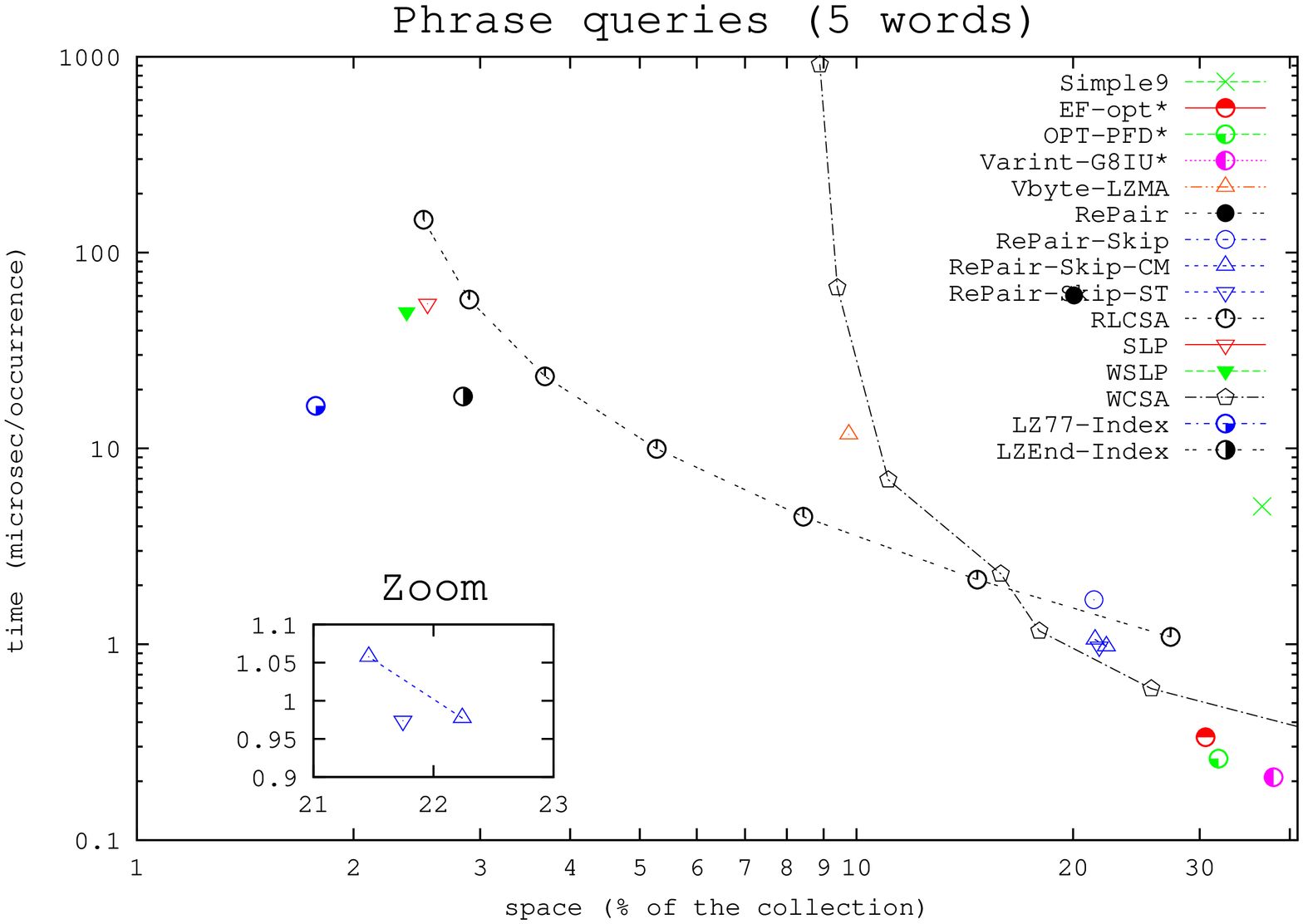}
\caption{Space/time tradeoffs for positional indexes. Logscale.}
\label{fig:pos2.2}
\end{center}
\end{figure}

We compare the best traditional inverted indexes with the variants we developed to exploit repetitiveness. In addition,
we include the self-indexes (tuned as shown above) in the comparison. Figure~\ref{fig:pos2.2} shows the results.

\repairNo\ and \repairSkip\ achieve almost the same space, %slightly below 30\%, 
close to 20\%,
and the latter is always faster for the same reasons as on
non-positional indexes. While for words \repairSkip\ is slower than the 
classical methods, its times become similar to those of \simplen\ on phrases. Adding sampling on 
top of \repairSkip\ clearly outperforms \simplen\ on phrases. Yet, \repairSkipST\ and \repairSkipCM\ are still clearly slower than the \efopt, \optpfd, and \varint, which obtain the best performance at phrase queries.

%obtain similar times to those of \vbyteCM\ and (only) 50-100\%
%slower than those of the fast \vbyteST.
%
The best space of inverted indexes is achieved by \vbyteLZMA, which reaches
a compression ratio near 10\% (half the space of \repairSkip\ variants). This represents a significant improvement upon the state of the art. Moreover, for single-word queries its times are
only slightly worse than those of \repairSkip, yet on phrase queries its need
to fully decompress the list makes it clearly slower (among the inverted indexes, only \repairNo\ performs worse than \vbyteLZMA\ in this scenario).

Self-indexes are able to use much less space. First, note that \wslp\ is only 
slightly smaller than \slp. This shows that grammar-based compressors do not gain
much from handling words instead of characters. They achieve around $3\%$
compression ratio. This important reduction in space compared to the 10\% of
\vbyteLZMA\ is paid with
a sharp increase in search times. On words, they are up to $100$--$150$ times slower 
than \vbyteLZMA. This gap, however, decreases to $12$ times on 2-word queries and to
$1.2$ times on 5-word queries. These self-indexes are mostly 
insensitive to the number of words in the query, whereas inverted indexes become 
much slower when looking for longer phrases. Thus, for long queries, 
\slp\ and \wslp\ are very attractive alternatives.

The \rlcsa\ offers a wide space/time tradeoff that goes from roughly the 
space of \lzendindex\ (where the latter is faster) to that reported by inverted 
indexes. Although these are clearly faster for word searches, differences are
reduced as the search phrase becomes longer. Note that \rlcsa\ reports
similar numbers than \repair-based inverted indexes for 5-word phrase queries.
Regarding \wcsa, it can be seen as a word-based variant of the \rlcsa, yet it 
is not so well optimized for highly repetitive sequences. As expected, \wcsa\
is far from the space reached by other self-indexes (its best space is
about $10\%$ of the original collection). On the contrary, it reports the
best self-index times for all types of queries when using sufficient space.
For that space, other inverted indexes are much faster on word queries, but 
the \wcsa\ retains a niche on phrase queries. In conclusion, \rlcsa\ and 
\wcsa\ build a bridge between self-index and inverted index tradeoffs,
opening an interesting area for future improvements.

Finally, LZ-based self-indexes report the best numbers in compression. \lzindex\ 
achieves the least space, overcoming its variant \lzendindex\ and also grammar-based 
compressors, both in time and space. The \lzindex\ takes less 
than 2\% space and answers queries in 8--10 $\mu s$ per occurrence. 
The \lzendindex\ needs about $2.4\%$ of the original space and solves queries
in 9--11 $\mu s$ per occurrence. Their performance is particularly interesting 
on 5-word phrase queries. In this case, they are faster than \vbyteLZMA,
and compete in the same order of magnitude of \simplen\ and all \repair-based 
inverted indexes. Thus, for long queries these self-indexes compete with 
traditional approaches, but use up to $10$--$20$ times less space.

\subsubsection{Text extraction} \label{exp:pos:extract}

Since self-indexes represent the text as a part of the index, it is relevant
to measure how fast they are at extracting an arbitrary text snippet. For
fairness we have added to our inverted indexes a \repair-compressed version
of the text. In order to support snippet extraction, we add a regular
sampling over the $C$ vector, which indicates the text position where the
corresponding symbol starts. For decompressing an arbitrary snippet we binary
search the rightmost preceding sample and decompress from there. This induces
a space/time tradeoff regarding the sampling step.

\begin{figure}[t]
\begin{center}
\includegraphics[angle=-90,width=0.49\textwidth]{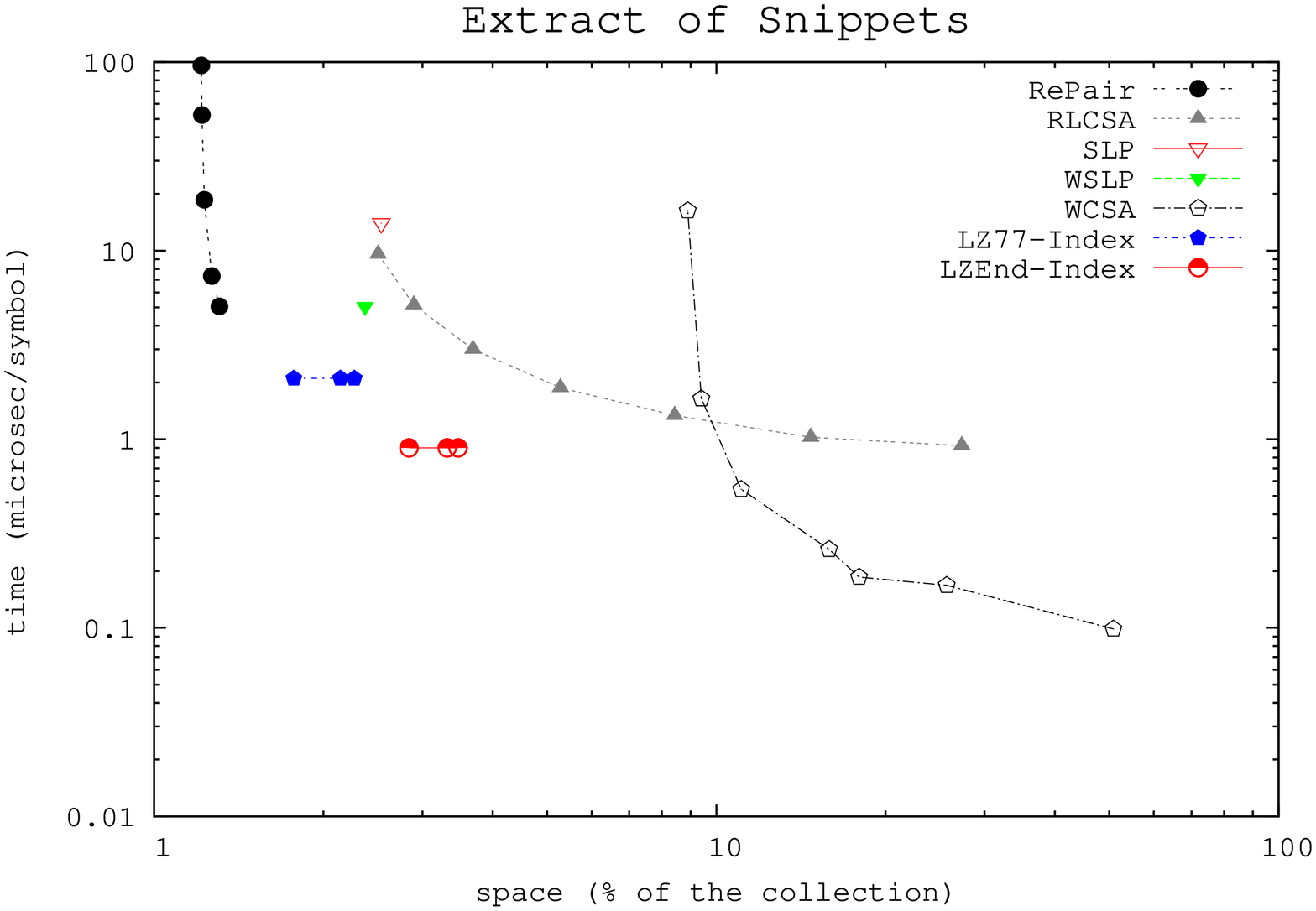}
\includegraphics[angle=-90,width=0.49\textwidth]{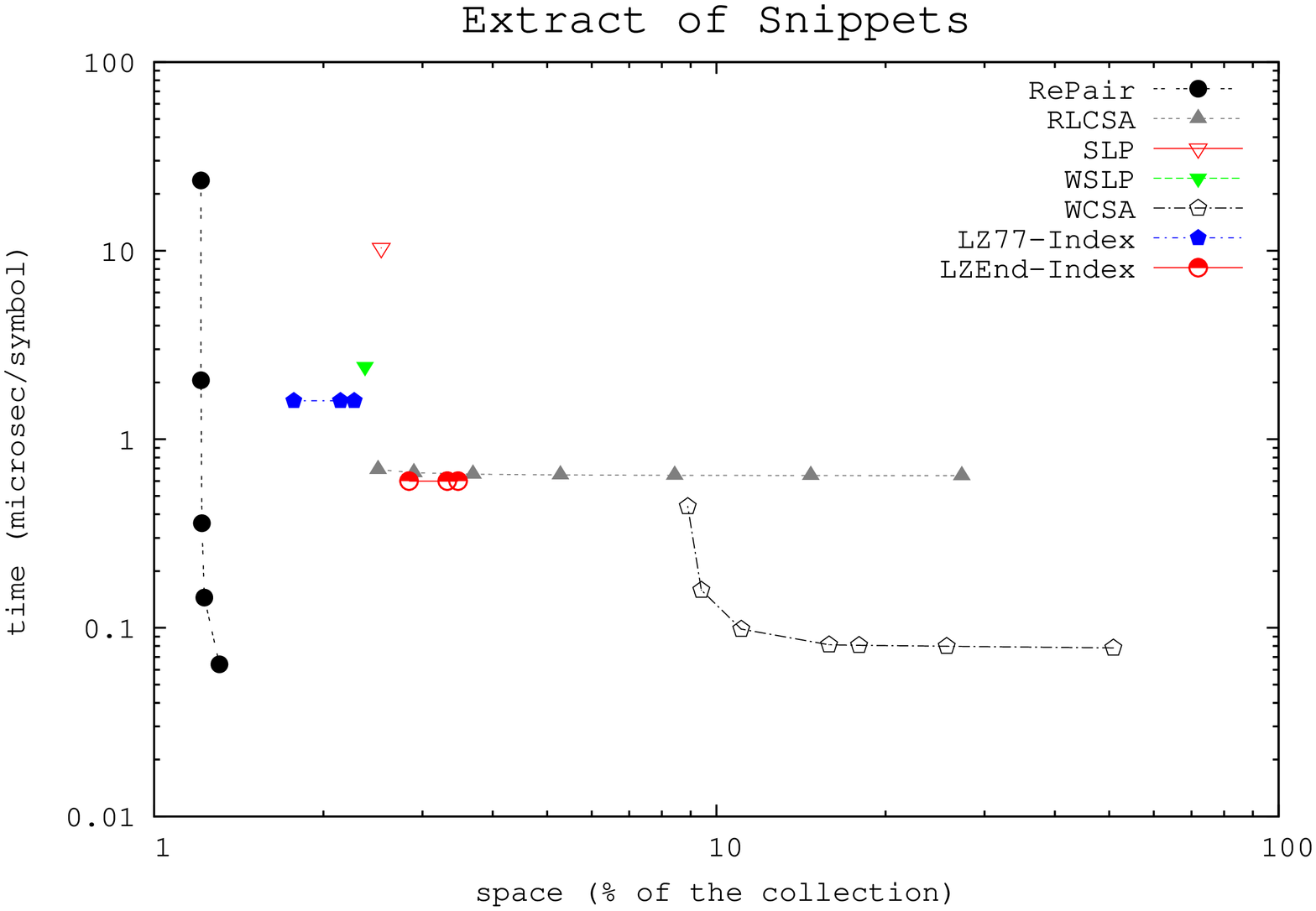}
\caption{Space/time tradeoffs for extraction. Logscale.}
\label{fig:extract}
\end{center}
\end{figure}

Figure~\ref{fig:extract} shows the results of extracting random snippets
of length 80 and 13,000 characters. Word-based indexes \wcsa\ and \wslp\ extract a number of
words equal to 80 or 13,000 divided by the average word length, to provide
a roughly comparable result. 

Here the word-based self-indexes \wcsa\ and \wslp\ are significantly faster than
their corresponding character-based counterparts \rlcsa\ and \slp. \wslp\ is
slightly faster than \lzindex, but it is always overcome by the \lzendindex. The 
fastest method overall is \wcsa, at $0.1 \mu s$ per extracted symbol, but at the price of using much space. The second-fastest is the \lzendindex, reaching $0.5$--$0.7 \mu s$ per 
extracted symbol ($1.4$--$2$ MB/sec) and much better space than \wcsa, yet still far 
from the smaller \lzindex, which takes $1.2$--$1.5 \mu s$ per 
extracted symbol. Finally, \rlcsa\ competes better for long snippets, but it 
nevers dominate the comparison. Note that the methods that offer a tradeoff 
are much more sensitive to a denser sampling when extracting short snippets.

The line marked \repairNo\ {\tt (text)} corresponds to the text represented with 
\repair\ plus sampling. It uses the least space but it must be added to an inverted 
index in order to support searches. \repairNo\ is very slow to display short 
contexts. When displaying long contexts, however, it is the fastest option, 
even beating \wcsa. This shows that the method is intrinsically fast, yet
it is costly to arrive at the right starting position to extract (in the
densest sampling shown, we sample every position in $C$).

%%%%%%%%%%%%%%%%%%%%%%%%%%%%%%%%%%%%%%%%%%%%%%%%%%%%%%%%%%%%%%%%%%%%%%%%%%%%%%%%%%%%%%%%% 
%%%%%%%%%%%%%%%%%%%%%%%%%%%%%%%%%%%%%%%%%%%%%%%%%%%%%%%%%%%%%%%%%%%%%%%%%%%%%%%%%%%%%%%%% 
\section{Conclusions}

%contribuciones
%  rle
%  lzma para nonpos e intersecc
%  repair
%  wslp
%  slp mejorada

We have studied the problem of indexing text collections that are highly
repetitive, that is, where most documents are very similar to others. 
Many of the fastest-growing text collections today are indeed repetitive, 
and therefore exploiting repetitiveness in order to store and index them 
within little space is the key to handle the huge collection sizes that are 
becoming commonplace. 

Repetitive collections may arise
in controlled scenarios like versioned document collections, where versions
have a known linear structure (such as Wikipedia) or a known tree structure
(such as software repositories), but also in less controlled scenarios like
DNA sequence collections of similar species or periodic publications, where
the repetition structure may be chaotic and unknown. Our main focus has been
natural language text collections, where the inverted index is the main
actor, but we have also studied self-indexes, a new family of structures that
apply on general string collections. 

We first studied how known non-positional inverted indexes perform on repetitive
collections. These indexes store the documents where each word appears.
While classical indexes require 3\%--6\% of the plain text size in these
collections, some of the more recent inverted indexes 
\cite{MS00,AM05,YDS09,OV14} reach just 0.5\%-1.0\% and are about as fast. This 
is, however, significant if we consider that a Lempel-Ziv compression on the
repetitive text reduces it to just 0.5\% of its size. Our new non-positional
inverted indexes, instead, reach 0.1\%--0.2\% of the plain text size, at the
price of being a few times slower. Previous work \cite{HZS10} obtains even less
space and time, but they can only be used if the versioning structure is formed
by known 
isolated documents and their versions. Our techniques, instead, are
universal: they do not need that a clear versioning structure is known or even
exists.

Positional inverted indexes also store the positions of the words in the
documents, and support phrase queries. Classical ones require more than 40\% of
the plain text size, whereas the more recent formats \cite{MS00,YDS09,OV14}
reach around 30\% and are about as fast. Our new repetition-tailored 
representations, instead require 10\%--20\% space and are a few times slower.
Self-indexes, which offer similar functionality, reach as little as 2\%--3\% 
space, but are orders of magnitude slower.

Our main technical novelty is to apply grammar-based compression (\repair, in 
particular) to the whole set of differentially-encoded inverted lists, and
enhance the grammar nonterminals with summary information that allows us 
intersecting the compressed inverted lists without fully decompressing them.
This is the key to obtain significant space reductions while only moderately
slowing down the operations. We also implement and tune several other simpler
or existing ideas, some of which also obtain relevant results.

In the case of inverted indexes, being a few times slower is not so important
if their lower space allows us holding the index in a faster and smaller memory,
for example in main memory instead of disk. If both structures must reside on
disk, the smaller size of our indexes allows retrieving the lists with fewer I/Os,
which blurs the relatively small CPU-time differences and plays in our favor.
For example, if our inverted index is 3--5 times smaller than a classical one,
then a long enough list will also be read into main memory 3--5 times faster.

In the case of self-indexes, even if they are orders of magnitude slower, they 
may still be convenient if their smaller size allows us fit them in main memory 
instead of on disk. In addition, self-indexes handle general string 
collections, not only natural language. On the other hand, self-indexes
do not perform well on disk.

These giant space/time differences clearly indicate that much more can be done
in this area. An interesting line of future work is to find further regularities
induced by repetitiveness in the inverted indexes, so as to match the space of
self-indexes while retaining the good time performance of inverted indexes.
We have shown that grammar-based compression exploits several inter- and 
intra-list regularities, while allowing for fast list processing, but the 
results on self-indexes show that we may be missing many others.

Another important line to explore is that of more sophisticated queries. 
One such query is to find the occurrences of a pattern within a range of 
document identifiers. This could correspond to a range or subtree of versions, 
or a temporal interval. While the task is relatively simple with an inverted 
index, self-indexes deliver the results out of order, and such a requirement 
is challenging for them \cite{HSTV12}. 

Another challenging family of queries for self-indexes are the 
document-oriented ones. The equivalent of a non-positional inverted index is
a self-index offering {\em document listing}, which lists the documents where
a pattern appears. Only very recent proposals 
exist for this problem on repetitive collections \cite{CM13,GKNPS13}, but
even the best implementations use 7\% space or more. More sophisticated
queries involve obtaining the $k$ most important documents where the pattern
appears, according to some definition of importance. There already exist some 
proposals for this problem in the repetitive scenario, but they are either
inverted indexes for the particular versioning structure described above
\cite{He:Sigir12}, or they are self-indexes using at least 20\% of space
\cite{NPS14}.

We have left our codes and experimental testbeds available at
{\tt https://github.com/migumar2/uiHRDC}.
%%%%%%%%%%%%%%%%%%%%%%%%%%%%%%%%%%%%%%%%%%%%%%%%%%%%%%%%%%%%%%%%%%%%%%%%%%%%%%%%%%%%%%%%% 
\section*{Acknowledgements}
We thank Jinru He, Junyuan Zeng, and Torsten Suel, for providing
their input collection and useful information on how to use their files.

%%% FALTAN ACKS
%%%%%%%%%%%%%%%%%%%%%%%%%%%%%%%%%%%%%%%%%%%%%%%%%%%%%%%%%%%%%%%%%%%%%%%%%%%%%%%%%%%%%%%%% 
%%%%%%%%%%%%%%%%%%%%%%%%%%%%%%%%%%%%%%%%%%%%%%%%%%%%%%%%%%%%%%%%%%%%%%%%%%%%%%%%%%%%%%%%% 

\bibliographystyle{abbrv}
\bibliography{paper}

\appendix

\section{Self-Indexes} \label{sec:self}

Self-indexes \cite{NM07} are an innovative approach for positional indexing. 
They process the concatenated text collection $\D$ to build compressed 
structures that integrate the text and a positional index in a single 
representation. It is worth noting that self-indexes are designed for general 
strings, not only natural language collections where only words and phrases 
can be sought.

Self-indexes provide efficient pattern searching and snippet extraction. In the first operation, 
the self-index will find all the occurrences of the pattern, yet usually not in order. This 
implies that the $o$ pattern occurrences must be sorted prior to translating them
into the document and offset format, using the technique described in Section~\ref{sec:lists}.
This leads a total translation cost of $O(o\log n)$. The extraction operation is implemented by decoding
the required text fragment from the compressed self-index, and takes proportional time to the length
of the snippet.

We consider different self-indexes to compare their space/ time tradeoffs with those reported by compressed
inverted indexes. On the one hand, we analyze character-oriented approaches, which regard the text as a 
sequence of characters and report any substring matching the pattern. We choose \rlcsa, \lzindex, 
\lzendindex, and \slp\ as examples of character-oriented self-indexes. We make some tuning and improvements on the 
last three techniques, outperforming the original implementations. These are
described in this section. On the other hand, we include two 
word-oriented self-indexes in our study: \wcsa\ and \wslp, the latter designed for 
this paper. These techniques map words and {\em separators} (maximal spaces between words) to integers, 
and the indexes represent the resulting integer sequence. We use the spaceless words model \cite{MNZBY00}, where 
the single whitespace separator is omitted and assumed by default. The mapping between integers and their
corresponding words is provided by a simple vocabulary representation.

\subsection{CSA-based self-indexes}
The {\em Compressed Suffix Array} (\verb|CSA|) of Sadakane \cite{Sad03} proposes a succinct
suffix array encoding. In short, the {\em suffix array} $A[1,n]$ for a text $\D[1,n]$, with
alphabet $\Sigma=[1,\sigma]$, is a permutation of $[1,n]$ so that $\D[A[i],n]$ is the $i$-th
lexicographically smallest suffix in the text. Note that the suffix array arranges all the suffixes starting with any given search pattern $P[1,m]$ in a contiguous range, so their occurrences can be binary 
searched in time $O(m \log n)$.

\verb|CSA| encodes $\D$ and $A$ using two main structures (plus other less important ones). 
A bitmap $B[1,n]$ marks where the first symbol of the suffixes changes in $A$. That is, $B[i]=1$ iff $i=1$ 
or $\D[A[i]] \neq \D[A[i-1]]$. The second structure is the array $\psi[1,n]$, 
defined so that $A[\psi[i]]=A[i]+1$. That is, if the suffix
$\D[j,n]$ is pointed from $A[i]=j$, then the next text suffix $\D[j+1,n]$ is pointed from $A[\psi[i]]=j+1$. Note that the first symbol of $\D[A[i],n]$ can be
recovered with $rank_1(B,i)$, the second with $rank_1(B,\psi[i])$, the third
with $rank_1(B,\psi[\psi[i]])$, and so on, which enables $O(m\log n)$ time
binary searching for the area of $A$ of the suffixes that start with $P[1,m]$.

Once we have found that the suffixes starting with $p$ are in the range 
$A[l,r]$, its occurrences are precisely the positions $A[l], A[l+1],\ldots,
A[r]$. To recover each such entry $A[j]$, we also use $\psi$. We sample the
text positions that are a multiple of some parameter $s$, and store the entries
of $A$ pointing to those in a sampled suffix array $A_S[1,n/s]$. We mark in a
bitmap $S[1,n]$ the positions of $A$ that are sampled, so that 
$A[j]=A_S[rank_1(S,j)]$ if $S[j]=1$. If $S[j]=0$, we take $\psi[j]$ for
increasing $k$ until it holds $S[\psi^k[j]]=1$ for some $k$. Then, by the 
properties of $\psi$, it holds $A[j] = A[\psi^k[j]]-k$. Note that it is
guaranteed that $k \le s$, thus each occurrence is reported in $O(s)$ time,
whereas the extra space for $A_S$ is $O((n/s)\log n)$ bits.

For extracting a text snippet $\D[x,y]$, note that we can obtain any text suffix
$\D[A[i],n]$ with $B$ and $\psi$. A second sampled array stores inverse suffix
array values: $A_s^{-1}[t]$ is the position of $A$ pointing to text position
$s\cdot t$. This allows us extracting $\D[x,y]$ in time $O(s+y-x)$, by
locating the latest sampled position before $x$ in $i = A_S^{-1}[\lfloor x/s
\rfloor]$, and then using $\psi$ and $B$ to extract the symbols 
from positions $A[i]= \lfloor x/s\rfloor \cdot s$ to $y$.

The remaining issue in terms of space is how to compress array $\psi$. It has
been shown that $\psi$ can be compressed to the statistical entropy of $\D$
\cite{Sad03,NM07}. The compression achieved, however, is not good enough for
highly repetitive collections. We now briefly describe the variants
\rlcsa\ and \wcsa, which extend \verb|CSA| from two different perspectives. 

\paragraph{RLCSA} The {\em Run-Length Compressed Suffix Array} (\rlcsa) \cite{MNSV09} was 
the first self-index optimized for highly repetitive collections. The authors show that
$\psi$ contains long runs of successive values in such class of collections, and \rlcsa\
exploits this fact. 

\rlcsa\ compresses $\psi$ by performing run-length encoding of the differences $\psi(i)-\psi(i-1)$. Regular samples on $\psi$ permit direct access to absolute
$\psi(i)$ values in competitive time and space. This sampling yields different 
space/time tradeoffs. However, the other sampling, related to the parameter $s$
used to build $A_S$ and $A_S^{-1}$, is the one yielding the most relevant 
tradeoffs. The most important drawback of \rlcsa\ is that these two sampled
arrays are not easy to compress, and their space become dominant in highly
repetitive scenarios.

\paragraph{WCSA} The {\em Word Compressed Suffix Array} (\wcsa) \cite{FBNCPR12} optimizes 
\verb|CSA| for natural language self-indexing, by regarding the text as a 
sequence of words instead of characters. It was not designed specifically to cope with 
high repetitiveness, but it fills a particular niche in the current comparison. \wcsa\ demands
more space than the other self-indexes, but outperforms them for all types of queries. Thus,
\wcsa\ can be considered as a bridge between self-indexes and inverted indexes.

\wcsa\ transforms the original text collection $\D$ into an integer one, $\D_w$, where each
position refers to a word/separator in the vocabulary. Therefore, the space/time
complexities we have given stay the same if we regard $n$ and $m$ as the number
of words in $\D$ and $P$, respectively. This shows why both space and time improve. The drawback is that, like the inverted indexes, \wcsa\ can only search for
whole words and phrases, not for any substring of characters.

\subsection{SLP-based self indexes}
\label{sec:self:slp}

Grammar-based compression is a good choice for posting list encoding (see Section 
\ref{sec:list-repair}), but it is also a promising alternative for self-indexing
purposes. \slp\ and \wslp\ are two grammar-based self-indexes built around the
notion of {\em straight-line program} (SLP). In short, an SLP is a restricted type 
of grammar which only allows two types of rules. On the one hand, rules of the form 
$X_i \rightarrow j$ mean that the terminal $j$ is generated from the rule $X_i$. On 
the other hand, rules $X_i \rightarrow X_lX_r$ expand $X_i$ as the concatenation of
$X_l$ and $X_r$.

Finding the smallest grammar for a given text is NP-hard \cite{CLLPSS05}, 
so heuristics must be used to efficiently build an SLP. We choose again Re-Pair as our
grammar-based compressor. RePair does not generate exactly an SLP, but it can
be easily adapted. We only need to enhance its set of rules with a subset of terminal
rules $X_i \rightarrow j$ for each symbol $j$ used in the text collection. 
%% Note that nonterminal rules: $X_i \rightarrow X_lX_r$ generated by Re-Pair are free of loops.
Note that RePair output comprises the grammar, but also the reduced sequence $C$,
which must be also indexed by SLP-based self-indexes. In the description that
follows,
we call $\mathcal{F}(X)$ the expansion of the rule $X$ into terminals, and 
$\mathcal{F}^{rev}(X)$ the corresponding reverse string (read backwards).

\paragraph{SLP} 
The \slp\ self-index \cite{CN11} is proved to require 
space proportional to that of an SLP compression of the text. It was 
implemented and tested on highly repetitive biological databases 
\cite{CFMPN10}. \slp\ indexes independently the resulting set of $n$ rules, $R$,
and the reduced sequence, $C[1,c]$, obtained by Re-Pair.

The set of rules is represented as a {\em labeled binary relation},
$LBR: A \times B \rightarrow \mathcal{L}$, where $A=[1,n]$, $B=[1,n]$, and 
$\mathcal{L}=[1,n]$ (we refer to $A$ as the rows and $B$ as the columns of the $LBR$,
respectively). This structure is populated as follows. For each nonterminal rule 
$X_i \rightarrow X_lX_r \in R$, we store the value $i$ in the cell $LBR[l,r]$. The rows
are sorted by lexicographic order of $\mathcal{F}^{rev}$, while the columns are
lexicographically sorted by $\mathcal{F}$. A permutation structure $\pi$ is used for mapping 
from rows to columns and vice versa (this inner structure will be tuned in Section
\ref{exp:pos:selfindexes}). This structure enables direct and reverse
access to the rules in $O(\log n)$ time per retrieved element:

\begin{itemize}
	\item $\mathcal{L}(l,r)$ returns $j$ if $X_j \rightarrow X_lX_r$ or $\perp$ otherwise.
	\item $R(l_1,l_2,r_1,r_2)$ retrieves all pairs $(l,r) \in R$ such that 
	  $l_1 \leq l \leq l_2$, $r_1 \leq r \leq r_2$.
	\item $\mathcal{L}(s)$ obtains the pair $(l,r)$ such that $X_s \rightarrow X_lX_r$.
\end{itemize}

$LBR$ is represented by traversing it by rows and writing two sequences,
$S_b$ and $S_l$, which concatenate respectively the $r$ and $j$ values for each 
cell $LBR[l,r]=j$. Two bitmaps, $X_A$ and $X_B$, encode in unary the cardinalities
of rows and columns. Sequence $S_b$ is represented using a wavelet tree without 
pointers \cite{CN08}, whereas $S_l$ uses a fast representation for large 
alphabets \cite{GMR06}. Both structures are tuned in 
Section~\ref{exp:pos:others}. 
%Bitmaps $X_A$ and $X_B$ are implemented using 
%a recent representation {\color{red} RGK bitmaps [cita?]}.

The \slp\ index performs three main operations to search for a given pattern
 $P[1,m]=p_1p_2 \ldots p_m$. First, all the {\em primary occurrences} of the 
pattern are located in the $LBR$.  Those are the occurreneces that appear when
two nonterminals are concatenated in the right hand side of a rule. We look for the $m-1$ possible pattern partitions of the form $P=P_<P_>$, such that 
$P_<=p_1p_2 \ldots p_k$ and $P_>=p_k+1 \ldots p_m, 1 \leq k < m$. For each 
partition, the rows and columns of $LBR$ are binary searched for the reversed 
$P_<$ and for $P_>$, respectively. This determines contiguous ranges of rows 
$[l_1,l_2]$ and columns $[r_1,r_2]$ such that $P_<$ is a suffix 
of $\mathcal{F}(X_l), l_1 \leq l \leq l_2$, and $P_>$ is a prefix of 
$\mathcal{F}(X_r), r_1 \leq r \leq r_2$. Thus, the primary occurrences of $P$ are inside of
all nonterminals $X_i$ which label $LBR$ pairs in $[l_1,l_2] \times [r_1,r_2]$. For each nonterminal $X_i$ found, 
we also store the offset of the ocurrence within it.

Our current \slp\ implementation introduces an improvement to speed up prefix 
and suffix searching. It indexes all the different prefixes of length $q$ in 
$\mathcal{F}^{rev}$ and $\mathcal{F}$ and, for each one, stores the first rule 
that expands to a string starting with the corresponding $q$-gram. This
significantly reduces the number of comparisons to be made on binary searches,
at the price of a small space overhead. These indexes are implemented using 
compressed string dictionaries \cite{MPBCCN:16}.

Each located primary occurrence $X_i$ leads to {\em secondary occurrences},
which are obtained whenever $X_i$ appears, directly or transitively, in other
right hands of rules. We must track all the nonterminals $X_s$
that use $X_i$ and then find further secondary occurrences from those $X_s$.
Thus, we locate all the labels $X_s$ such that 
$X_s \rightarrow X_iX_*$ or $X_s \rightarrow X_*X_i$. We then proceed 
recursively from those $X_s$ until no more nonterminals use them. We also 
maintain the offset where $P$ occurs inside each retrieved nonterminal. The
same nonterminal $X_s$ may be reported several times, but with different 
offsets, which leads to different occurrences.

For each nonterminal $X_s$ where $P$ appears, we locate all its occurrences in
the reduced sequence $C$. This is easily implemented using {\em select} on $C$:
$select_{X_s}(C,i)$ retrieves the $i$-th occurrence of $X_s$ in $C$. The original
description \cite{CN11} represented $C$ with a wavelet tree without pointers, 
which provides {\em select} queries in time $O(\log n)$, but we currently 
discard it, as explained in the next paragraph.

We still have to locate, however,
those occurrences that cross more than one symbol in $C$, that is, pattern occurrences that are 
expanded from consecutive nonterminals in the reduced sequence. For that purpose, we need a second 
$LBR$, which relates the set of $n$ rules with the $c$ positions of the sequence. That is, for
$C=s_1s_2 \ldots s_c$, the $LBR$ relates, with label $i$, the suffix from $s_{i+1}$ (sorted in 
lexicographic order of $\mathcal{F}(s_i)$) and $s_i$ (sorted in lexicographic order of 
$\mathcal{F}^{rev}(s_i)$). This $LBR$ configuration is a novelty compared to the original \slp,
because we now represent a relation of {\em suffixes $\times$ rules} instead of the original 
{\em rules $\times$ suffixes}. This subtle change makes the $select$ structure for $C$ unnecesary because
{\em select} operations can now be resolved using the $S_b$ wavelet tree. 
The occurrences that cross symbols in $C$ are reported just like the primary
occurrences using $LBR$.

Finally, to convert positions in $C$ (and their offsets) into actual positions
of $\D$, we set up a bitmap $B[1,n]$ that marks the positions of $\D$ where the
symbols of $C$ begin. This bitmap is sparse, and thus represented by 
gap-encoding the distances between consecutive 1s, plus a sampling structure 
that stores regular positions of $B$ \cite{KNtcs12}. This sampling value is also
studied in Section~\ref{exp:pos:selfindexes}.

The extraction of snippets is easily implemented by exploting the self-indexing
capabilities. To extract $\D[x,y]$, we find with $rank_1(B,x)$ 
the first symbol of $C$ to be extracted, and then decode sequentially until 
reaching the symbol encoding $\D[y]$.  

\paragraph{WSLP}
The {\em Word-oriented SLP} (\wslp) is variant of $\slp$ designed for
this paper. It follows the same principles, but it adapts the structures and
algorithms to perform on a word-based (i.e., integer) representation instead of a character-based one.
\wslp\ also preprocesses $\D$ to transform it into the integer sequence $\D_w$. In this 
case, all the different words in the text collection are appended as terminals to the set of 
rules. Another difference is that \wslp\ does not use $q$-gram indexes, because of the space that 
would require, so prefix and suffix searches are directly performed on rows and columns of 
the $LBR$.

\subsection{LZ-based self-indexes}

LZ-based self-indexes build on an LZ77-like parsing, such as LZ77 itself or 
LZ-End (see Section~\ref{sec:lzend}). In short, an LZ77-like parsing of a text collection 
$\D[1,n]$ is a sequence $Z[1,n']$ of phrases such that $\D=Z[1]Z[2] \ldots Z[n']$. Each 
phrase encodes the first occurrence of a text substring and concatenates a {\em source} 
(a maximal substring previously seen in $\D$) and a trailing 
character.
Both \lzindex\ and \lzendindex\ \cite{KNtcs12} use the following
data structures to
encode $Z$ for efficient pattern searching and snippet extraction:

\begin{itemize}
	\item $S[1,n+n']$ is a bitmap that encodes the structure of the phrase sources.
	  We traverse the text from $\D[1]$ to $\D[n]$ and encode the number of sources that
	  start at each position: if $k$ sources start from $\D[i]$, we append 
	  $1^k0$ to $S$ (note that $k$ may be $0$). We consider that the empty-string sources start just before $\D[1]$. Thus, 0-bits 
	  encode text positions, whereas 1s encode the succesive sources. $S$ is stored in
	  gap-encoded form to exploit its sparseness.
	\item $\pi[1,n']$ is a permutation that maps phrases to sources, that is, $\pi[i]=j$
	  means that the source of the $i$-th phrase starts at the position corresponding to 
	  the $j$-th 1 in $S$. This is implemented using the approach from Munro
	  et al.~\cite{MRRR03}, and it is one of the structures tuned in Section~\ref{exp:pos:selfindexes}.
	\item $B[1,n]$ is a gap-encoded bitmap that marks the positions of $\D$ 
	  where each phrase ends.
	\item $L[1,n']$ is an array encoding the trailing characters added at the end of each phrase. 
\end{itemize}

We report each occurrence of a pattern in time $O(\log n')$. For pattern
searching, we also distinguish {\em primary} 
and {\em secondary occurrences}. In this case, we consider primary occurrences those overlapping more 
than one phrase or ending at a phrase boundary.

A relation analogous to $LBR$, of size $n' \times n'$, is used to find the 
primary occurrences. It stores a pair $(i,j)$ if the $i$-th phrase (in reverse
lexicographic order) is followed by the suffix starting at the $j$-th phrase 
(in lexicographic order of those suffixes). Thus, if we partition $P=P_< P_>$ 
as for SLPs, and find the ranges $[l_1,l_2]$ of phrases terminated with $P_<$ 
and $[r_1,r_2]$ of phrase-aligned suffixes starting with $P_>$, then all the 
points in $[l_1,l_2] \times [r_1,r_2]$ are primary occurrences. We find
them all by trying the $m-1$ possible partitions $P=P_< P_>$. Since this 
relation has no labels, we implement it as a wavelet tree $R[1,n']$, with
$R[i]=j$ if $(i,j)$ is a point (note there is only one $j$ per $i$ 
value). This provides $O(\log n')$-time access to any $R[i]$ and $R^{-1}[j]$.

Since accessing the strings is generally more expensive from the LZ-index
structures than from SLPs, it is worth replacing the binary searches on the
strings by trie structures built on the reversed phrases or the phrase-aligned
suffixes. In this second trie, the leaves store the identifiers of the 
corresponding phrases, $id[1,n']$. The tries are implemented as
{\em Patricia trees} \cite{Pat68}, succintly encoded as labeled trees using 
{\em dfuds} \cite{BDMRRR05}. The characters labeling the edges are stored  in
plain form, and the skip values of the Patricia tree are stored using
{\em Directly Addressable Codes} (DACs) \cite{BLNipm12}, since most skip values
are small. Once we arrive at the node corresponding to $P_>$, covering the
leaf range $[r_1,r_2]$, the corresponding phrase numbers are found in 
$id[r_1,r_2]$. An analogous Patricia tree can be stored to search for $P_<$,
but this time the identifiers $rid[1,n']$ need not be stored, as we can find 
them using $rid[j] = id[R^{-1}[j]]-1$.

Before delivering the primary occurrences associated with $P_< P_>$, the 
Patricia trees require a validation, because all or none of the strings they
find may be actual occurrences. Thus we extract one of the reported results 
and compare it with $P$ in order to determine that either all or none of the 
answers are valid. This costs $O(mh)$ time for \lzindex\ and $O(m+h)$ for 
\lzendindex, where $h$ is the length of the largest phrase in the parsing. 
If the occurrences are valid, then for each occurrence with partition 
$P_< = p_1 \ldots p_k$ found at $id[i]=j$, its original position in the text 
is $select_1(B,j-1)-k+1$.

The secondary occurrences triggered by each primary occurrence $\D[t,t+m-1]$
are found using $S$, $B$, and $\pi$. Every $1$ in $S$ before the $t$-th $0$ is
a phrase starting at $t$ or earlier. If it is the $j$-th $1$, then it is
copied to the $i$-th target phrase, for $i=\pi^{-1}[j]$. Its length is
$select_1(B,j)-select_1(B,j-1)$, and with this we know whether the source
covers the primary occurrence, and where is it copied. We report that new
(secondary) occurrence and also recursively find new secondary occurrences from
that one. 

To extract snippets $\D[x,y]$ we also use $S$, $B$, $\pi$, and $L$. With $B$ we
find the last phrase covering position $y$, and then extract its contents, plus
those of preceding phrases, until covering the area to extract. Each final 
character is found in $L$, and the rest is recursively obtained from the source
of the phrase. The source of the $i$-th phrase corresponds to the $j$-th $1$ in
$S$, for $j=\pi[i]$, and its starting position in $\D$ is found with $S$. 
For a snippet of length $l$, the extraction
takes time $O(l + h)$ for \lzendindex\ and $O(lh)$ for \lzindex.

In Section~\ref{exp:pos:selfindexes} we study five different configurations of
this index, regarding whether we use Patricia trees or binary search for each
dimension of the binary relation, and whether or not we store $rid$ explicitly.

\begin{comment}
The bitmap storing the positions where each document begins is stored as a set 
of pointers in the case of \repair, solving $select_1$ in $O(1)$ time and 
$rank_1$ by performing a binary search. In the case of \lzend\ a delta-encoded
 version \cite{KNtcs12} is used, achieving logarithmic time.
%OJO no deberia usarse nada, ya que los indices dan posiciones absolutas
\end{comment}

\end{document}